%% file: thesis.tex
\documentclass[12pt, a4paper,openany,oldfontcommands]{memoir}

\usepackage{graphicx}
\usepackage{epsfig}
\usepackage{amsmath}
\usepackage{amssymb}
\usepackage{amsthm}
\usepackage{booktabs}
\usepackage{url}
\usepackage{caption}
\usepackage{longtable}
\usepackage[figuresright]{rotating}
\usepackage{latexsym}
\usepackage{textcomp}
\usepackage[vcentermath]{youngtab}
\usepackage{type1cm}
\usepackage{eso-pic}
\usepackage{mathrsfs}
\DisemulatePackage{setspace}
\usepackage{setspace}
\usepackage{graphics}
\usepackage{times}
\usepackage[]{geometry}
\usepackage{lipsum}
\usepackage{cite}
\usepackage{rotating}
\usepackage{float}
\usepackage{bookmark}
\usepackage{ctable}
\usepackage{hhline}
\usepackage{hyperref}
\usepackage[sort&compress,numbers]{natbib}

\usepackage[normalem]{ulem}  

\def\bea{\begin{eqnarray}}
\def\eea{\end{eqnarray}}
\DeclareMathAlphabet\mathbfcal{OMS}{cmsy}{b}{n}

\makeindex

\usepackage{xcolor}
\definecolor{mycolor}{cmyk}{.4,1,0,0}    
\definecolor{mycolor1}{cmyk}{1,.4,0,.45}
\definecolor{mycolor2}{cmyk}{0,1,1,.2}
\definecolor{greenyellow}   {cmyk}{0.15, 0   , 0.69, 0   }
\definecolor{yellow}        {cmyk}{0   , 0   , 1   , 0   }
\definecolor{goldenrod}     {cmyk}{0   , 0.10, 0.84, 0   }
\definecolor{dandelion}     {cmyk}{0   , 0.29, 0.84, 0   }
\definecolor{apricot}       {cmyk}{0   , 0.32, 0.52, 0   }
\definecolor{peach}         {cmyk}{0   , 0.50, 0.70, 0   }
\definecolor{melon}         {cmyk}{0   , 0.46, 0.50, 0   }
\definecolor{yelloworange}  {cmyk}{0   , 0.42, 1   , 0   }
\definecolor{orange}        {cmyk}{0   , 0.61, 0.87, 0   }
\definecolor{burntorange}   {cmyk}{0   , 0.51, 1   , 0   }
\definecolor{bittersweet}   {cmyk}{0   , 0.75, 1   , 0.24}
\definecolor{redorange}     {cmyk}{0   , 0.77, 0.87, 0   }
\definecolor{mahogany}      {cmyk}{0   , 0.85, 0.87, 0.35}
\definecolor{maroon}        {cmyk}{0   , 0.87, 0.68, 0.32}
\definecolor{brickred}      {cmyk}{0   , 0.89, 0.94, 0.28}
\definecolor{red}           {cmyk}{0   , 1   , 1   , 0   }
\definecolor{orangered}     {cmyk}{0   , 1   , 0.50, 0   }
\definecolor{rubinered}     {cmyk}{0   , 1   , 0.13, 0   }
\definecolor{wildstrawberry}{cmyk}{0   , 0.96, 0.39, 0   }
\definecolor{salmon}        {cmyk}{0   , 0.53, 0.38, 0   }
\definecolor{carnationpink} {cmyk}{0   , 0.63, 0   , 0   }
\definecolor{magenta}       {cmyk}{0   , 1   , 0   , 0   }
\definecolor{violetred}     {cmyk}{0   , 0.81, 0   , 0   }
\definecolor{rhodamine}     {cmyk}{0   , 0.82, 0   , 0   }
\definecolor{mulberry}      {cmyk}{0.34, 0.90, 0   , 0.02}
\definecolor{redviolet}     {cmyk}{0.07, 0.90, 0   , 0.34}
\definecolor{fuchsia}       {cmyk}{0.47, 0.91, 0   , 0.08}
\definecolor{lavender}      {cmyk}{0   , 0.48, 0   , 0   }
\definecolor{thistle}       {cmyk}{0.12, 0.59, 0   , 0   }
\definecolor{orchid}        {cmyk}{0.32, 0.64, 0   , 0   }
\definecolor{darkorchid}    {cmyk}{0.40, 0.80, 0.20, 0   }
\definecolor{purple}        {cmyk}{0.45, 0.86, 0   , 0   }
\definecolor{plum}          {cmyk}{0.50, 1   , 0   , 0   }
\definecolor{violet}        {cmyk}{0.79, 0.88, 0   , 0   }
\definecolor{royalpurple}   {cmyk}{0.75, 0.90, 0   , 0   }
\definecolor{blueviolet}    {cmyk}{0.86, 0.91, 0   , 0.04}
\definecolor{periwinkle}    {cmyk}{0.57, 0.55, 0   , 0   }
\definecolor{cadetblue}     {cmyk}{0.62, 0.57, 0.23, 0   }
\definecolor{cornflowerblue}{cmyk}{0.65, 0.13, 0   , 0   }
\definecolor{midnightblue}  {cmyk}{0.98, 0.13, 0   , 0.43}
\definecolor{navyblue}      {cmyk}{0.94, 0.54, 0   , 0   }
\definecolor{royalblue}     {cmyk}{1   , 0.50, 0   , 0   }
\definecolor{blue}          {cmyk}{1   , 1   , 0   , 0   }
\definecolor{cerulean}      {cmyk}{0.94, 0.11, 0   , 0   }
\definecolor{cyan}          {cmyk}{1   , 0   , 0   , 0   }
\definecolor{processblue}   {cmyk}{0.96, 0   , 0   , 0   }
\definecolor{skyblue}       {cmyk}{0.62, 0   , 0.12, 0   }
\definecolor{turquoise}     {cmyk}{0.85, 0   , 0.20, 0   }
\definecolor{tealblue}      {cmyk}{0.86, 0   , 0.34, 0.02}
\definecolor{aquamarine}    {cmyk}{0.82, 0   , 0.30, 0   }
\definecolor{bluegreen}     {cmyk}{0.85, 0   , 0.33, 0   }
\definecolor{emerald}       {cmyk}{1   , 0   , 0.50, 0   }
\definecolor{junglegreen}   {cmyk}{0.99, 0   , 0.52, 0   }
\definecolor{seagreen}      {cmyk}{0.69, 0   , 0.50, 0   }
\definecolor{green}         {cmyk}{1   , 0   , 1   , 0   }
\definecolor{forestgreen}   {cmyk}{0.91, 0   , 0.88, 0.12}
\definecolor{pinegreen}     {cmyk}{0.92, 0   , 0.59, 0.25}
\definecolor{limegreen}     {cmyk}{0.50, 0   , 1   , 0   }
\definecolor{yellowgreen}   {cmyk}{0.44, 0   , 0.74, 0   }
\definecolor{springgreen}   {cmyk}{0.26, 0   , 0.76, 0   }
\definecolor{olivegreen}    {cmyk}{0.64, 0   , 0.95, 0.40}
\definecolor{rawsienna}     {cmyk}{0   , 0.72, 1   , 0.45}
\definecolor{sepia}         {cmyk}{0   , 0.83, 1   , 0.70}
\definecolor{brown}         {cmyk}{0   , 0.81, 1   , 0.60}
\definecolor{tan}           {cmyk}{0.14, 0.42, 0.56, 0   }
\definecolor{gray}          {cmyk}{0   , 0   , 0   , 0.50}
\definecolor{black}         {cmyk}{0   , 0   , 0   , 1   }
\definecolor{white}         {cmyk}{0   , 0   , 0   , 0   }

\setlrmargins{*}{*}{1}
\setulmargins{*}{*}{1}
\setmarginnotes{5mm}{45.23mm}{\onelineskip}
\checkandfixthelayout

\makechapterstyle{mychapterstyle}{%
    \renewcommand{\chapnamefont}{\color{black}\LARGE\rmfamily\bfseries}%
    \renewcommand{\chapnumfont}{\LARGE\rmfamily\bfseries}%
    \renewcommand{\printchapternum}{%
        \chapnumfont\thechapter%
        }%
}

\makeatletter
\newcommand\thickhrulefill{\leavevmode \leaders \hrule height 1ex \hfill \kern \z@}
\makechapterstyle{VZ14}{

\renewcommand\chapnamefont{\Large\scshape}
\renewcommand\printchapternum{%
\chapnamefont\null\thickhrulefill\quad
\@chapapp\space\thechapter\quad\thickhrulefill}

}
\makeatother
\chapterstyle{VZ14}

\setsecheadstyle{\color{black}\Large \bfseries}
\setsubsecheadstyle{\color{black}\Large\bfseries}
\setsubsubsecheadstyle{ \color{black}\normalfont\bfseries}
\setparaheadstyle{\normalfont\sffamily}
\makeevenhead{headings}{\thepage}{}{\footnotesize\bfseries\leftmark}
\makeoddhead{headings}{\footnotesize\bfseries\rightmark}{}{\thepage}

\setsecnumdepth{subsection}
 \maxsecnumdepth{subsection}
\settocdepth{subsection}
\maxtocdepth{subsection}

\begin{document}
\doublespacing
\pagenumbering{gobble}
\noindent
\begin{center}
\Huge
   \textsc{\bfseries Constraining the density dependence of symmetry energy \\ using mean field models}
\end{center} 

\begin{center}
\vskip 0.70cm
{\bf {\em By}} 
\vskip -0.2cm
{\bf {\large CHIRANJIB MONDAL}}
\vskip 0.0cm
{\bf {\large PHYS05201204016 }}
\vskip 0.5cm
{\bf {\large Saha Institute of Nuclear Physics, Kolkata}}
\vskip 0.cm
{ {\em {\large A thesis submitted to the
\vskip 0.05cm
Board of Studies in Physical Sciences
\vskip 0.05cm
In partial fulfillment of requirements
\vskip 0.05cm
For the Degree of 
}}}
\vskip 0.05cm
{\bf {\large DOCTOR OF PHILOSOPHY}}
\vskip 0.1cm
{ {\em of}}
\vskip 0.1cm
{\bf {\large HOMI BHABHA NATIONAL INSTITUTE}}
\vfill
\begin{figure}[hbt]
\begin{center}
\includegraphics[scale=0.25]{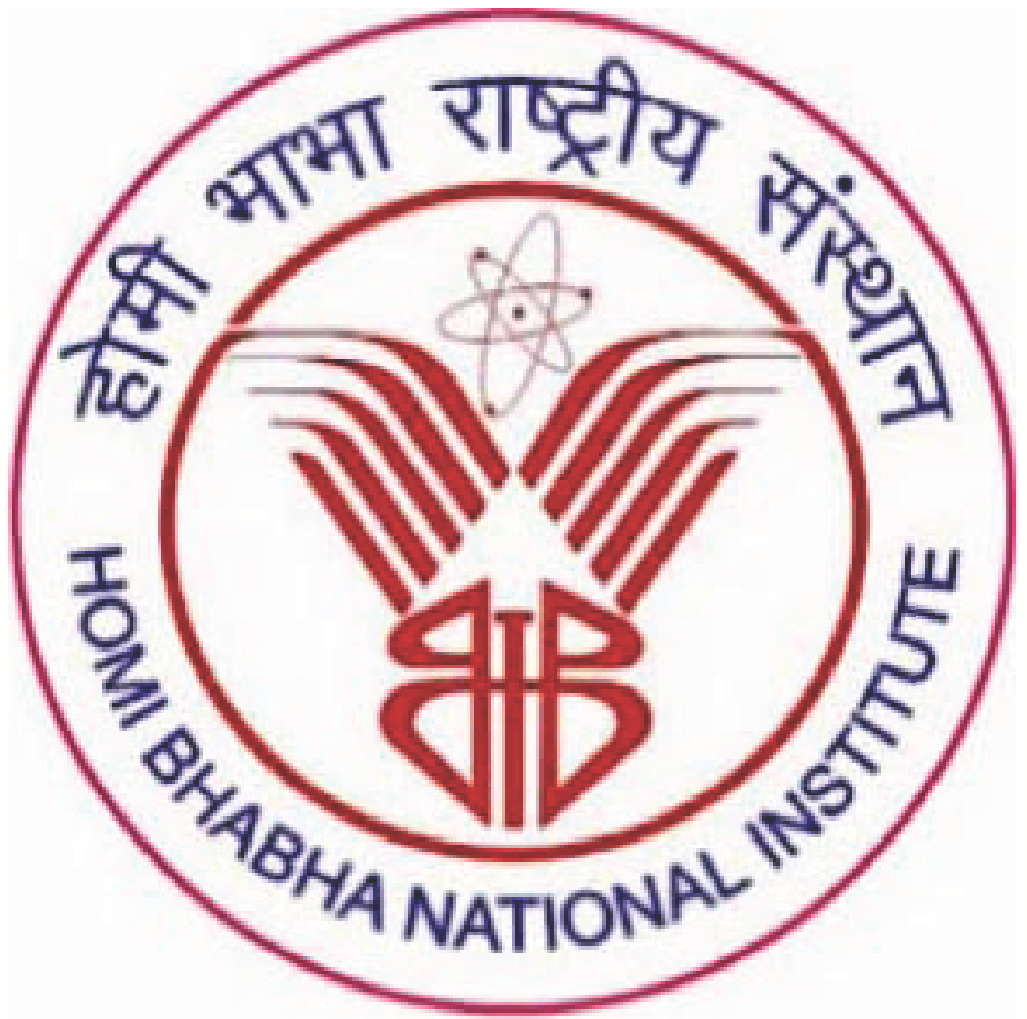}
\end{center}
\end{figure}
%
%
{\bf {\large October, 2017}}
\vfill
\end{center}
\cleardoublepage

\newpage

%
%
\centerline{{\bf{\LARGE Homi Bhabha National Institute}}}
\vskip 0.2cm
\centerline{{\bf {\large Recommendations of the Viva Voce Committee}}}
\vskip 0.2cm
As members of the Viva Voce Committee, we certify that we have read the
dissertation prepared by {\bf{Chiranjib Mondal}} entitled {\bf{``Constraining the density dependence of symmetry energy 
using mean-field models"}} and
recommend that it maybe accepted as fulfilling the thesis requirement for the award of Degree of Doctor of Philosophy.
\vskip 0.05cm
\underline{\hspace{12.0cm}} Date:
\vskip -0.1cm 
Chairman - Prof. Munshi Golam Mustafa
\vskip 0.05cm
\underline{\hspace{12.0cm}} Date:
\vskip -0.1cm 
Guide / Convener - Prof. Bijay Kumar Agrawal
\vskip 0.05cm
\underline{\hspace{12.0cm}} Date:
\vskip -0.1cm 
Co-guide - (if any) 
\vskip 0.05cm
\underline{\hspace{12.0cm}} Date:
\vskip -0.1cm 
External Examiner - Prof. Praveen C. Srivastava 
\vskip 0.05cm
\underline{\hspace{12.0cm}} Date:
\vskip -0.1cm 
Member 1 - Prof. Debades Bandyopadhyay 
\vskip 0.05cm
\underline{\hspace{12.0cm}} Date:
\vskip -0.1cm 
Member 2 - Prof. Maitreyee Saha Sarkar 
\vskip -0.1cm
\rule{14cm}{1pt}
\vskip -0.1cm
\hspace{0.2cm} Final approval and acceptance of this thesis is contingent upon the candidate's
submission of the final copies of the thesis to HBNI.
\vskip -0.1cm
\hspace{0.2cm} I/We hereby certify that I/we have read this thesis prepared under my/our
direction and recommend that it may be accepted as fulfilling the thesis requirement.

\vskip 0.0cm 
{\bf Date:} 
\vskip -0.2cm 
{\bf {Place:} \hspace{6cm} Guide:{\underline{ \hspace{5.0cm}}}
\hspace{8.65cm}}
\cleardoublepage

\newpage
\centerline{{\bf {\large STATEMENT BY AUTHOR}}}
\vskip 1.00cm
%
%
This dissertation has been submitted in partial fulfillment of
requirements for an advanced degree at Homi Bhabha National Institute
(HBNI) and is deposited in the Library to be made available to borrowers
under rules of the HBNI.
\vskip 0.6cm
Brief quotations from this dissertation are allowable without special
permission, provided that accurate acknowledgement of source is made.
Requests for permission for extended quotation from or reproduction of
this manuscript in whole or in part may be granted by the Competent
Authority of HBNI when in his or her judgment the proposed use of the
material is in the interests of scholarship. In all other instances,
however, permission must be obtained from the author.

\vskip 1.5cm


$~$\hspace{11.2cm}Chiranjib Mondal
\newpage
\cleardoublepage
\vskip 1.2cm
\centerline{{\bf{\large{DECLARATION}}}}
\vskip 1.2cm
I, hereby declare that the investigation presented in the thesis has been
carried out by me. The work is original and has not been submitted
earlier as a whole or in part for a degree / diploma at this or any
other Institution / University.
\vskip 2.0cm
%
%
\rightline{Chiranjib Mondal \hspace{0.9cm}}
%
\newpage
\cleardoublepage
\chapter*{List of Publications Arising from the Thesis}
\phantomsection
\begin{flushleft}
 \textbf{Peer reviewed journal:}
\end{flushleft}

\begin{enumerate}
\item Constraining the symmetry energy content of nuclear matter from nuclear masses: A covariance analysis.\\
 Chiranjib Mondal, Bijay Agrawal and J. N. De;\\
\textit{Physical Review C},  \textbf{2015}, \textit{92, 024302} \href{https://arxiv.org/pdf/1507.05384.pdf}{[arXiv:1507.05384]}
\item Sensitivity of elements of the symmetry energy of nuclear matter to the properties of neutron-rich systems.\\
 Chiranjib Mondal, Bijay Agrawal, J. N. De and S. K. Samaddar;\\
\textit{Physical Review C},  \textbf{2016}, \textit{93, 044328} \href{https://arxiv.org/pdf/1603.08645.pdf}{[arXiv:1603.08645]}
\item Model dependence of the neutron-skin thickness on the symmetry energy.\\
 Chiranjib Mondal, B. K. Agrawal, M. Centelles, G. Col\`o, X. Roca-Maza, N. Paar, X. Vi\~nas, S. K. Singh and S. K. Patra;\\
\textit{Physical Review C}, \textbf{2016}, \textit{93, 064303} \href{https://arxiv.org/pdf/1605.05048.pdf}{[arXiv:1605.05048]}
\item Interdependence of different symmetry energy elements.\\
 Chiranjib Mondal, B. K. Agrawal, J. N. De, S. K. Samaddar, M. Centelles \\and X. Vi\~nas;\\
\textit{Physical Review C (Rapid Communication)}, \textbf{2017}, \textit{96, 021302} 
\href{https://arxiv.org/pdf/1708.03846.pdf}{[arXiv:1708.03846]}
\end{enumerate}
\vspace{0.001cm}

\begin{flushleft}
 \textbf{Contribution other than thesis}:
\end{flushleft}

\begin{enumerate}
\item Limiting symmetry energy elements from empirical evidence.\\
 Bijay Agrawal, J. N. De, S. K. Samaddar, Chiranjib Mondal and Subhranil De;\\
\textit{International Journal of Modern Physics E},  \textbf{2017}, \textit{26, 1750022}
\href{https://arxiv.org/pdf/1703.03549.pdf}{[arXiv:1703.03549]}
\end{enumerate}

\vspace{0.001cm}
\begin{flushleft}
 \textbf{Conference proceeding}:
\end{flushleft}

\begin{enumerate} 
\item  {Model dependence in the density content of nuclear symmetry energy}\\ 
Chiranjib Mondal, S. K. Singh, B. K. Agrawal, M. Centelles, G. Col\`o, X. Roca-Maza, N. Paar, S. K. Patra and X. Vi\~nas; \\ 
\textit{Proceedings of the DAE-BRNS Symp. on Nucl. Phys.}, \textbf{2014}, \href{https://www.sympnp.org/proceedings/59/A9.pdf}{59, 66-67}.
\item  {Information content of nuclear masses: A covariance analysis}\\
Chiranjib Mondal, Bijay Agrawal and J. N. De;\\
\textit{Proceedings of the DAE-BRNS Symp. on Nucl. Phys.}, \textbf{2015}, \href{https://www.sympnp.org/proceedings/60/A1.pdf}{60, 56-57}.
\item  {Sensitivity analysis of optimized nuclear energy density functional}\\
Chiranjib Mondal, Bijay Agrawal, J. N. De and S. K. Samaddar;\\
\textit{Proceedings of the DAE-BRNS Symp. on Nucl. Phys.}, \textbf{2016}, \href{https://www.sympnp.org/proceedings/61/A3.pdf}{61, 66-67}.
\end{enumerate}


\vspace{1cm}

\begin{flushright}
Chiranjib Mondal
\end{flushright}

\cleardoublepage

\newpage
\cleardoublepage
\vskip 2 cm
\rightline{\bf \Large To\ \ \   My Family}
\rightline{\bf \Large and\ \ \ Ajit da ...}
\cleardoublepage
\vskip 1.0cm

\centerline{{\bf{\large ACKNOWLEDGEMENTS}}}
\vspace{1cm}

\vskip 0.5cm
%
Probably no gratitude is enough for the constant support provided by Prof. Bijay Kumar
Agrawal, who has been the supervisor for this project. I was introduced to the subject of
theoretical nuclear physics by him. The amount of inquisitiveness and enthusiasm he has
to offer towards the subject is truly inspiring. I am forever in debt to him for making me
love the subject. The relationship with him was never limited to the discussion of Physics.
The biggest lesson I learned from him is that the path of morale and hardship is always the
right path in life. I also acknowledge the fatherly guidance and care of Prof. Jadunath De
and Prof. Santosh Samaddar, with whom I was so privileged to collaborate for a number of
occasions. The wisdom and the stories what they shared, will always bring a joyous smile
in my face. I also acknowledge the help provided by Mrs. Tanuja Agrawal during any
technical difficulties. She has been a true inspiration for solving the puzzle called `life'.

I take great pleasure to express my gratitude towards Prof. Xavier Vi ̃nas and Prof.
Mario Centelles for collaborating at different phases of this thesis work, who had also
hosted me at University of Barcelona offering very warm hospitality during April, 2017.
I take this opportunity to thank all my other collaborators Prof. Gianluca Col`o, Xavier
Roca-Maza, Nils Paar, Prof. Suresh Kumar Patra, Shailesh Kumar Singh and Subhranil
De. Name of Tuhin Malik needs a special mention, as he is not only a collaborator but also
a good friend. I also want to thank the members of my doctoral committee Prof. Debades
Bandyopadhyay, Prof. Munshi Golam Mustafa and Prof. Maitreyee Saha Sarkar for giving
very useful suggestions. I also acknowledge the careful guidance of Prof. Asimananda
Goswmai during my Post-MSc project.

I gratefully acknowledge the financial support provided by Department of Atomic Energy, 
Govt. of India throughout the whole project. The support was also extended for
attending several school, workshop and conferences.

I started the journey of my PhD with a very vibrant set of people at Saha Institute
of Nuclear Physics (SINP) back in 2012, the so called ‘Post MSc’ batch. It has been a
memorable last five years with these wonderful people. I thank all of you from the bottom
of my heart. I am sure the journey with most of you will be life-long. Achyut da, Kuntal da,
Suvankar, Gouranga da, I will cherish the friendship what we have developed during last
five years and beyond. Naosad da, Satyajit, Pankaj, Sayanee, Sanjib da, Sukanta, Mily and
all the other students of our ’Post MSc’ year, I acknowledge the presence of you around me
whenever I needed it. Thank you for all those trips together, what brought a lot of oxygen
into life.

I thank all the present and past members of ‘Theory Division’, who actively helped in
this project. To name a few, Parijat di, Prasanta da, Satya da, Goutam da, thank you for
your elderly suggestions at the time of my need. Thank you Aminul da for sharing your
thoughts about life. I also want to record my appreciation for Kumar da, Augniva, Avik
(Jr), Aranya, Sukannya, Mugdha and all the other members of 3319. A significant part
of the numerical calculations of this project was carried out by using the computer cluster
facility of Theory Division, SINP. I also appreciate the official jobs taken care by Pradyut
da, Dola di and Sangita di and the other non-academic members of the Theory division. I
heartily recognize the friendships what I have developed with people outside the Theory
division, to name a few, Samrat, Abhishek, Sudesna, Maireyee, Samik and many others.
Suparna, I will cherish all the wonderful conversations we had during last two years or so.

Life has blessed me with lot of good friends, who often showed the best ways to deal
with a problem, becoming the best teachers. Ananta, Suchandan, Amitava and Buddha,
without you life would not be this beautiful. Gagan and Arshiya, it has been a pleasure having you in my life.

Now comes the members of MSA-II, with whom I spent the last five years of my life.
I have very fond memories of those birthday celebrations with Kuntal, Amrita, Barnamala,
Binita, Sanjukta, Tirthankar and Aritra. Discussions with Tirthankar (Tirtha) probably
brought out the best of my way of life. Whatever doubts or questions I used to have, Tirtha
always had the best words to describe it. Sanjukta, I always dream to be as compassionate
as you. I have no other words to describe you. Aritra, I will cherish your witty silences and
silly jokes. Tirtha, Sanjukta and Aritra, it was one hell of a journey together. Kuntal, 
the memories we created together won't fade easily.

I express my deepest love and respect towards my family for being at my
side, no matter what paths I have chosen so far in my life. Believe me,
{\it Baba-Ma}, I wish I could show my admiration more! {\it Didibhai},
thank you for all your support.

Before I finish, I want to pay my tribute to {\it Ajit da}. Your words 
still inspire me, ``{\it sobai bole choroibeti, choroibeti! Ami boli 
thamte sekh.}"

\vskip 1.0cm
\rightline{Chiranjib Mondal \hspace{0.9cm}}
\newpage
\pagenumbering{roman}
\pagestyle{plain}
\setcounter{page}{1}

\normalfont
\clearpage
\begin{KeepFromToc}
  \tableofcontents
\end{KeepFromToc}
\clearpage


\chapter*{Synopsis} 
\addcontentsline{toc}{chapter}{Synopsis}

Apart from a very few light nuclei, all terrestrial finite nuclei are
asymmetric. In other words, for most of the finite nuclei, number of
neutrons is higher than protons. Competition between Coulomb
energy and symmetry energy makes them asymmetric. On the other
extreme, astrophysical objects like neutron stars are also highly asymmetric.
However, the reason behind asymmetry in neutron star is attributed
to charge neutrality and beta equilibrium of the system. The density
associated with the center of the nucleus is very close to saturation
density ($\rho_0=0.16$ fm$^{-3}=2.7\times10^{14}$ gm/cm$^3$) of
infinite nuclear matter. Density at the core of a neutron star
is four to five times $\rho_0$. The symmetry energy
controls the radii of neutron stars, the thicknesses of their
crusts, the rate of cooling of neutron stars, and the properties of
nuclei involved in r-process nucleosynthesis.
Studying symmetry energy and its density dependence over a wide
range of density is thus a major topic of research for past few
decades. Presently, several laboratories around
the world are set up to test the limits of stability of nuclei towards
the neutron drip-line or super-heavy region. A precise understanding of
density dependence of symmetry energy can facilitate to explore new
areas of research, which might help to understand the isovector part of the
effective nucleon-nucleon interaction inside the nucleus, which
is still not known accurately.

Density dependence of symmetry energy can be characterized
essentially by three quantities, namely, symmetry energy $J$,
slope parameter $L$ and curvature parameter $K_{sym}$;
all of these quantities pertain to infinite nuclear matter at the
density $\rho_0$.
To find reliable constraints on $J$, $L$ or $K_{sym}$ one needs to
relate them with experimental observables of finite nuclei or
neutron stars since infinite nuclear matter can not be accessed in laboratories.
Due to computational limitations, starting from
no-core shell model, finding even the ground state properties of
finite nuclei e.g. binding energy, charge radii etc beyond
$^{40}$Ca is yet far fetched. Over the years mean
field models with very few parameters thus became a viable
alternative to calculate the properties of finite nuclei
spanning the entire periodic table as well as of neutron stars.

Fitting few thousand observed nuclear masses within a finite range
droplet model (FRDM) or taking double differences of nuclear masses
the estimated value of symmetry energy $J$ at saturation is $\sim$ 32 MeV
with an accuracy of 1-2 MeV \cite{Moller12, Jiang12}. Droplet model (DM) suggests that
neutron-skin thickness $\Delta r_{np}$ (difference between root
mean square radii of neutron and proton distribution) of a heavy
nucleus is linearly correlated to the slope parameter $L$ \cite{Myers69, Myers80}. This
correlation was verified by using a representative set of relativistic
and non-relativistic mean field models. There have been several attempts
to measure the neutron-skin thickness of $^{208}$Pb. However, the
most model independent measurement at Jefferson lab (Lead
radius experiment or PREX), based on weak interaction, predicts a value with
very large uncertainty \cite{Horowitz01, Abrahamyan12, Dubach89}.
There have also been attempts to look for alternative
isovector probes e.g isovector
giant dipole resonance (IVGDR) \cite{Tamii11, Rossi13, Roca-Maza13, 
Roca-Maza15}, isospin diffusion \cite{Chen05},
$\pi^+-\pi^-$ ratio etc. The uncertainty
associated with the value of slope parameter $L$ still remains large.
The shrouds of uncertainty looms even larger when
one tries to constrain curvature parameter $K_{sym}$.

In this dissertation, our primary goal is to constrain the density
dependence of symmetry energy by obtaining tighter bounds on $L$ and
$K_{sym}$ using some relativistic and non-relativistic mean
field models. The investigation has been carried out by using two
different methods. Firstly, covariance analysis is employed to
study the relations among different experimental
observables and parameters of a relativistic mean field model.
We attempt to constrain the density dependence of symmetry energy
by incorporating in the fit-data the binding energies of some highly
asymmetric nuclei; binding energies of finite nuclei are the
most accurately measured quantities in nuclear physics.
Secondly, using different mean field models existing
in the literature, we tried to explore new model independent
correlations among different isovector sensitive quantities.

In the study based on covariance method, we observed that parameters
of mean field models obtained by fitting binding energies and
charge radii of few closed shell nuclei predict a wide range of
values for the slope parameter $L$ \cite{Dutra12, 
Dutra14, Centelles09}. However, macroscopic FRDM model obtained
by fitting binding energies of few thousand nuclei predicts a
quite restricted value of $L$ \cite{Moller12}. Inspired by this
result we incorporated for the first time binding energies
of some highly asymmetric nuclei ($^{24}$O, $^{30}$Ne), where
the neutron number is twice to that of protons, in the fit-data
in order to optimize the parameters of a relativistic mean field
model. Our detailed investigation clearly reveals that the inclusion of highly
asymmetric nuclei in the fitting protocol reduces the
uncertainty on the symmetry energy elements significantly.
A sensitivity analysis is performed by including further in
the fitting protocol the binding energies of few more highly asymmetric
nuclei ($^{36}$Mg, $^{58}$Ca) together with the measured
maximum mass of neutron star \cite{Mondal16a}. Such an analysis reveals quantitatively the
sensitivity of binding energies of highly asymmetric nuclei to
the symmetry energy parameters. It also shows that maximum mass
of neutron star has some sensitivity to the symmetry energy
parameters.

Using a representative set of different mean-field models,
we found that correlation between neutron-skin thickness
$\Delta r_{np}$ of $^{208}$Pb with slope parameter $L$ is
model dependent \cite{Mondal16b}. A model independent correlation
was found between slope parameter $L$ and bulk part of the neutron
skin thickness $\Delta r_{np}^{bulk}$ of $^{208}$Pb conjectured by DM.
Models from different families
predicting similar values of $L$, show a variation in $\Delta 
r_{np}$ which is few times higher than what is predicted by DM.
We defined an effective value of slope parameter $L_{eff}$ within
local density approximation, pertaining to the average density of a heavy nucleus.
Variation in $L_{eff}$ seems to be in harmony with variation
in $\Delta r_{np}$ predicted by DM.

Having constrained the slope of symmetry energy $L$, we also
explored the possibility of constraining the symmetry curvature
parameter $K_{sym}$. Considering a general form of the density
dependent nucleon nucleon interaction along with the Gibbs-Duhem
relation we found an analytical relation connecting curvature
parameter $K_{sym}$, slope parameter $L$ and symmetry energy $J$.
Using five hundred mean field models both relativistic and
non-relativistic, this correlation was realized between $K_{sym}$
with linear combination of $L$ and $J$. The correlation stood
out further for few realistic as well as finite range Gogny
interactions. The universality in the correlation of $K_{sym}$
with linear combination of $L$ and $J$ strongly suggests that
a tight bound on $K_{sym}$ can be obtained from bounds on $L$ and $J$.

\clearpage
\listoffigures
\clearpage
\listoftables
\cleardoublepage
\pagenumbering{arabic}
\pagestyle{plain}

\setcounter{page}{1}

\include{chap1}
\include{chap2}
\include{chap3}
\include{chap4}
\include{chap5}
\include{chap6}
\include{chap7}
\bibliographystyle{apsrev4-1}
\nocite{apsrev41Control}

%

\end{document}

%% file: chap1.tex
\chapter{Introduction}\label{ch1}
An atom possesses a tiny positive core surrounded by negatively 
charged electrons. This positive core is coined as nucleus. Although, the 
most of the mass of an atom is carried by its nucleus, the size of an 
atom is orders of magnitude higher compared to its nucleus. The length scale 
associated to an atom is few angstroms ($10^{-10}$ meter) whereas size of a 
nucleus is few femtometers ($10^{-15}$ meter). The nucleus not 
only contains positively charged protons, but also electrically neutral 
neutrons. Due to the presence of protons, which are positively charged, 
binding of a nucleus has to overcome Coulomb repulsion. That is why nuclear 
force needs to be very strong in nature. The tiny size of the nucleus 
further suggests that the nuclear force is a short-range one in nature. 
Unlike atoms, a nucleus is a self-bound many-body system. Perturbing 
a nucleus by colliding a particle or by means of electromagnetic 
probes gives rise a plethora of phenomena like rotation, giant collective 
vibration, deformation and unique nuclear phenomena like fission 
or fusion. That is why even after 
a century of discovery of nucleus by Rutherford, understanding 
the nature of nuclear force inside the medium presents various unique 
challenges both in theoretical and experimental studies. 

\begin{figure}[]{}
\centering
\includegraphics[width=0.8\textwidth]{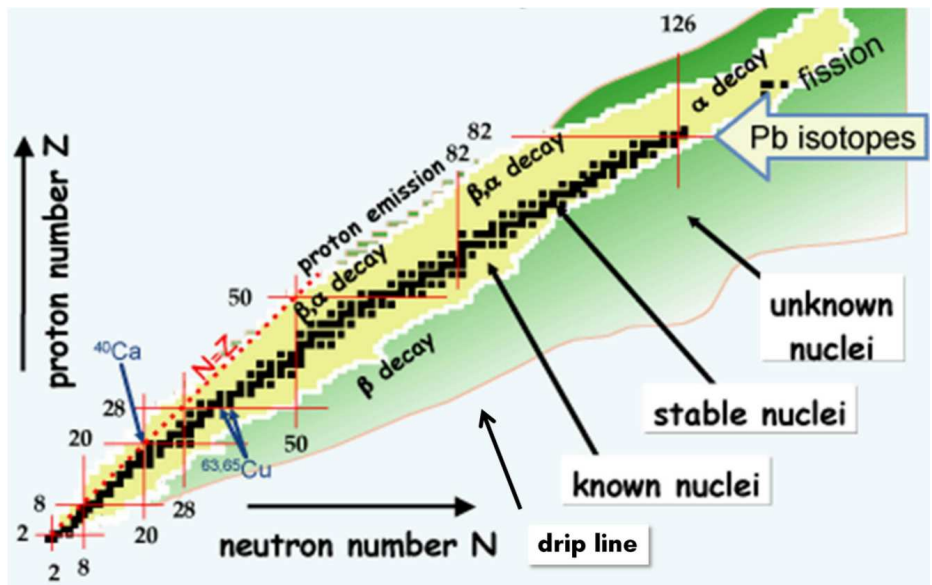}
\caption{\label{ch1_fig1}
Nuclear chart for $\sim6000$ nuclei in the $N$-$Z$ plane. See text for details.}
\end{figure}
Since the discovery of neutron by Chadwick in 1932, the study of nuclei
has become even more fascinating. Based on the number of protons ($Z$)
inside a nucleus, chemical nature of the corresponding atom (element)
changes. However, it is the number of neutrons ($N$) in a nucleus
which plays a decisive role in its binding and eventually determining
the stability and abundance of different isotopes ($A=N+Z$). Based
on the knowledge so far, there exist $\sim118$ different elements with
$\sim300$ stable isotopes. One should keep in mind that by ``stable''
it is meant that the half-life of decay for these nuclei are of the
order of the age of the earth. In the nuclear chart depicted in Fig.
\ref{ch1_fig1}, these stable isotopes are marked as black dots.  Through
various experiments performed over the years, existence of $\sim2700$
unstable nuclei are also found. These unstable nuclei are displayed
by yellow region in Fig. \ref{ch1_fig1}. There exist a few nuclei with
certain number of neutrons and/or protons (magic numbers), which show a
greater amount of stability compared to their neighbours. These nuclei
are called magic nuclei, which are marked with red horizontal (magic
$Z$) and vertical (magic $N$) lines in Fig. \ref{ch1_fig1}. The green
region, largely occupying the neutron-rich region depicts the nuclei
whose existence are predicted by different theoretical conjectures, but
not yet found experimentally.  During the evolution of a star, heavier
nuclei are thought to be formed through rapid neutron-capture process or
r-process, which involves the nuclei lying in the green region shown in
Fig. \ref{ch1_fig1}.  Beyond this green region, the boundary given by red
line depicts the neutron drip line. It signifies that beyond this line
adding or subtracting a neutron from a nucleus does not cost any energy.
The neutron drip line and the red dotted line representing $N=Z$ points
out that apart from a very few light nuclei, most of the terrestrial
finite nuclei are asymmetric.

Upon discovery of neutron star in 1967, a new dimension opened up in the
research of systems made up of nucleons i.e. neutrons and protons.  When a
massive star with mass greater than 10 times the solar mass exhausts its
nuclear fuel, it starts collapsing under gravity resulting in a supernova
explosion \cite{Glendenning00}. The remnant of this explosion further
collapses under gravity and end up in one of the most compact objects
in the universe, called neutron stars.  Due to the enormous amount of
gravitational pull, the electrons inside the stellar matter collide with
the protons forming neutrons. Further gravitational pull tries to bring
these neutrons closer and closer. However, Pauli's exclusion principle
restricts two neutrons to occupy the same quantum state resulting in
an opposing pressure to that of the gravitational pull.  In the steady
state, the matter inside this compact object is in beta equilibrium and
electrically neutral. This results in the primary constituent of these 
compact objects being neutrons and hence the name neutron star.

\begin{figure}[]{}
\centering
\includegraphics[width=0.8\textwidth]{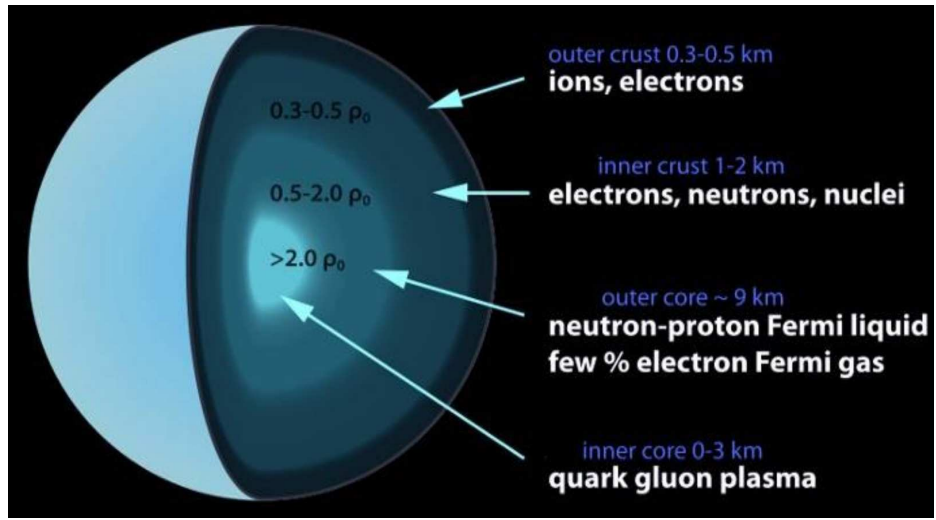}
\caption{\label{ch1_fig2} Structure of a neutron star depicting 
constituents of different layers along with the density and range 
associated with the corresponding layers. The density $\rho_0$ depicted 
in the figure is the saturation density of symmetric infinite nuclear 
matter.}
\end{figure}
The heaviest neutron star observed so far has twice the mass of
sun \cite{Demorest10, Antoniadis13} and radius of about 10 to 15
kilometers. The compactness can be understood by looking into the fact
that the radius of sun is $\sim6.9\times10^5$ kilometers. The generic
features of a typical neutron star is depicted in Fig. \ref{ch1_fig2}.
It is conjectured that at the very core of a neutron star there might
be a possibility of finding the matter in its most fundamental form
i.e. quark-gluon plasma.  The outer core, which constitutes the major
part of the volume of a neutron star is thought be beta-equilibrated,
electrically neutral nuclear matter with a typical ratio of $N/Z\approx6$
and small amount of electron Fermi gas.  The crust of the neutron star
is can be imagined to be made of an inner and outer crust. In the inner
crust, there might be still some possibility of occurring the usual
nuclear reactions what happens inside a burning star. Possibly very
neutron rich nuclei along with free neutrons and electrons are thought
to be the constituents of the inner crust. The outer crust is mainly
made of cold ions and electrons.

Terrestrial finite nuclei and the neutron stars are complex many-body
systems governed by the strong force. Quantum chromodynamics (QCD) is
the most fundamental theory which can explain the properties of these
nuclear systems. However, tremendous amount of challenges are faced when
one tries to solve the theory in non-perturbative regime for a complex
many-body system. Only very recently, the state-of-the-art computational
facilities made it possible for theorists to develop some {\it ab initio}
approaches to calculate the ground state properties of nuclei like binding
energy or charge radius, based on QCD \cite{Barrett13,Hagen14}. However,
it has only been possible for lighter nuclei. To calculate the properties
of finite nuclei as well as of neutron star in the same framework has
only been possible for effective mean-field theories with a few degrees
of freedom.  These theories are developed on the basis of an {\it energy
density functional} (EDF) with few parameters. There are mainly two types
EDFs used in the literature: non-relativistic and relativistic. Skyrme
functional is the most popular EDF in the non-relativistic domain where
the nucleons interact through local effective potentials.  Relativistic
mean field models provide a covariant description of the nuclear system
which is based on quantum field theory. A comprehensive discussion
on the mean-field models can be found in Ref. \cite{Bender03}. The
parameters of the mean-field models are obtained by fitting gross
nuclear properties which contain the many-body correlations. That is
why, even if the mean-field EDFs are constructed by one body densities,
the many-body effects get embedded in the parameters of the model. At a
nominal computational cost, mean-field models based on EDFs provide a
very high accuracy on the global nuclear properties both in the domain
of finite nuclei and neutron stars.

From the discussions of Figs. \ref{ch1_fig1} and \ref{ch1_fig2} it is 
clear that most of the nuclear systems are asymmetric across a scale 
of 18 orders of magnitude (length scale associated to finite nuclei is 
$10^{-15}$ meter and for neutron star it is $10^3$ meter). Inside a nucleus, 
strength of the interaction of a neutron ($n$) - proton ($p$) pair is stronger 
compared to a $n$-$n$ or $p$-$p$ pair. In other words, in the absence of 
any Coulomb force, all finite nuclei would be symmetric with same number 
of neutrons and protons. In reality, there is always a competition between 
the Coulomb energy and the symmetry energy what makes most of
the finite nuclei asymmetric. In neutron star, however, the requirement
of charge neutrality and beta equilibrium is primarily responsible for
the asymmetry. Nuclear symmetry energy, characterized by the variation
of the energy of a nuclear system with the change in the ratio of its
neutron and proton content, plays a crucial role in the binding of the
corresponding nuclear system. Symmetry energy also plays crucial roles
in controlling the radii of neutron stars, the thickness of their crusts
and the properties of nuclei produced in the r-process nucleosynthesis 
\cite{Baldo16}.

The density associated with the center of a nucleus is $\sim$
0.16 fm$^{-3}$ or 2.7$\times 10^{14}$ gm/cm$^3$. Density at the
core of a neutron star is 5-6 times to this density. A microscopic
description of symmetry energy along with its density dependence
over a wide range of density is thus a major topic of research in
nuclear physics. Moreover, across several laboratories nuclei are being
synthesized near the drip-lines. Nuclei near neutron-drip lines carry
valuable informations regarding the r-process nucleosynthesis in the
stellar matter. Understanding of terrestrial nuclei from the conventional
theories often fail to explain the properties of nuclides in these
extremely asymmetric regions. A precise knowledge of symmetry energy and
its density dependence is thus inevitable in order to explore this
region of ``{\it terra-incognita}''. Conversely, inputs from experiments
performed in these asymmetric nuclei hold the key to improve the existing
nuclear theories to be applied to the physics near drip-lines.

To understand the behavior of different nuclear systems, a hypothetical
system is defined which is called infinite nuclear matter. It is a
system of infinite number of neutrons and protons which is uniform in
nature with no boundary and Coulomb interaction. This simplified system
helps to understand the bare nucleon-nucleon interaction. In general,
nuclear matter can be asymmetric. The energy density of such a system
can be decomposed into a symmetric and a purely asymmetric part. The
symmetric matter saturates at a certain density where the energy density
corresponding to it becomes minimum. The density is called as saturation
density $\rho_0$. One should note that the density at the center of a 
heavy nucleus is very close to $\rho_0$. The asymmetric part of the energy is
essentially called the symmetry energy of nuclear matter. Symmetry energy
is mainly characterized by three parameters namely, symmetry energy
coefficient $C_2^0$, its density slope $L_0$ and curvature parameter
$K_{sym}^0$; all of these quantities are defined at $\rho_0$. Proper
definitions of these nuclear matter properties are given in Section
\ref{sec_INM} of Chapter \ref{ch2}.

Constraining quantitatively the symmetry energy parameters $C_2^0$,
$L_0$ and $K_{sym}^0$ has been a major focus of research in present
day nuclear physics.  One should keep in mind that these parameters are
defined in the domain of infinite nuclear matter. As nuclear matter is not
accessible in laboratories one needs to connect these quantities to actual
experimentally measurable entities. These connections often imply finding
correlation between the symmetry energy parameters and some experimentally
measurable quantities using theoretical models. Conventionally, the
correlations can be studied using different models. A classic example
would be studying correlation between $L_0$ and neutron-skin thickness
$\Delta r_{np}$ of a heavy nucleus, where $\Delta r_{np}$ is defined by
the difference between the root mean square radii of the neutron and
proton distributions inside a nucleus \cite {Centelles09}. The models
which are explored in the present thesis work, are primarily based on
mean-field approach. Over the years mean-field models, both relativistic
and non-relativistic, proved to be very successful at very reasonable
computing costs. The accuracies over extracted quantities, however,
are at par with theories based on more fundamental approaches e.g.
{\it ab-initio} method or configuration-interaction method. A theoretical
edifice to calculate the ground state properties like binding energy
and charge radius of closed-shell spherical nuclei using mean-field
models is given in Chapter \ref{ch2}. It includes discussion of Skyrme
models based on a non-relativistic mean-field as well as a relativistic
mean-field model.  Chapter \ref{ch2} also incorporates a discussion on
properties of infinite nuclear matter which are extracted used different
men-field models in the present thesis work. Different correlations
mentioned above can also be studied using a single model by applying the
method of covariance analysis. Apart from the correlations one can also
study the errors on the model parameters and calculated observables. These
informations are very important to benchmark the findings of a theoretical
model. How one can study the errors on quantities of interest along with
different correlations are discussed in Chapter \ref{ch3}.

Out of the three symmetry energy parameters mentioned above, $C_2^0$ has
been known to lie in good confidence in the range $\sim32\pm4$ MeV from
different experimental data. Using the finite range droplet model \cite
{Moller12} or studying the double differences of experimental nuclear
masses \cite {Jiang12}, the error bar is further reduced to $\sim0.5$
MeV. Exploration on the slope parameter $L_0$, however, shows a wide
variation. The value of $L_0$ may lie in the range from 20 MeV - 120
MeV \cite {Centelles09}.  There have been a tremendous amount of effort
to constrain the value of $L_0$ from several experimental findings
in finite nuclei and astrophysical observations \cite {Centelles09,
Steiner12, Shetty07, Famiano06, Li08, Trippa08, Roca-Maza13a, Roca-Maza13,
Carbone10, Chen11, Tsang12, Niksic08, Zhao10, Chabanat98, Roca-Maza12,
Roca-Maza11}. Among different experimental data, binding energies are
the most accurately known experimental quantities. However, using
binding energies of few closed shell spherical nuclei, mean-field
models fail to constrain the value of $L_0$ (c.f. Fig. \ref{prl102})
in a narrow range. In Chapter \ref{ch4} we reconcile this view by using
binding energies of some extremely asymmetric nuclei, where number of
neutrons is twice to that of protons, to obtain the parameters of a
relativistic mean-field model \cite {mondal15, Mondal16}. A covariance
analysis accompanied by a sensitivity analysis is performed further to
find the merits of incorporating these highly asymmetric nuclei in the
fitting protocol. Correlations existing between different quantities
are also explored.

As mentioned above, $\Delta r_{np}$ of a heavy nucleus like $^{208}$Pb
is correlated to $L_0$, which was originally proposed in Ref. \cite
{Centelles09} by using Droplet Model \cite{Myers69, Myers80}. This
correlation was realized by using $\sim40$ mean-field models. We find
this correlation has some degree of model dependence \cite{Mondal16b}
which is discussed in Chapter \ref{ch5} in view of Droplet Model.

Compared to $L_0$, the uncertainty creeps in by even in larger amount for
the case of curvature parameter $K_{sym}^0$. Across several mean-field
models both relativistic and non-relativistic, the values of $K_{sym}^0$
lie within a huge range, $-700$ MeV $< K_{sym}^0 < 400$ MeV. The value of
$K_{sym}^0$ plays quite a significant role in determining the symmetry
energy behavior of highly asymmetric dense matter e.g. in neutron
star or supernova explosion. There has been no attempt till date to
constrain the value of $K_{sym}^0$ from experimental data. Theoretical
calculations connecting poorly known $K_{sym}^0$ to other comparatively
better known nuclear matter properties may hold the key to pin down the
value of $K_{sym}^0$ \cite{Dong12, Mondal17b}. In Chapter \ref{ch6}
a simple model based on fundamental laws of statistical mechanics
is proposed. Analytical relations between different symmetry energy
parameters are derived further. Special importance is given to the
relatively poorly known quantity $K_{sym}^0$. A linear correlation is
suggested by the simple model between $K_{sym}^0$ and other nuclear
matter properties. The correlation was realized by using 500 mean-field
models used in the literature which shows the near-universality in the
correlation which is proposed.

In a nutshell, this thesis aims towards constraining the different
symmetry energy parameters using different mean-field models. Contents of
different chapters are described very briefly above. Special attention
was given to those parameters which controls the density dependence of
symmetry energy. A brief summary and future outlooks are discussed in
Chapter \ref{ch7}.

%% file: chap2.tex
\chapter{Mean Field Models}\label{ch2}
\section{Introduction}
{\it Ab initio} methods based quantum chromodynamics (QCD) or configuration 
interaction are the most fundamental theories to describe the many-body 
nuclear systems like finite nuclei or neutron stars \cite{Barrett13,Hagen14}. 
However, the computational cost is too high to calculate even the ground state 
properties of lighter nuclei (mass number $A\le40$) with these methods.
Unless there is an unprecedented improvement in the
modern day computing facilities, it will be extremely unlikely to provide
even the ground state description of heavy or super-heavy nuclei with
these {\it ab-initio} approaches. On the contrary, mean field models provide a
coherent description of astrophysical objects like neutron stars as well 
as terrestrial finite nuclei throughout the whole nuclear chart at a very 
nominal computational cost with quite high accuracy.

In this thesis work mainly two class of mean-field models are employed
namely, a non-relativistic variant based on zero range Skyrme force
\cite{Skyrme56,Skyrme59} and a relativistic one formulated on the basis
of interaction between nucleons through mesons \cite{Walecka74,Boguta77}. The
parameters of these two variants of mean-filed models are obtained by
fitting experimental data on binding energy and charge radii of closed
shell nuclei. In this chapter, we discuss the method to obtain the binding 
energy and charge radii for closed shell nuclei along with the 
properties of infinite nuclear matter in the mean-field framework.

\section{Finite nuclei}
In the present thesis work, mostly we have dealt with closed shell spherical 
nuclei. The ground state properties are calculated within Hartree-Fock 
approximation both with the non-relativistic Skyrme and relativistic 
formalism. Pairing for nucleons is also discussed in the constant BCS 
approximation \cite{Greiner96}.
\subsection{Skyrme Formalism}
For Hartree-Fock (HF) description \cite{Vautherin72,Chabanat98,Dhiman07}
of ground state configuration, the many body wave function $\psi$ of a
nucleus of mass number $A$ can be given by the Slater determinant as,
\begin{eqnarray}
\label{Slater}
\psi(x_1,x_2,\cdots,x_A)=\frac{1}{\sqrt{A!}}\left|\begin{array}{ccccc}
\phi_1(x_1) & \phi_1(x_2) & \phi_1(x_3) & \cdots & \phi_1(x_A)\\
\phi_2(x_1) & \phi_2(x_2) & \cdots      & \cdots & \cdots     \\
\cdots      &             &             &        &            \\
\phi_A(x_1) & \phi_A(x_2) & \phi_A(x_3) & \cdots & \phi_A(x_A)
\end{array}\right|,
\end{eqnarray}
where $\phi_i$ (with $i=1,2,\cdots A$) denotes the occupied single particle states and $x$'s run over
${\bf r}$, spin $\sigma$ and isospin $q$ ($=p$ for proton and $n$ for
neutron).  Then the expectation value of the total energy of system can
be written as,
\begin{eqnarray}
\label{skm_etot}
E&=&\langle \psi|(T+V)|\psi\rangle\nonumber\\
&=&\sum_i\langle\phi_i\left|\frac{p^2}{2m}\right|\phi_i\rangle
+\frac{1}{2}\sum_{ij}\langle\phi_i\phi_j\left|\overline{v}_{12}\right|
\phi_i\phi_j\rangle+\frac{1}{6}\sum_{ijk}\langle\phi_i\phi_j\phi_k\left|
\overline{v}_{123}\right|\phi_i\phi_j\phi_k\rangle\nonumber\\
&=&\int\mathcal{H}\left({\bf r}\right) d^3{\bf r}.
\end{eqnarray}
Here, $\overline{v}$'s denote the two-body and three body antisymmetrized
matrix elements. For Skyrme interaction the Hamiltonian density
$\mathcal{H}$ is algebraic function of nucleon densities $\rho_q$, kinetic
energy densities $\tau_q$ and spin densities ${\bf J}_q$. With some
phenomenological corrections \cite{Chabanat98,Dhiman07} over the actual
interaction given by Skyrme \cite{Skyrme56,Skyrme59}, the interaction
part of the Hamiltonian density can be given by,
\begin{eqnarray}
\label{skm_interaction}
V({\bf r}_1,{\bf r}_2)&=&t_0(1+x_0P_{\sigma})\delta({\bf r})+\frac{1}{2}t_1(1+x_1P_{\sigma})
\left[\delta({\bf r}){{\bf k}^{\prime}}^2+{\bf k}^2\delta({\bf r})\right]
+t_2(1+x_2P_{\sigma}){\bf k}^{\prime}\cdot\delta({\bf r}){\bf k}\nonumber\\
&&+t_3\rho^{\alpha}(1+x_3P_{\sigma})\delta({\bf r})+iW_0{\boldsymbol\sigma}\cdot
\left[{\bf k}^{\prime}\times\delta({\bf r}){\bf k}\right],
\end{eqnarray}
where, ${\bf r}={\bf r}_1-{\bf r}_2$, ${\bf k}=\frac{{\boldsymbol\nabla}_1-{\boldsymbol\nabla}_2}{2i}$,
${\bf k}^{\prime}$ is complex conjugate of ${\bf k}$ acting on the left, ${\boldsymbol\sigma}=
{\boldsymbol\sigma}_1+{\boldsymbol\sigma}_2$, and the spin exchange operator $P_{\sigma}=
\frac{1}{2}(1+{\boldsymbol\sigma}_1\cdot{\boldsymbol\sigma}_2)$.
Consequently following Eq. (\ref{skm_etot}) the Energy density functional (EDF) or 
the Hamiltonian density is given by,
\begin{eqnarray}
\mathcal{H}&=&\mathcal{K}+\mathcal{H}_0+\mathcal{H}_3+\mathcal{H}_{\text{eff}}+\mathcal{H}_{\text{fin}}
+\mathcal{H}_{\text{SO}}+\mathcal{H}_{\text{sg}}+\mathcal{H}_{\text{Coul}}+\mathcal{H}_{\text{pair}}.
\end{eqnarray}
Here, kinetic energy is denoted by, $\mathcal{K}=\frac{\hbar^2}{2m}\tau$, Coulomb 
energy for the protons and the pairing energy are given by $\mathcal{H}_{\text{Coul}}$ 
and $\mathcal{H}_{\text{pair}}$, respectively. The rest of the terms are coming from the Skyrme interaction 
given in Eq. (\ref{skm_interaction}), which are written in terms of different 
densities as, 
\begin{eqnarray}
\label{e_skm}
\mathcal{H}_0({\bf r})&=&\frac{1}{4}t_0\left[\left(2+x_0\right)\rho^2-\left(2x_0+1\right)\left(\rho_p^2+\rho_n^2\right)\right],\nonumber\\
\mathcal{H}_3({\bf r})&=&\frac{1}{24}t_3\rho^{\alpha}\left[\left(2+x_3\right)\rho^2-\left(2x_3+1\right)\left(\rho_p^2+
\rho_n^2\right)\right],\nonumber\\
\mathcal{H}_{\text {eff}}({\bf r})&=&\frac{1}{8}\left[ t_1\left(2+x_1\right)+t_2\left(2+x_2\right)\right]\tau\rho
+\frac{1}{8}\left[ t_2\left(2x_2+1\right)-t_1\left(2x_1+1\right)\right]\left( \tau_p\rho_p+\tau_n\rho_n\right),\nonumber\\
\mathcal{H}_{\text{fin}}({\bf r})&=&\frac{1}{32}\left[3t_1\left(2+x_1\right)-t_2\left(2+x_2\right)
\right]\left({\boldsymbol\nabla}\rho\right)^2,\nonumber\\
&&-\frac{1}{32}\left[3t_1\left(2x_1+1\right)+t_2\left(2x_2+1\right)\right]
\left[\left({\boldsymbol\nabla}\rho_p\right)^2+\left({\boldsymbol\nabla}\rho_n\right)^2\right],\nonumber\\
\mathcal{H}_{\text {SO}}({\bf r})&=&\frac{1}{2}W_0\left[{\bf J}\cdot{\boldsymbol\nabla}\rho+{\bf J}\cdot{\boldsymbol\nabla}\rho_p+
{\bf J}\cdot{\boldsymbol\nabla}\rho_n\right],\nonumber\\
\mathcal{H}_{\text {sg}}({\bf r})&=&\frac{1}{16}\left( t_1-t_2\right)\left[{\bf J}_p^2+{\bf J}_n^2\right]-\frac{1}{16}\left( t_1x_1+
t_2x_2\right){\bf J}^2.
\end{eqnarray}
The Coulomb energy for the protons is given by,
\begin{eqnarray}
\label{e_coul}
\mathcal{H}_{\text{Coul}}({\bf r})&=&\frac{1}{2}e^2 \rho_p({\bf r})\int d^3{\bf r}^{\prime}\frac{\rho_p({\bf r}^{\prime})}
{|{\bf r}-{\bf r}^{\prime}|}-\frac{3}{4}e^2\rho_p({\bf r})\left(\frac{3\rho_p({\bf r})}{\pi}\right)^{1/3}.
\end{eqnarray}
In Eqs. (\ref{e_skm}) and (\ref{e_coul}), the various densities are given by,
\begin{eqnarray}
\label{density_skm}
\rho_q\left({\bf r}\right)&=&\sum_{i,\sigma}|\phi_i\left({\bf r},\sigma,q\right)|^2,\nonumber\\
\tau_q\left({\bf r}\right)&=&\sum_{i,\sigma}|{\boldsymbol\nabla}\phi_i\left({\bf r},\sigma,q\right)|^2,\nonumber\\
{\bf J}_q\left({\bf r}\right)&=&-i\sum_{i,\sigma,\sigma^{\prime}}\phi_i^*\left({\bf r},\sigma,q\right)
\left[{\boldsymbol\nabla}\phi_i\left({\bf r},\sigma^{\prime},q\right)\times \langle\sigma | {\boldsymbol\sigma}| 
\sigma^{\prime}\rangle\right].
\end{eqnarray}
The total densities $\rho$, $\tau$ and ${\bf J}$ are calculated by summing 
over $q=n$ and $p$.

Now the HF equations are obtained by writing E stationary with respect to
variation of individual single particle states $\phi_i$ with the added condition 
that $\phi_i$'s are normalized,
\begin{equation}
\label{variation_phi}
\frac{\delta}{\delta\phi_i}\left( E-\sum_i \epsilon_i \int |\phi_i\left({\bf r}\right)|^2 d^3r\right)=0.
\end{equation}
An equivalent description can be obtained if the variation is performed with 
respect to the densities $\tau_q$, $\rho_q$ and ${\bf J}_q$ instead of $\phi_i$. Then the 
variational equation takes the form
\begin{eqnarray}
\label{variation_den}
\left[-{\boldsymbol\nabla}\frac{\hbar^2}{2m^*_q({\bf r})}\cdot{\boldsymbol\nabla}
+U_q({\bf r})-i{\bf W}_q({\bf r})\cdot\left({\boldsymbol\nabla}\times{\boldsymbol\sigma}
\right)\right]\phi_i({\bf r},q)=\epsilon_i\phi_i({\bf r},q).
\end{eqnarray}
Eq. (\ref{variation_den}) represents the Schr{\"o}dinger equation for the single 
particle states $\phi_i$. The different coefficients $\frac{\hbar^2}{2m^*}$, 
$U_q$ and ${\bf W}_q$ determine the effective mass, central potential and the 
spin-orbit potential, respectively. The effective mass $m^*_q$ is given by,
\begin{eqnarray}
\label{var_eff}
\frac{\delta \mathcal{H}}{\delta \tau_q({\bf r})}=\frac{\hbar^2}{2m^*_q({\bf r})}
&=&\frac{\hbar^2}{2m}+\frac{1}{8}\left[t_1(2+x_1)+t_2(2+x_2)\right]\rho({\bf r})\nonumber\\
&&+\frac{1}{8}\left[t_2(2x_2+1)-t_1(2x_1+1)\right]\rho_q({\bf r}).
\end{eqnarray}
The central, spin-orbit and Coulomb potentials are given by,
\begin{eqnarray}
\label{var_uqwq}
\frac{\delta \mathcal{H}}{\delta \rho_q({\bf r})}=U_q({\bf r})&=&\frac{1}{2}t_0\left[
(2+x_0)\rho({\bf r})-(1+2x_0)\rho_q({\bf r})\right]\nonumber\\
&+&\frac{1}{24}t_3\left\{(2+x_3)
(2+\alpha)\rho^{\alpha+1}({\bf r})\right.\nonumber\\
&&\ \ \ \ \ \ \ \left.-(2x_3+1)\left[2\rho^{\alpha}({\bf r})\rho_q(
{\bf r})+\alpha\rho^{\alpha-1}({\bf r})\left(\rho_p^2({\bf r})+\rho_n^2({\bf r})\right)\right]\right\}\nonumber\\
&+&\frac{1}{8}\left[t_1(2+x_1)+t_2(2+x_2)\right]\tau({\bf r})+\frac{1}{8}\left[t_2(2x_2+1)-
t_1(2x_1+1)\right]\tau_q({\bf r})\nonumber\\
&+&\frac{1}{16}\left[t_2(2+x_2)-3t_1(2+x_1)\right]{\boldsymbol\nabla}^2\rho({\bf r})\nonumber\\
&&\ \ \ \ \ \ \ +\frac{1}{16}\left[3t_1(2x_1+1)+t_2(2x_2+1)\right]{\boldsymbol\nabla}^2\rho_q({\bf r})\nonumber
\end{eqnarray}
\begin{eqnarray}
&-&\frac{1}{2}W_0\left({\boldsymbol\nabla}\cdot{\bf J}({\bf r})+{\boldsymbol\nabla}\cdot{\bf J}_q({\bf r})\right)
+\delta_{q,p}V_{\text{Coul}}({\bf r}),\\
\frac{\delta \mathcal{H}}{\delta {\bf J}_q({\bf r})}={\bf W}_q({\bf r})&=&
\frac{1}{2}W_0\left({\boldsymbol\nabla}\rho({\bf r})+{\boldsymbol\nabla}\rho_q({\bf r})\right)
+\frac{1}{8}(t_1-t_2){\bf J}_q({\bf r})\nonumber\\
&&\ \ \ \ \ \ \ -\frac{1}{8}(t_1x_1+t_2x_2){\bf J}({\bf r}),\\
V_{\text{Coul}}({\bf r})&=&\frac{1}{2}e^2 \int d^3{\bf r}^{\prime}\frac{\rho_p({\bf r}^{\prime})}
{|{\bf r}-{\bf r}^{\prime}|}-\frac{3}{4}e^2\left(\frac{3\rho_p({\bf r})}{\pi}\right)^{1/3}.
\end{eqnarray}
For doubly closed shell nuclei the HF equations can be derived using, 
\begin{eqnarray}
\label{wfn_spherical}
\phi_i({\bf r},\sigma,\boldsymbol\tau)&=&\frac{R_{\alpha}(r)}{r}\mathcal{Y}_{ljm}
(\hat{r},\sigma)\chi_q(\boldsymbol\tau),\nonumber\\
\mathcal{Y}_{ljm}(\hat{r},\sigma)&=&\sum_{m_l m_s}\langle l \frac{1}{2} m_l m_s|jm\rangle
Y_{lm_l}(\hat{r})\chi_{m_s}(\sigma),
\end{eqnarray}
where, $R_{\alpha}(r)$ is the radial part of the wave-function and $\mathcal{Y}_{ljm}$ 
is the spherical harmonics representing the spin and angular part of the 
wave-function. In Eq. (\ref{wfn_spherical}) $i\equiv (q,n,l,j,m)$ and $\alpha\equiv (n,l,j)$ 
convention was used, where, $q$ is the charge, the principal quantum number $n$, 
orbital angular momentum $l$, total single-particle angular momentum $j$ and magnetic quantum 
number $m$. Consequently the densities in Eq. (\ref{density_skm}) get modified 
as \cite{Vautherin72}, 
\begin{eqnarray}
\label{density_skm_sph}
\rho_q(r)&=&\frac{1}{4\pi r^2}\sum_{\alpha}(2j_{\alpha}+1)R_{\alpha}^2(r)\chi_q^2,\nonumber\\
\tau_q(r)&=&\frac{1}{4\pi}\sum_{\alpha}(2j_{\alpha}+1)\left[\left(\frac{d\varphi_{\alpha}}{dr}
\right)^2+\frac{l_{\alpha}(l_{\alpha}+1)}{r^2}\varphi_{\alpha}^2\right]\chi_q^2,\nonumber\\
{\bf J}_q({\bf r})&=&\frac{{\bf r}}{r}J_q(r),\ \ \ \ {\text{with}}\nonumber\\
J_q(r)&=&\frac{1}{4\pi r^3}\sum_{\alpha}(2j_{\alpha}+1)\left[j_{\alpha}(j_{\alpha}+1)-
l_{\alpha}(l_{\alpha}+1)-\frac{3}{4}\right]R_{\alpha}^2(r)\chi_q^2,
\end{eqnarray}
where, $\varphi_{\alpha}=\frac{R_{\alpha}(r)}{r}$. Due to the symmetry in ${\bf J}_q$ in 
Eq. (\ref{density_skm_sph}), expression of ${\bf W}_q$ in Eq. (\ref{variation_den}) 
gets modified as,
\begin{eqnarray}
\label{pot_wq}
{\bf W}_q({\bf r})&=&\frac{1}{r}W_q(r)\vec{l}\cdot{\boldsymbol\sigma}\ \ \ \ {\text{with}}\nonumber\\
W_q(r)&=&\frac{1}{2}W_0\frac{d}{dr}(\rho+\rho_q)+\frac{1}{8}(t_1-t_2)J_q(r)
-\frac{1}{8}(t_1x_1+t_2x_2)J(r).
\end{eqnarray}
Consequently, for closed shell spherical nuclei Eq. (\ref{variation_den}) takes 
the form,
\begin{eqnarray}
\label{variation_sph}
-\frac{\hbar^2}{2m_q^*}\nabla^2\phi_i-\left({\boldsymbol\nabla}\frac{\hbar^2}
{2m_q^*}\right)\cdot{\boldsymbol\nabla}\phi_i+\left(U_q+\frac{1}{r}W_q\vec{l}\cdot
{\boldsymbol\sigma}\right)\phi_i=\epsilon_i\phi_i.
\end{eqnarray}
Using the expressions for gradient and Laplacian operators in spherical coordinate as,
\begin{eqnarray}
\label{operator_sph}
\nabla^2&\equiv&\frac{1}{r}\frac{\partial^2}{\partial r^2}r-\frac{\vec{l}^2}{r^2},\nonumber\\
{\boldsymbol\nabla}\frac{\hbar^2}{2m_q^*}&=&\frac{{\bf r}}{r}\frac{d}{dr}\frac{\hbar^2}{2m_q^*},
\end{eqnarray}
one can write the coupled Schr{\"o}dinger equation for radial part of the wave-function 
$R_{\alpha}(r)$ as,
\begin{eqnarray}
\label{schrodinger_coupled}
&&\frac{\hbar^2}{2m_q^*}\left[-\frac{d^2 R_{\alpha}(r)}{dr^2}+\frac{l_{\alpha}(l_{\alpha}+1)}
{r^2}R_{\alpha}(r)\right]-\frac{d}{dr}\left(\frac{\hbar^2}{2m_q^*}\right)\frac{dR_{\alpha}(r)}{dr}\nonumber\\
&+&\left\{U_q(r)+\frac{1}{r}\frac{d}{dr}\left(\frac{\hbar^2}{2m_q^*}\right)+\left[
j_{\alpha}(j_{\alpha}+1)-l_{\alpha}(l_{\alpha}+1)-\frac{3}{4}\right]\times\frac{1}{r}W_q(r)\right\}
R_{\alpha}(r)=\epsilon_{\alpha}R_{\alpha}(r)\nonumber\\.
\end{eqnarray}

The solutions for $R_{\alpha}$ in Eq. (\ref{schrodinger_coupled}) are obtained 
iteratively by solving self-consistently the Eqs. (\ref{var_eff}), 
(\ref{var_uqwq}) and (\ref{pot_wq}). First a guess solution for the unknown 
$R_{\alpha}$ is taken as Harmonic-oscillator or Wood-Saxon wave functions along 
with a particular occupation probability distributions of the single particle 
levels. Inserting them in Eq. (\ref{density_skm_sph}) expression for densities are 
obtained. Consequently, the expressions for $\frac{\hbar^2}{2m_q^*}$, $U_q$ 
and $W_q$ are obtained from Eqs. (\ref{var_eff}), (\ref{var_uqwq}) and 
(\ref{pot_wq}) respectively. Putting those values in Eq. (\ref{schrodinger_coupled}) 
the next guess for $R_{\alpha}$ is obtained. In case of presence of a pairing 
interaction, the occupation probability distribution of the single particle levels 
are calculated satisfying the number of particle for a nucleus. This process
iterated until a consistent solution is obtained.

\subsection{Relativistic Mean Field formalism}
The effective Lagrangian for the relativistic mean field (RMF) model
employed in the present thesis work is similar to that of the FSU
one \cite{Todd-Rutel05}.  The system contains the nucleonic field
$\psi$ and three different types of mesons, which mediate the force, 
namely, isoscalar-scalar $\sigma$, isoscalar-vector $\omega$ and
isovector-vector $\boldsymbol\rho$ (field denoted by ${\bf R}_{\mu}$)
\cite{Reinhard86,Reinhard89,Estal01,Bender03}. The protons also interact
through an electromagnetic field $A^{\mu}$. The total Lagrangian can be
decomposed into different components as,
\begin{eqnarray}
\label{lag_tot}
\mathcal {L}=\int d^3 {\bf r}\Big\{\mathcal{L}_{\text{BM}}+\mathcal{L}_{\sigma}+\mathcal{L}_{\omega}
+\mathcal{L}_{\boldsymbol\rho}+\mathcal{L}_{\omega\boldsymbol\rho}
+\mathcal{L}_{\text{em}}\Big\}.
\end{eqnarray}
The baryonic-mesonic Lagrangian containing the Yukawa couplings between 
the nucleon and the mesons is given by, 
\begin{eqnarray}
\label{lag_bm}
\mathcal{L}_{\text{BM}}=\overline{\psi}\left[i\gamma^{\mu}\partial_{\mu}-
\left(M-g_{\sigma}\sigma\right)-\gamma^{\mu}\left(g_{\omega}\omega_{\mu}+\frac{1}{2}g_{\mathbf{\boldsymbol\rho}}
\boldsymbol\tau \cdot{\bf R}_{\mu}\right)\right]\psi.
\end{eqnarray}
The parameters $g_{\sigma}$, $g_{\omega}$ and $g_{\boldsymbol\rho}$ describe 
the strength of the couplings of $\psi$ with $\sigma$, $\omega$ and ${\boldsymbol\rho}$ mesons 
respectively. $M$ is the free nucleon mass and the Dirac effective mass is 
denoted by $M^*_{\text{Dir}}=(M-g_{\sigma}\sigma)$. The Lagrangian for the 
mesons including the self-interaction terms are given by,
\begin{eqnarray}
\label{lag_meson}
\mathcal{L}_{\sigma}&=&\frac{1}{2}\left(\partial_{\mu}\sigma\partial^{\mu}\sigma-m_{\sigma}^2
\sigma^2\right)-\frac{{\kappa_3}}{3!M} g_{\sigma}m_{\sigma}^2\sigma^3-\frac{{\kappa_4}}
{4!M^2}g_{\sigma}^2 m_{\sigma}^2\sigma^4,\nonumber\\
\mathcal{L}_{\omega}&=&-\frac{1}{4}\omega_{\mu\nu}\omega^{\mu\nu}+\frac{1}{2}m_{\omega}^2
\omega_{\mu} \omega^{\mu}+\frac{1}{4!}\zeta_0 g_{\omega}^{2}\left(\omega_{\mu}\omega^{\mu}\right)^2,\nonumber\\
\mathcal{L}_{\boldsymbol\rho}&=&-\frac{1}{4}{\bf R}_{\mu\nu}{\bf R}^{\mu\nu}
+\frac{1}{2}m_{\boldsymbol\rho}^2 {\bf R}_{\mu} {\bf R}^{\mu}.
\end{eqnarray}
Field tensors for $\omega$ and ${\boldsymbol\rho}$ mesons are given by, $\omega_{\mu\nu}=
\partial_{\mu}\omega_{\nu}-\partial_{\nu}\omega_{\mu}$ and ${\bf R}_{\mu\nu}=
\partial_{\mu}{\bf R}_{\nu}-\partial_{\nu}{\bf R}_{\mu}$. The cross-coupling 
between the $\omega$ and the $\boldsymbol\rho$ mesons is given by,
\begin{eqnarray}
\label{lag_cross}
\mathcal{L}_{\omega\boldsymbol\rho}=\frac{\eta_{2\boldsymbol\rho}}{4M^2}g_{\omega}^2m_{\boldsymbol\rho
}^{2}\omega_{\mu}\omega^{\mu}\bf R_{\nu}\bf R^{\nu}.
\end{eqnarray}
The electromagnetic interaction between the protons is given by,
\begin{eqnarray}
\label{lag_em}
\mathcal{L}_{\text{em}}=-\frac{1}{4}F_{\mu\nu}F^{\mu\nu}-e
{\overline\psi}\gamma^{\mu}\frac{1+\tau_0}{2}A_{\mu}\psi,
\end{eqnarray}
where $F_{\mu\nu}=\partial_{\mu}A_{\nu}-\partial_{\nu}A_{\mu}$, $e$ 
is the charge of proton and $\tau_0=1$ for protons and $=-1$ for neutrons.

In the covariant formalism the Euler-Lagrange equation for a field $\varphi$ is 
given by,
\begin{eqnarray}
\label{euler_lagrange}
\partial_{\mu}\left(\frac{\partial\mathcal{L}}{\partial(\partial_{\mu}\varphi)}\right)=
\frac{\partial\mathcal{L}}{\partial\varphi}\ \ .
\end{eqnarray}
Immediately the equation of motion of the single particle wave-functions $\phi$ 
for the nucleons is given by the Dirac equation as,
\begin{eqnarray}
\label{dirac_eq}
i\frac{\partial}{\partial t}\phi_{\alpha}=
\gamma^0 \left[-i{\boldsymbol\gamma}\cdot{\boldsymbol\nabla}+\gamma^{\mu}\left(g_{\omega}\omega_{\mu}
+\frac{1}{2}g_{\boldsymbol\rho}{\tau}_i \cdot{R}_{i,\mu}
+e\frac{1+\tau_0}{2}A_{\mu}\right)+\left(M-g_{\sigma}\sigma\right)\right]
\phi_{\alpha}.\nonumber\\
\end{eqnarray}
Here the nucleon field operator $\hat{\psi}$ 
is already expanded over the single particle states $\phi$ by the relation, 
$\hat{\psi}=\sum_{\alpha} \phi_{\alpha}{a}_{\alpha}$.  The quantity $|a_{\alpha}|^2$ 
determines the probability of finding a particle in single particle state $\phi_{\alpha}$. 
Similarly, Euler-Lagrange equations for the mesons and the electromagnetic field 
$A^{\mu}$ are given by,
\begin{eqnarray}
\label{eom_field}
\frac{\partial^2}{\partial t^2}\sigma&=&\left(\Delta-m_{\sigma}^2\right)\sigma+g_{\sigma}\rho_{\text{s}}
-\frac{{\kappa_3}}{2M} g_{\sigma}m_{\sigma}^2\sigma^2-\frac{{\kappa_4}}
{3!M^2}g_{\sigma}^2 m_{\sigma}^2\sigma^3,\nonumber\\
\frac{\partial^2}{\partial t^2}\omega_{\mu}&=&\left(\Delta-m_{\omega}^2\right){\omega}_{\mu}+g_{\omega}\rho_{\mu}
-\frac{1}{3!}\zeta_0 g_{\omega}^{2}{\omega}_{\nu}{\omega}^{\nu}{\omega}_{\mu}-\frac{\eta_{2\boldsymbol\rho}}{2M^2}g_{\omega}^2
m_{\boldsymbol\rho}^{2}{R}_{i, \nu}{R}_i^{\nu}{\omega}_{\mu},\nonumber\\
\frac{\partial^2}{\partial t^2}R_{i,\mu}&=&\left(\Delta-m_{\boldsymbol\rho}^2\right)R_{i,\mu}+\frac{1}{2}g_{\boldsymbol\rho} \rho_{i,\mu}
-\frac{\eta_{2\boldsymbol\rho}}{2M^2}g_{\omega}^2m_{\boldsymbol\rho}^{2}{\omega}_{\nu}{\omega}^{\nu}R_{i,\mu},\nonumber\\
\frac{\partial^2}{\partial t^2}A_{\mu}&=&\Delta A_{\mu}+e\rho_{p,\mu},
\end{eqnarray}
where the different densities are given by,
\begin{eqnarray} 
\label{density_rmf}
\rho_{\text{s}} &=&\sum_{\alpha=-\infty}^{\infty}w_{\alpha}\overline{\phi}_{\alpha}\phi_{\alpha},\nonumber\\
\rho_{\mu}&=&\sum_{\alpha=-\infty}^{\infty}w_{\alpha}\overline{\phi}_{\alpha}\gamma_{\mu}\phi_{\alpha},\nonumber\\
\rho_{i,\mu}&=&\sum_{\alpha=-\infty}^{\infty}w_{\alpha}\overline{\phi}_{\alpha}\tau_i\gamma_{\mu}\phi_{\alpha},\nonumber\\
\rho_{p,\mu}&=&\sum_{\alpha=-\infty}^{\infty}w_{\alpha}\overline{\phi}_{\alpha}\frac{1+\tau_0}{2}
\gamma_{\mu}\phi_{\alpha},
\end{eqnarray}
where, $w_{\alpha}=1$ for levels below Fermi surface including both the positive 
and negative energy states and $w_{\alpha}=0$ for levels above Fermi surface. 

For ground state properties of finite nuclei, only the static solutions are 
relevant. Due to this reason all the mesonic fields are time-independent and 
the nucleon wave-function is determined by the single particle energies $\epsilon_{\alpha}$. 
Moreover, $\psi$ is even under time reversal, meaning the vector currents e.g. 
$\rho_{\mu}$ or $\rho_{i,\mu}$ only survive by their $\mu=0$ components. 
Further due to isospin symmetry, only the component $\rho_{0,0}$ survives 
in the current $\rho_{i,0}$. So the equation of motion for the nucleons,  
mesons and the electromagnetic field take the form, 
\begin{eqnarray}
\label{eom_rmf}
\epsilon_{\alpha}\phi_{\alpha}&=&
\gamma^0 \left[-i{\boldsymbol\gamma}\cdot{\boldsymbol\nabla}+\gamma^{0}\left(g_{\omega}\omega_{0}
+\frac{1}{2}g_{\boldsymbol\rho}{\tau}_0 \cdot{R}_{0,0}
+e\frac{1+\tau_0}{2}A_{0}\right)+\left(M-g_{\sigma}\sigma\right)\right]
\phi_{\alpha},\nonumber\\
\left(-\Delta+m_{\sigma}^2\right)\sigma&=&g_{\sigma}\rho_{\text{s}}
-\frac{{\kappa_3}}{2M} g_{\sigma}m_{\sigma}^2\sigma^2-\frac{{\kappa_4}}
{3!M^2}g_{\sigma}^2 m_{\sigma}^2\sigma^3,\nonumber\\
\left(-\Delta+m_{\omega}^2\right){\omega}_{0}&=&g_{\omega}\rho_{0}
-\frac{1}{3!}\zeta_0 g_{\omega}^{2}{\omega}_{0}^3-\frac{\eta_{2\boldsymbol\rho}}{2M^2}g_{\omega}^2
m_{\boldsymbol\rho}^{2}{R}_{0,0}^2{\omega}_0,\nonumber\\
\left(-\Delta+m_{\boldsymbol\rho}^2\right)R_{0,0}&=&\frac{1}{2}g_{\boldsymbol\rho} \rho_{0,0}
-\frac{\eta_{2\boldsymbol\rho}}{2M^2}g_{\omega}^2m_{\boldsymbol\rho}^{2}{\omega}_0^2R_{0,0},\nonumber\\
-\Delta A_0&=&e\rho_{p,0}.
\end{eqnarray}
The sum for all the densities given in Eq. (\ref{density_rmf}) still runs over 
both the positive and negative energy spectrum of the Dirac equation. 
The full summation is too difficult to handle 
numerically. In "No-Sea" approximation the sum runs over few positive energy 
bound state or so to say the number of shell model states ($=\Omega$) included 
in the numerical calculation, where, 
\begin{eqnarray}
\label{w_alpha}
\sum_{\alpha=1}^{\Omega} w_{\alpha}=\begin{cases}N\\ Z \end{cases}\ .
\end{eqnarray}
Depending on the neutron or proton occupation probability, the summation gives 
the total number of particles i.e. $N$ for neutrons and $Z$ for protons. In these 
set of approximations the different densities take the form,
\begin{eqnarray} 
\label{density_rmf2}
\rho_{\text{s}} &=&\sum_{\alpha=1}^{\Omega}w_{\alpha}\overline{\phi}_{\alpha}\phi_{\alpha},\nonumber
\end{eqnarray}
\begin{eqnarray}
\rho_{0}&=&\sum_{\alpha=1}^{\Omega}w_{\alpha}\overline{\phi}_{\alpha}\gamma_{0}\phi_{\alpha},\nonumber\\
\rho_{0,0}&=&\sum_{\alpha=1}^{\Omega}w_{\alpha}\overline{\phi}_{\alpha}\tau_i\gamma_{0}\phi_{\alpha},\nonumber\\
\rho_{p,0}&=&\sum_{\alpha=1}^{\Omega}w_{\alpha}\overline{\phi}_{\alpha}\frac{1+\tau_0}{2}
\gamma_{0}\phi_{\alpha}.
\end{eqnarray}
In the covariant formalism the stress-energy tensor $T^{\mu\nu}$ for a field 
$\varphi$ is given by,
\begin{eqnarray}
\label{stress_energy}
T^{\mu\nu}=\frac{\partial\mathcal{L}}{\partial\left(\partial_{\mu}\varphi\right)}
\partial^{\nu}\varphi-g^{\mu\nu}\mathcal{L}\ \ ,
\end{eqnarray}
where $g^{\mu\nu}$ are the components of metric tensor given by $g^{\mu\nu}=\text
{Diag}[1\ -1\ -1\ -1]$. The component $T^{00}$ gives the energy of the system. So,
the mean-field energy is then given by,
\begin{eqnarray} 
\label{rmf_etot}
E&=&\int d^3 r \Bigg\{\sum_{\alpha}w_{\alpha}\overline{\phi}_{\alpha}
\left[-i{\boldsymbol\gamma}\cdot{\boldsymbol\nabla}+\gamma^{0}\left(g_{\omega}\omega_{0}
+\frac{1}{2}g_{\boldsymbol\rho}{\tau}_0 \cdot{R}_{0,0}
+e\frac{1+\tau_0}{2}A_{0}\right)+\left(M-g_{\sigma}\sigma\right)\right]\phi_{\alpha}\nonumber\\
&&+\frac{1}{2}\left[\left({\boldsymbol\nabla}\sigma\right)^2+m_{\sigma}^2\sigma^2\right]
+\frac{\kappa_3}{3!M}g_{\sigma}m_{\sigma}^2\sigma^3+\frac{{\kappa_4}}{4!M^2}g_{\sigma}^2 m_{\sigma}^2
\sigma^4\nonumber\\
&&-\frac{1}{2}\left[\left({\boldsymbol\nabla}\omega_0\right)^2+m_{\omega}^2\omega_0^2
+\left({\boldsymbol\nabla}R_{0,0}\right)^2+m_{\boldsymbol\rho}^2 R_{0,}^2+
\left({\boldsymbol\nabla}A_0\right)^2\right]-\frac{1}{4!}\zeta_0 g_{\omega}^{2}\omega_0^4
-\frac{\eta_{2\boldsymbol\rho}}{4M^2}g_{\omega}^2 m_{\boldsymbol\rho}^{2}
\omega_0^2 R_{0,0}^2\Bigg\}\nonumber\\
&=&\sum_{\alpha}w_{\alpha}\epsilon_{\alpha}+\int d^3r\frac{1}{2}\Bigg[g_{\sigma}\rho_{\text{s}}\sigma
-\frac{\kappa_3}{6M}g_{\sigma}m_{\sigma}^2\sigma^3-\frac{{\kappa_4}}{6M^2}g_{\sigma}^2 m_{\sigma}^2
\sigma^4\nonumber\\
&&-g_{\omega}\rho_0\omega_0+\frac{1}{6}\zeta_0 g_{\omega}^{2}\omega_0^4-\frac{1}{2}g_{\boldsymbol\rho}
\rho_{0,0}R_{0,0}+\frac{\eta_{2\boldsymbol\rho}}{2M^2}g_{\omega}^2 m_{\boldsymbol\rho}^{2}
\omega_0^2 R_{0,0}^2-e\rho_{p,0}A_0\Bigg].
\end{eqnarray}
For spherically symmetric mean-fields i.e. $\sigma=\sigma(|r|)$, $\omega_0=
\omega_0(|r|)$ and so on the two component nucleon wave-function can be 
expressed as,
\begin{eqnarray} 
\label{dirac_spinor}
\phi_{\alpha}=\left(\begin{array}{c}
i\frac{G_{\alpha}(r)}{r}\mathcal{Y}_{j_{\alpha}l_{\alpha}m_{\alpha}}\\ 
\frac{F_{\alpha}(r)}{r}\frac{\boldsymbol\sigma\cdot{\bf p}}{r}
\mathcal{Y}_{j_{\alpha}l_{\alpha}m_{\alpha}}
\end{array}\right),
\end{eqnarray}
where $\mathcal{Y}_{jlm}$ denotes the spinor spherical harmonics and $G_{\alpha}$ 
and $F_{\alpha}$ are the radial parts of the two component nuclear 
wave-function (similar to Eq. (\ref{wfn_spherical})). $G_{\alpha}$ and $F_{\alpha}$ 
are further subject to normalization,
\begin{eqnarray} 
\label{spinor_norm}
\int_0^{\infty}dr\left\{|G_{\alpha}|^2+|F_{\alpha}|^2\right\}=1.
\end{eqnarray}
Here, both $G_{\alpha}$ and $F_{\alpha}$ can be considered as real. Then the 
densities in Eq. (\ref{density_rmf2}) take the form,
\begin{eqnarray} 
\label{density_rmf_sph}
\rho_{\text{s}} &=&\frac{1}{4\pi r^2}\sum_{\alpha}w_{\alpha}(2j_{\alpha}+1)(G_{\alpha}^2-F_{\alpha}^2),\nonumber\\
\rho_{0}&=&\frac{1}{4\pi r^2}\sum_{\alpha}w_{\alpha}(2j_{\alpha}+1)(G_{\alpha}^2+F_{\alpha}^2),\nonumber\\
\rho_{0,0}&=&\frac{1}{4\pi r^2}\sum_{\alpha}w_{\alpha}(2j_{\alpha}+1)\tau_{0\alpha}(G_{\alpha}^2+F_{\alpha}^2),\nonumber\\
\rho_{p,0}&=&\frac{1}{2}\left(\rho_0+\rho_{0,0}\right).
\end{eqnarray}
With these form of the densities, the Euler-Lagrange equations for the mesons 
and the electromagnetic field look like Laplace equations,
\begin{eqnarray}
\label{eom_rmf_sph}
\left(-\frac{d^2}{dr^2}+m_{\sigma}^2\right)\sigma&=&\left\{g_{\sigma}\rho_{\text{s}}
-\frac{{\kappa_3}}{2M} g_{\sigma}m_{\sigma}^2\sigma^2-\frac{{\kappa_4}}
{3!M^2}g_{\sigma}^2 m_{\sigma}^2\sigma^3\right\},\nonumber\\
\left(-\frac{d^2}{dr^2}+m_{\omega}^2\right){\omega}_{0}&=&\left\{g_{\omega}\rho_{0}
-\frac{1}{3!}\zeta_0 g_{\omega}^{2}{\omega}_{0}^3-\frac{\eta_{2\boldsymbol\rho}}{2M^2}g_{\omega}^2
m_{\boldsymbol\rho}^{2}{R}_{0,0}^2{\omega}_0\right\},\nonumber\\
\left(-\frac{d^2}{dr^2}+m_{\boldsymbol\rho}^2\right)R_{0,0}&=&\left\{\frac{1}{2}g_{\boldsymbol\rho} \rho_{0,0}
-\frac{\eta_{2\boldsymbol\rho}}{2M^2}g_{\omega}^2m_{\boldsymbol\rho}^{2}{\omega}_0^2R_{0,0}\right\},\nonumber\\
-\frac{d^2}{dr^2} A_0&=&e\rho_{p,0}.
\end{eqnarray}
Similarly, the coupled Dirac equation for the radial wave-functions are 
given by,
\begin{eqnarray}
\label{dirac_coupled}
\epsilon_{\alpha}G_{\alpha}&=&\left(-\frac{d}{dr}+\frac{a_{\alpha}}{r}\right)F_{\alpha}
+\left(M-g_{\sigma}\sigma+g_{\omega}\omega_{0}+\frac{1}{2}g_{\boldsymbol\rho}{\tau}_{0\alpha} 
\cdot{R}_{0,0}+e\frac{1+\tau_{0\alpha}}{2}A_{0}\right)G_{\alpha},\nonumber\\
\epsilon_{\alpha}F_{\alpha}&=&\left(\frac{d}{dr}+\frac{a_{\alpha}}{r}\right)G_{\alpha}
-\left(M-g_{\sigma}\sigma-g_{\omega}\omega_{0}-\frac{1}{2}g_{\boldsymbol\rho}{\tau}_{0\alpha} 
\cdot{R}_{0,0}-e\frac{1+\tau_{0\alpha}}{2}A_{0}\right)F_{\alpha},\nonumber\\
\end{eqnarray}
where
\begin{eqnarray}
a_{\alpha}=\begin{cases}
-\left(j_{\alpha}+\frac{1}{2}\right)\ \ \ \ \text{for}\ j=l+\frac{1}{2}\\
+\left(j_{\alpha}+\frac{1}{2}\right)\ \ \ \ \text{for}\ j=l-\frac{1}{2}\end{cases}\ \ .
\end{eqnarray}
One can eliminate $F_{\alpha}$ in Eq. (\ref{dirac_coupled}) and expression for 
$G_{\alpha}$ can be obtained as,
\begin{eqnarray}
\label{dirac_schrodinger}
\epsilon_{\alpha}G_{\alpha}&=&-\left(\frac{d}{dr}-\frac{a_{\alpha}}{r}\right)M_{\text{eff}}^{-1}
\left(\frac{d}{dr}+\frac{a_{\alpha}}{r}\right)G_{\alpha}+U_{\text{eff}}G_{\alpha},\\
\text{with},\ \ M_{\text{eff}}&=&\epsilon_{\alpha}
+M-g_{\sigma}\sigma-g_{\omega}\omega_{0}-\frac{1}{2}g_{\boldsymbol\rho}{\tau}_{0\alpha} 
\cdot{R}_{0,0}-e\frac{1+\tau_{0\alpha}}{2}A_{0},\nonumber\\
U_{\text{eff}}&=&M-g_{\sigma}\sigma+g_{\omega}\omega_{0}+\frac{1}{2}g_{\boldsymbol\rho}{\tau}_{0\alpha} 
\cdot{R}_{0,0}+e\frac{1+\tau_{0\alpha}}{2}A_{0}\nonumber.
\end{eqnarray}
Eq. (\ref{dirac_schrodinger}) looks very similar to Eq. (\ref{schrodinger_coupled}). 
$F_{\alpha}$ in Eq. (\ref{dirac_coupled}) is then reconstructed by,
\begin{eqnarray}
\label{bottom_spinor}
F_{\alpha}=M_{\text{eff}}^{-1}\left(\frac{d}{dr}+\frac{a_{\alpha}}{r}\right)G_{\alpha}.
\end{eqnarray}

The solution for $F_{\alpha}$ and $G_{\alpha}$ can be obtained by the following 
way. First a guess solution, typically a Wood-Saxon or Harmonic oscillator 
function is taken for $F_{\alpha}$ and $G_{\alpha}$ with a corresponding set 
of $w_{\alpha}$'s. A guess solution for $\sigma$ field is taken as a Fermi 
function. Now, the different densities are calculated accordingly following 
Eq. (\ref{density_rmf_sph}). Upon that, the solutions for other mesons and 
electromagnetic field are found from Eq. (\ref{eom_rmf_sph}). Consequently, 
$M_{\text{eff}}$ and $U_{\text{eff}}$ are calculated from Eq. 
(\ref{dirac_schrodinger}). Then new solution of $G_{\alpha}$ is found by 
using Eq. (\ref{dirac_schrodinger}) subject to the normalization,
\begin{eqnarray} 
\label{norm_sph}
\int dr\left\{G_{\alpha}G_{\beta}+F_{\alpha}F{\beta}\right\}=\delta_{\alpha\beta},
\end{eqnarray}
where, $F_{\alpha}$ is obtained by using Eq. (\ref{bottom_spinor}). With these 
new $G_{\alpha}$ and $F_{\alpha}$, new set of $\epsilon_{\alpha}$'s are obtained 
by solving the coupled Dirac equation as,
\begin{eqnarray} 
\label{new_dirac}
\epsilon_{\alpha}=\int_0^{\infty}dr\left\{F_{\alpha}\left(\frac{d}{dr}+
\frac{a_{\alpha}}{r}\right)G_{\alpha}+G_{\alpha}\left(-\frac{d}{dr}+
\frac{a_{\alpha}}{r}\right)F_{\alpha}+G_{\alpha}U_{\text{eff}}G_{\alpha}-F_{\alpha}
\left(M_{\text{eff}}-\epsilon_{\alpha}\right)F_{\alpha}\right\}.\nonumber\\
\end{eqnarray}
In the right hand side of Eq. (\ref{new_dirac}) old set of $\epsilon_{\alpha}$'s 
are used. In presence of pairing interaction, the $w_{\alpha}$'s calculated now
using Eq. (\ref{w_alpha}). This whole process is repeated until a self-consistent 
solution is obtained.

\subsection{Pairing in BCS approximation}
Pairing plays a very important role in the occupation probability of
the single particle levels near the Fermi surface. Well below the Fermi
surface the occupation probability of a single-particle state is unity
and it becomes nearly zero well above the Fermi surface. However, near the Fermi
surface it becomes a fractional number lying between zero and unity. It
can be understood by means of a coupling between a single particle state
with its time-reversed partner. In this thesis work pairing for finite
nuclei is treated in BCS approximation which was originally given by
Bardeen, Cooper, and Schrieffer \cite{Bardeen57,Bardeen57a} for electronic
systems. A more general version of the BCS pairing can be applied
using Lipkin-Nogami pairing model \cite{Lipkin60,Nogami64}. However,
here pairing is restricted to a constant gap model.

At second quantization Hamiltonian of nuclear system can be written as, 
\begin{eqnarray} 
\label{pairing_hamiltonian}
\hat{H}=\sum_{\alpha} \epsilon_{\alpha}^0 \hat{a}_{\alpha}^{\dagger}\hat{a}_{\alpha}+\sum_{{\alpha},{\alpha}^{\prime}> 0}
\left<{\alpha},-{\alpha}\left|v\right|{\alpha}^{\prime},-{\alpha}^{\prime}\right>\hat{a}_{\alpha}^{\dagger}\hat{a}_{-{\alpha}}^{\dagger}
\hat{a}_{-{\alpha}^{\prime}}\hat{a}_{{\alpha}^{\prime}},
\end{eqnarray}
where $\hat{a}_{\alpha}^{\dagger}$ creates a particle in single particle state 
$|\alpha\rangle$ and $\hat{a}_{\alpha}$ annihilates a particle in quantum state 
$|\alpha\rangle$. $|-\alpha\rangle$ corresponds to a time reversed state 
of $|\alpha\rangle$ with opposite spin. The first term in the 
right hand side of Eq. (\ref{pairing_hamiltonian}) represents the sum over all the 
occupied states below the Fermi surface. The second term represents the residual 
interaction essentially between a state with its time-reversed partner. 
Considering a constant matrix element $-G$ for the interaction in the second term, 
the pairing Hamiltonian in Eq. (\ref{pairing_hamiltonian}) can be written as, 
\begin{eqnarray} 
\label{const_hamiltonian}
\hat{H}=\sum_{\alpha} \epsilon_{\alpha}^0 \hat{a}_{\alpha}^{\dagger}\hat{a}_{\alpha}-G\sum_{{\alpha},{\alpha}^{\prime}> 0}
\hat{a}_{\alpha}^{\dagger}\hat{a}_{-{\alpha}}^{\dagger}
\hat{a}_{-{\alpha}^{\prime}}\hat{a}_{{\alpha}^{\prime}}.
\end{eqnarray}
An analytic solution to this equation is not available. An approximate solution 
is provided by the BCS state which is given by,
\begin{eqnarray} 
\label{BCS_state}
|\varphi_{\text{\tiny BCS}}\rangle=\prod_{{\alpha}>0}^{\infty}\left(u_{\alpha}+v_{\alpha}\hat{a}_{\alpha}^{\dagger}
\hat{a}_{-{\alpha}}^{\dagger}\right)|0\rangle .
\end{eqnarray}
It signifies that the state $(\alpha,-\alpha)$ is occupied with probability 
$|v_{\alpha}|^2$ and is vacant with probability $|u_{\alpha}|^2$. In 
practical purpose $u_{\alpha}$ and $v_{\alpha}$ are considered to be 
real numbers. The normalization condition is given by, 
\begin{eqnarray} 
\label{norm_BCS}
\langle \varphi_{\text{\tiny BCS}}|\varphi_{\text{\tiny BCS}} \rangle
&=&\prod_{{\alpha}>0}^{\infty}\left(u_{\alpha}^2+v_{\alpha}^2\right),\nonumber
\end{eqnarray}
\begin{eqnarray}
u_{\alpha}^2+v_{\alpha}^2&=&1.
\end{eqnarray}
In Eq. (\ref{norm_BCS}), the quantity $v_{\alpha}$ can be identified with 
$w_{\alpha}$ in Eq. (\ref{w_alpha}). The expectation value of the number operator 
is given by,
\begin{eqnarray} 
\label{number_BCS}
N=\langle \varphi_{\text{\tiny BCS}}|\hat{N}|\varphi_{\text{\tiny BCS}} \rangle&=&\langle \varphi_{\text{\tiny BCS}}|\sum_{{\alpha}> 0}\left(\hat{a}_{\alpha}^{\dagger}\hat{a}_{\alpha}
+\hat{a}_{-{\alpha}}^{\dagger}\hat{a}_{-{\alpha}}\right)|\varphi_{\text{\tiny BCS}} \rangle\nonumber\\
&=&\sum_{{\alpha}> 0}2v_{\alpha}^2.
\end{eqnarray}
Clearly, $N$ is not a good quantum number for the Hamiltonian in Eq. 
(\ref{pairing_hamiltonian}), which can be understood by studying the 
particle number uncertainty as,
\begin{eqnarray} 
\label{number_uncertainty}
\Delta N^2&=&\langle \varphi_{\text{\tiny BCS}}|\hat{N}^2|\varphi_{\text{\tiny BCS}}\rangle-\langle \varphi_{\text{\tiny
BCS}}|\hat{N}|\varphi_{\text{\tiny BCS}}\rangle^2\nonumber\\
&=&\Big(4\sum_{\substack{{\alpha}\ne {\alpha}^{\prime}\\ {\alpha}{\alpha}^{\prime}>0}}v_{\alpha}^2 v_{{\alpha}^{\prime}}^2+4\sum_{{\alpha}>0} 
v_{\alpha}^2\Big)-\Big(\sum_{{\alpha}> 0}2v_{\alpha}^2\Big)^2\nonumber\\
&=&4\sum_{{\alpha}>0}u_{\alpha}^2 v_{\alpha}^2\ne 0\ \ (\text{if}\ v_{\alpha}\ne 0\ \text{or}\ 1).
\end{eqnarray}
Clearly, $N$ becomes a good quantum number for $v_{\alpha}=0$ or 1. For fractional 
occupation probability the values of $u_{\alpha}$ and $v_{\alpha}$ are found by solving 
a variational equation taking a product of Lagrange multiplier $\lambda$ and $N$ and 
subtracting it from the Hamiltonian in Eq. (\ref{pairing_hamiltonian}) as,
\begin{eqnarray} 
\label{variation_BCS}
\delta\langle \varphi_{\text{\tiny BCS}}|\hat{H}&-&\lambda\hat{N}|\varphi_{\text{\tiny BCS}}\rangle=0\nonumber\\
\Rightarrow\frac{\partial}{\partial v_{\alpha}}\langle \varphi_{\text{\tiny BCS}}|\sum_{\alpha}
(\epsilon_{\alpha}^0-\lambda)\hat{a}_{\alpha}^{\dagger}\hat{a}_{{\alpha}}
&-&G\sum_{{\alpha},{\alpha}^{\prime}> 0}\hat{a}_{\alpha}^{\dagger}\hat{a}_{-{\alpha}}^{\dagger}\hat{a}_{-{\alpha}^{\prime}}
\hat{a}_{{\alpha}^{\prime}}|\varphi_{\text{\tiny BCS}}\rangle=0.
\end{eqnarray}
Now, using the normalization condition $u_{\alpha}^2+v_{\alpha}^2=1$ one can write,
\begin{eqnarray} 
\label{var_operator}
\frac{\partial}{\partial v_{\alpha}}=\left.\frac{\partial}{\partial v_{\alpha}}\right|_{u_{\alpha}}-\frac{v_{\alpha}}{u_{\alpha}}
\left.\frac{\partial}{\partial v_{\alpha}}\right|_{v_{\alpha}}\ \ .
\end{eqnarray}
Further exploiting few expectation values,
\begin{eqnarray} 
\label{expectation_operator}
\langle \varphi_{\text{\tiny BCS}}|\hat{a}_{\alpha}^{\dagger}\hat{a}_{{\alpha}}|\varphi_{\text{\tiny BCS}}
\rangle&=&v_{\alpha}^2,\nonumber\\
\langle \varphi_{\text{\tiny BCS}}|\hat{a}_{\alpha}^{\dagger}\hat{a}_{-{\alpha}}^{\dagger}
\hat{a}_{-{\alpha}^{\prime}}\hat{a}_{{\alpha}^{\prime}}
|\varphi_{\text{\tiny BCS}}\rangle&=&
\begin{cases}u_{\alpha}v_{\alpha}u_{{\alpha}^{\prime}}v_{{\alpha}^{\prime}}\ \text{for}\ {\alpha}\ne {\alpha}^{\prime}\\
v_{\alpha}^2\ \ \ \ \ \ \ \ \ \ \ \ \text{for}\ {\alpha}={\alpha}^{\prime}\end{cases}\ ,
\end{eqnarray}
the second matrix element in Eq. (\ref{variation_BCS}) can be written as,
\begin{eqnarray} 
\label{var_mat_element}
\langle \varphi_{\text{\tiny BCS}}|-G\sum_{{\alpha},{\alpha}^{\prime}> 0}\hat{a}_{\alpha}^{\dagger}
\hat{a}_{-{\alpha}}^{\dagger}\hat{a}_{-{\alpha}^{\prime}}
\hat{a}_{{\alpha}^{\prime}}|\varphi_{\text{\tiny BCS}}\rangle&=&-G\sum_{\substack{{\alpha}\ne {\alpha}^{\prime}\\
{\alpha}{\alpha}^{\prime}>0}}u_{\alpha}v_{\alpha}
u_{{\alpha}^{\prime}}v_{{\alpha}^{\prime}}-G\sum_{{\alpha}>0}v_{\alpha}^2\nonumber\\
&=&-G\left(\sum_{{\alpha}>0}u_{\alpha}v_{\alpha}\right)^2-G\sum_{{\alpha}>0}v_{\alpha}^4\ .
\end{eqnarray}
So implementing the variational equation in Eq. (\ref{variation_BCS}) and using the 
derivative in Eq. (\ref{var_operator}) one arrives at,
\begin{eqnarray} 
\label{}
\frac{\partial}{\partial v_{\alpha}}\left[2\sum_{{\alpha}>0} \left(\epsilon_{\alpha}^0-\lambda\right)v_{\alpha}^2\right.&-&\left.
G\left(\sum_{{\alpha}>0}u_{\alpha}v_{\alpha}\right)^2-G\sum_{{\alpha}>0}v_{\alpha}^4\right]=0\nonumber\\
4\left(\epsilon_{\alpha}^0-\lambda\right)v_{\alpha}-2G\left(\sum_{{\alpha}^{\prime}}u_{{\alpha}^{\prime}}v_{{\alpha}^{\prime}}\right)
u_{\alpha}&-&4Gv_{\alpha}^3-\frac{v_{\alpha}}{u_{\alpha}}\left[-2G\left(\sum_{{\alpha}^{\prime}>0}u_{{\alpha}^{\prime}}v_{{\alpha}^{\prime}}\right)v_{\alpha}\right]=0.\nonumber\\
\end{eqnarray}
Now with some definitions,
\begin{eqnarray} 
\label{def_gap}
\Delta&=&G\sum_{{\alpha}^{\prime}>0}u_{{\alpha}^{\prime}}v_{{\alpha}^{\prime}},\nonumber\\
\epsilon_{\alpha}&=&\epsilon_{\alpha}^0-\lambda-Gv_{\alpha}^2,
\end{eqnarray}
the variational equation [Eq. (\ref{variation_BCS})] takes the form,
\begin{eqnarray} 
\label{variation_BCS2}
2\epsilon_{\alpha}v_{\alpha}u_{\alpha}+\Delta\left(v_{\alpha}^2-u_{\alpha}^2\right)=0.
\end{eqnarray}
In Eq. (\ref{def_gap}), $\Delta$ is called as pairing gap and $\lambda$ is 
identified as the chemical potential. Now, putting the 
condition $\epsilon_{\alpha}\rightarrow\infty\Rightarrow v_{\alpha}=0$ one gets,
\begin{eqnarray} 
\label{v_u_alpha}
v_{\alpha}^2&=&\frac{1}{2}\left(1-\frac{\epsilon_{\alpha}}{\sqrt{\epsilon_{\alpha}^2+\Delta^2}}\right),\nonumber\\
u_{\alpha}^2&=&\frac{1}{2}\left(1+\frac{\epsilon_{\alpha}}{\sqrt{\epsilon_{\alpha}^2+\Delta^2}}\right).
\end{eqnarray}
So the gap-equation is given by,
\begin{eqnarray} 
\label{gap_equation}
\Delta&=&G\sum_{{\alpha}>0}u_{{\alpha}}v_{{\alpha}}\nonumber\\
&=&\sum_{{\alpha}>0}\frac{G}{2}\sqrt{1-\frac{\epsilon_{\alpha}^2}{\epsilon_{\alpha}^2+\Delta^2}}\nonumber\\
\Rightarrow\Delta&=&\frac{G}{2}\sum_{{\alpha}>0}\frac{\Delta}{\sqrt{\epsilon_{\alpha}^2+\Delta^2}}.
\end{eqnarray}

In the present thesis work, constant gap
(i.e. $\Delta=\frac{11.2}{\sqrt{A}}$ MeV \cite{Reinhard86}) BCS
approximation for the pairing is used. For the self-consistent
determination of the radial wave functions in Skyrme or RMF formalism,
first a set of single particle energies along with a guess value of
$\lambda$ and $G$ enter into the pairing calculation, separately
for neutrons and protons. First value of $\lambda$ is determined
self-consistently by using equation for $\epsilon_{\alpha}$ in
Eq. (\ref{def_gap}) and $v_{\alpha}^2$ in Eq.  (\ref{v_u_alpha})
fulfilling the number equation in Eq. (\ref{number_BCS}).

Depending on the isospin of the particles concerned, $N$ becomes number of neutrons or 
protons. Finally, the value of $G$ is calculated from Eq. (\ref{gap_equation}). 
The iteration to determine $v_{\alpha}$ and pairing strength $G$ runs inside 
the iteration process of Skyrme of RMF formalism. At the final step while 
determining the energy in Eq. (\ref{skm_etot}) or (\ref{rmf_etot}), pairing 
energy is added separately as, 
\begin{eqnarray} 
\label{pairing_energy}
E_{q,\text{pair}}=\frac{-\Delta_q^2}{G_q},
\end{eqnarray}
where, $q$ is either neutron or proton.

\section{Infinite Nuclear Matter}\label{sec_INM}
Infinite nuclear matter is a hypothetical isotropic system of infinite 
number of nucleons with no boundary and Coulomb interaction. 
The energy per nucleon for infinite nuclear matter with density $\rho=(\rho_n+
\rho_p)$ and isospin asymmetry $\delta=\left(\frac{\rho_n-\rho_p}{\rho}\right)$ 
can be written as a Taylor's expansion as,
\begin{eqnarray} 
\label{ener_nm}
\mathcal{E}(\rho,\delta)&\approx&\mathcal{E}(\rho,\delta=0)+\frac{1}{2}
\left(\frac{\partial^2\mathcal{E}(\rho,\delta)}{\partial\delta^2}\right)_{\delta=0}\delta^2
+\frac{1}{4!}\left(\frac{\partial^4\mathcal{E}(\rho,\delta)}{\partial\delta^4}\right)_{\delta=0}\delta^4\nonumber\\
&\approx&\mathcal{E}(\rho,\delta=0)+C_2(\rho)\delta^2+C_4(\rho)\delta^4,
\end{eqnarray}
where, $\mathcal{E}(\rho,\delta=0)$ represents the energy per nucleon
for symmetric nuclear matter and $C_2(\rho)$ is the symmetry energy. As
the nuclear force symmetric under the exchange of neutrons and protons,
the expansion contains only the even powers of $\delta$. For finite
nuclear systems the effect of asymmetry is quite less compared to the
symmetric part. The maximum asymmetry associated with a nucleus is $\delta 
\sim0.33$. So, the third term in the RHS of Eq. (\ref{ener_nm}) has very less 
contribution to the binding energy of a nucleus. 
However, for dense asymmetric
systems like  neutron star, where the concerned asymmetry $\delta\sim0.7$, 
$C_4(\rho)$ may contribute non-vanishingly to the system. 
Throughout this thesis work the expansion is thus restricted up to order of $\delta^2$. 
The symmetric nuclear matter attains a saturation where
the energy minimizes. The corresponding density is coined as "saturation
density" $(=\rho_0)$. At the center of the nuclei the density associated
with finite nuclei is very close to this density. Thus, the
saturation density is a very important quantity to
estimate from theoretical models.  All quantities characterizing infinite
nuclear matter are evaluated at saturation density from a theoretical model, which
are eventually used to construct the equation of state relevant for dense
matter, subjected to heavy ion collision experiments or astrophysical
observations.

Now, energy for symmetric matter $\mathcal{E}(\rho,\delta=0)$ or $\mathcal{E}(\rho,0)$ can be
expanded around $\rho_0$ as,
\begin{eqnarray} 
\label{ener_snm}
\mathcal{E}(\rho,0)\approx\mathcal{E}(\rho_0)&+&\frac{1}{2}K_0\left(\frac{\rho_0-\rho}{3\rho_0}\right)^2
-\frac{1}{6}Q_0\left(\frac{\rho_0-\rho}{3\rho_0}\right)^3,\nonumber\\
\text{with},\ \ K_0&=&9\rho_0^2\left(\frac{\partial^2\mathcal{E}(\rho,0)}{\partial\rho^2}\right)_{\rho_0},\nonumber\\
Q_0&=&27\rho_0^3\left(\frac{\partial^3\mathcal{E}(\rho,0)}{\partial\rho^3}\right)_{\rho_0}.
\end{eqnarray}
In the right hand side of Eq. (\ref{ener_snm}), the first derivative vanishes 
as the energy attains its minimum at $\rho_0$. The quantity $K_0$ and $Q_0$ are 
called the incompressibility and skewness parameter for symmetric matter. 
Similar to $\mathcal{E}(\rho,0)$ in Eq. (\ref{ener_snm}), one can also 
expand $C_2(\rho)$ in Eq. (\ref{ener_nm}) around $\rho_0$ as, 
\begin{eqnarray} 
\label{sym_expand}
C_2(\rho)\approx C_2^0-L_0\left(\frac{\rho_0-\rho}{3\rho_0}\right)
&+&\frac{1}{2}K_{sym}^0\left(\frac{\rho_0-\rho}{3\rho_0}\right)^2
-\frac{1}{6}Q_{sym}^0\left(\frac{\rho_0-\rho}{3\rho_0}\right)^3,\nonumber\\
\text{with},\ \ C_2^0&=&C_2(\rho_0),\nonumber\\
L_0&=&3\rho_0\left(\frac{\partial C_2(\rho)}{\partial\rho}\right)_{\rho_0},\nonumber\\
K_{sym}^0&=&9\rho_0^2\left(\frac{\partial^2 C_2(\rho)}{\partial\rho^2}\right)_{\rho_0},\nonumber\\
Q_{sym}^0&=&27\rho_0^3\left(\frac{\partial^3 C_2(\rho)}{\partial\rho^3}\right)_{\rho_0}.
\end{eqnarray}
Here, $L_0$ is slope parameter of symmetry energy, $K_{sym}^0$ the 
symmetry energy curvature parameter and $Q_{sym}^0$ the symmetry 
energy skewness parameter. These quantities play very important roles 
in the study of asymmetric systems e.g. nuclei
near drip line or astrophysical objects like neutron star.

For nuclear matter, single particle levels are of no interest. In
mean-field formalism, with a given set of parameters, saturation density
$\rho_0$ can be determined  by minimizing the energy for symmetric matter 
with respect to particle density $\rho$. To do so, different type of densities 
in Eq. (\ref{density_skm}) and (\ref{density_rmf}) are taken to be 
equal for neutrons and protons. Moreover, infinite nuclear matter
is uniform by definition. So gradients of all type of densities
vanish identically inside nuclear matter. Other
symmetric nuclear matter properties defined in Eq. (\ref{ener_snm}) are
then calculated accordingly by performing numerical derivatives of 
energy density with respect to $\rho$. For asymmetric nuclear matter,
symmetry energy parameters defined in Eq. (\ref{sym_expand}) can be calculated by taking 
numerical derivatives of energy with respect to corresponding powers of 
 $\delta$ and $\rho$.

%% file: chap3.tex
\chapter{Error Analysis}\label{ch3}
\section{Introduction}
In 1976, George E. P. Box commented on 'Science and statistics' as,
"{\it Since all models are wrong the scientist can not obtain a
"correct" one by excessive elaboration}" \cite{Box76}. Probably,
nothing can put more aptly the hazards of extrapolating theoretical
models. Any experimental measurement is not acceptable without
specification on the uncertainties. However, "{\it its all too often
the case that the numerical results are presented without uncertainty
estimates}"; as pointed out by the editors of Physical Review A
\cite{Editor11}. More often than not, theoretical models involve
prediction of observables beyond its domain of validity. For example,
models based on non-relativistic Skyrme force or relativistic mean field
(RMF) obtained by fitting experimental data on finite nuclei are often
used to predict neutron-star properties. So, estimating the statistical
uncertainties for nuclear models is inevitable.

As the basic nuclear interaction between two nucleons is not known
exactly, uncertainties are bound to creep in for the nuclear models
obtained by fitting properties of finite nuclei and neutron stars.
If the models were exact, any prediction by the models would match
exactly with the experiment or observation, leaving no room for new
measurements to help in any new understanding. On the contrary, for a
model built without any preconceived fundamental knowledge, all the
measured or observed quantities would be independent to each other,
resulting in zero predictive power. The real scenario lies in between
these two extreme situations. Few recent calculations \cite{Kortelainen10,
Reinhard10, Dobaczewski14, Piekarewicz15, Erler15} put forward an
extensive importance on the error estimation in theoretical models,
which includes error on the optimized parameters as well as on the
predicted or estimated experimental and empirical observables. Moreover,
statistical analyses address how fast the objective function (typically
a $\chi^2$ function) moves away from its minimum value when one perturbs
the optimized parameter set. Depending on the set of observables one uses
to optimize the parameter space, correlation may exist among different
observables and parameters. Studying the correlations among different
observables and parameters offers a load of information which can't be
comprehended otherwise.

\section{Covariance analysis}
To obtain the optimized model parameters, a set of experimental data is 
fitted. First a suitable objective function is minimized, which is 
defined as \cite{Reinhard10, Dobaczewski14},
\begin{eqnarray}
\label{chidef}
\chi^2({\bf p})&=&\sum\limits_{i=1}^{N_d}\left (\frac{\mathcal{O}_i^{th}
-\mathcal{O}_i^{exp}}{\Delta\mathcal{O}_i}\right)^2.
\end{eqnarray}
In Eq. (\ref{chidef}), ${\bf p}\ (p_1,\ p_2\ ....\ p_{N_p} )$ is the
parameters set with typically $N_d\sim10$ for nuclear models. $N_d$ is
the number of data used to fit the parameters. $\mathcal{O}^{th}$ is the
value of an observable calculated theoretically with the parameter set
${\bf p}$. $\mathcal {O}^{exp}$ is experimental value of the corresponding
observable. The quantity $\Delta\mathcal{O}$ represents the adopted error
on any observable $\mathcal{O}$ which is given by $\Delta\mathcal{O}^2=
(\Delta \mathcal{O}^{th})^2+ (\Delta \mathcal{O}^{exp})^2+ (\Delta
\mathcal{O}^{num})^2$, where contributions come from theory, experiment
as well as numerical methods associated with the analysis. Out of
these three contributions the most undetermined one is the theoretical
error. One needs to be very careful to estimate the adopted theoretical
error. Often nuclear models miss certain many body correlations in its
formulation. Demanding too much accuracy on certain channels may end
up in erroneous optimizations. Following the definition of $\chi^2$
in Eq. (\ref{chidef}), one can also define the likelihood function of
a parameter set by,
\begin{eqnarray}
\label{likelihood}
 {L}({\bf p})&=&N \text {exp}\left [{-\frac{1}{2}\chi^2({\bf p})}\right ].
\end{eqnarray}
As the name suggests, it determines the likelihood of a parameter 
set ${\bf p}$ to reproduce the experimental data. Immediately the average 
value of a quantity $A({\bf p})$ can be obtained as,
\begin{eqnarray}
\label{average_a}
\overline{A}&=& \int {L} ({\bf p})\cdot A({\bf p})d{\bf p}.
\end{eqnarray}
Using the likelihood function, variance on $A$ and covariance between 
$A$ and $B$ can be expressed as, 
\begin{eqnarray}
\label{var_covar}
\overline{(\Delta A)^2}&=& \int {L} ({\bf p})\cdot \left(A({\bf p})- \overline{A}\right)^2d{\bf p},\nonumber\\
\overline{(\Delta A \Delta B)}&=&\int {L} ({\bf p})\cdot \left(A({\bf p})- \overline{A}\right)
\left(B({\bf p})- \overline{B}\right) d{\bf p}.
\end{eqnarray}
The correlation coefficient between $A$ and $B$ can be obtained 
as \cite{Brandt97},
\begin{eqnarray}
\label{corr}
{C}_{AB} &=&
  \frac{\overline{\Delta {A}\,\Delta {B}}}
{\sqrt{\overline{\Delta {A}^2}\;\overline{\Delta {B}^2}}}.
\end{eqnarray}

The ideal way to obtain computationally the average or variance of any quantity and 
covariance between two quantities would to make samples over the whole 
parameter space using a Metropolis-Monte-Carlo algorithm originally 
given by Metropolis et al \cite{Metropolis53}. In a Monte-Carlo approach 
over a huge sample parameter space the definitions of average, variance of 
a quantity and covariance between two quantities would be redefined as,
\begin{eqnarray}
\label{metropolis}
\overline{A}&=&\displaystyle\lim_{M\to\infty} \frac{1}{M}\sum_{m=1}^{M} A({\bf p}_m) \nonumber\\
\overline{(\Delta A)^2}&=&\displaystyle\lim_{M\to\infty} \frac{1}{M}\sum_{m=1}^{M} (A({\bf
p}_m)- \overline{A})^2\nonumber\\
\overline{(\Delta A \Delta B)}&=&\displaystyle\lim_{M\to\infty} \frac{1}{M}\sum_{m=1}^{M} 
(A({\bf p}_m)- \overline{A})  (B({\bf p}_m)- \overline{B})
\end{eqnarray}
Typical calculation of the objective function $\chi^2$ with $\sim$
10 parameters in a nuclear model takes few minutes of computation
time. Sampling over several thousands of parameter sets would end up
taking few months of computation time to estimate the uncertainties. So a
more efficient method is essential to perform statistical error analysis
involving nuclear models.

\begin{figure}[]{}
\centering
\includegraphics[width=0.7\textwidth]{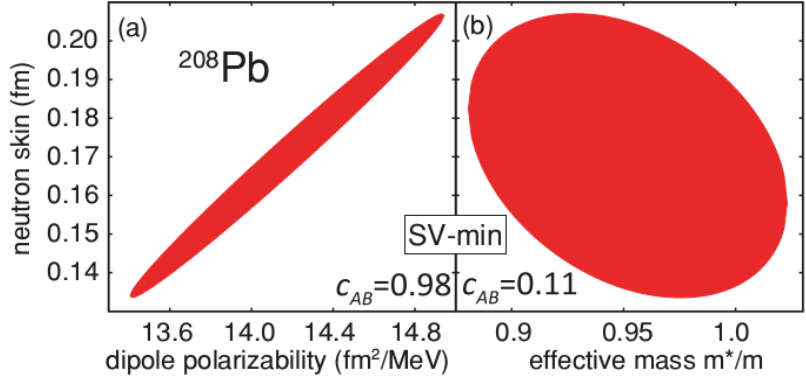}
\caption{\label{ch3_fig1}
Covariance ellipsoids for two pairs of observables as given in Ref. \cite{Reinhard10}. 
See text for details.} 
\end{figure}

The most tractable method to calculate the statistical uncertainties is
{\it covariance analysis} \cite{Brandt97}. First the $\chi^2$ function
is minimized following a derivative method. A typical method would be
Levenberg-Marquardt method. Once the minimum of the $\chi^2$ function is
obtained, it can be expanded around the minimum by Taylor's expansion.
Keeping upto quadratic terms the expansion can be approximated as,
\begin{eqnarray}
\label{chiquad}
\chi^2({\bf p}) &\approx& \chi^2({\bf p}_0)+ \sum_{i=1}^{N_p} (p-p_{0})_i \left(\frac{\partial \chi^2({\bf p})}
{\partial {{p}_i}}\right)_{\bf p_0}+\frac{1}{2} \sum_{i,j=1}^{N_p} (p-p_{0})_i \left(\frac{\partial^2 
\chi^2({\bf p})}{\partial {{p}_i}\partial {{p}_j}}\right)_{\bf p_0} (p-p_{0})_j, \nonumber\\
\chi^2({\bf p}) &\approx& \chi^2({\bf p}_0)+ 0 + \sum_{i,j=1}^{N_p}(p-p_{0})_i\mathcal{M}_{ij}(p-p_{0})_j\ \ .
\end{eqnarray}
The first derivative of $\chi^2$ vanishes as the objective function is 
at minimum. Here, Hessian matrix is given by,
\begin{eqnarray}
\mathcal{M}_{ij}=\frac{1}{2}\left(\frac{\partial^2 \chi^2({\bf p})}{\partial {{p}_i}\partial {{p}_j}}\right)_{\bf p_0}.
\end{eqnarray}
Now following Eq. (\ref{likelihood}), the likelihood function in 
quadratic approximation takes the form,
\begin{eqnarray}
\label{likeliquad}
{L}({\bf p})&=&N\ \text {exp}\left [{-\frac{1}{2}\chi^2({\bf p})}\right ]\approx N^\prime \ \text{exp}\left[{-\frac{1}{2} 
\sum_{i,j=1}^n(p-p_{0})_i\mathcal{M}_{ij}(p-p_{0})_j}\right],
\end{eqnarray}
where, contribution of $\chi^2({\bf p}_0)$ is absorbed in the
constant $N^\prime$.  In Eq. (\ref{likeliquad}), if the quantity
inside the exponential remains constant over a sample parameter
sets, it forms an ellipsoid of constant probability surface over
the multidimensional parameter space. It can also be understood
from the the Eq. (\ref{chiquad}) by putting the first term in the
right to the left hand side.  Upon projecting this multidimensional
ellipsoid along any two parameters, one can study correlation between
two parameters. One can also explore correlation between a pair of
observables by exploiting the covariance ellipsoid for parameters. For that 
one needs to calculate the values of the observables of interest using the 
parameters lying within the domain of covariance ellipsoid. A
representative example of the covariance ellipsoid is depicted in
Fig. \ref{ch3_fig1} taken from Ref. \cite{Reinhard10}, where neutron-skin
thickness $\Delta \text{r}_{\text{np}}$ of $^{208}$Pb is plotted against
dipole polarizability and effective mass $\frac{m^*}{m}$ of nucleon. The
thinner shape of the ellipsoid in the ``$\Delta \text{r}_{\text{np}}$
- dipole polarizability'' plane contrasting to that in the ``$\Delta
\text{r}_{\text{np}}$ - $\frac{m^*}{m}$'' plane depicts stronger
correlation in the former compared to the latter.

Now further, integration of the likelihood function over the whole parameter space gives 
unity,
\begin{eqnarray}
\label{intlikeli}
\int_{- \infty}^{+ \infty} {L}({\bf p}) d{\bf p} &=& N^\prime \int_{- \infty}^{+ \infty} \text{exp} \left [-\frac{1}{2}  
({\bf p}-{\bf p_0})^T \mathbfcal{M}({\bf p}-{\bf p_0})\right ] d{\bf p}  =1 .
\end{eqnarray}
Here all the bold notations signify matrices. If there exists an orthogonal 
transformation so that,
\begin{eqnarray}
\mathbfcal{M}&=& {\bf O}^T \tilde {\mathbfcal{M}} {\bf O},
\end{eqnarray}
where, $\tilde{\mathbfcal{M}} = \text{diag}\left [\tilde{\mathcal{M}}_1 \qquad \tilde{\mathcal{M}}_2 \qquad .....
\tilde{\mathcal{M}}_{N_p}\right]$, multidimensional ellipsoid equation can be rewritten as, 
\begin{eqnarray}
\label{multiellipse}
({\bf p}-{\bf p_0})^T \mathbfcal{M} ({\bf p}-{\bf p_0})
&=&({\bf p}-{\bf p_0})^T {\bf O}^T \tilde {\mathbfcal{M}} {\bf O}  ({\bf p}-{\bf p_0})\nonumber\\
&=& ({\bf O} ({\bf p}-{\bf p_0}))^T \tilde {\mathbfcal{M}}{\bf O}  ({\bf p}-{\bf p_0})\nonumber\\
&=&  \tilde {\bf p}^T\tilde{\mathbfcal{M}} \tilde {\bf p}\nonumber\\
&=&\sum_{k=1}^{N_p} \tilde {p}_k \tilde {\mathcal{M}_k} \tilde {p}_k
=\sum_{k=1}^{N_p} \tilde {\mathcal{M}}_k \tilde {p}_k^2\ \ \ .
\end{eqnarray}
So, using Eq. (\ref{intlikeli}) and (\ref{multiellipse}) value of $N^\prime$ 
can be obtained as,
\begin{eqnarray}
\label{gaussconst}
\int_{- \infty}^{+ \infty} {L}({\bf p}) d{\bf p}&=&N^\prime \int_{- \infty}^{+ \infty} \text{exp}
\left [-\frac{1}{2} \sum_{k=1}^{N_p} \tilde {\mathcal{M}}_k \tilde {p}_k^2 d\tilde{p}\right]\nonumber\\
&=& N^\prime \prod_{k=1}^{N_p} \int_{- \infty}^{+ \infty} \text{exp} \left
[-\frac{1}{2} \tilde {\mathcal{M}}_k \tilde {p}_k^2 \right ]d\tilde{p}_k\nonumber\\
&=& N^\prime \prod_{k=1}^{N_p} \left [\frac{2 \pi } {\tilde {\mathcal{M}}_k} \right ]^{1/2}\nonumber\\
\Rightarrow 1&=& N^\prime  (2 \pi) ^{\frac{N_p}{2}}\ \text{Det}^{-\frac{1}{2}} [\tilde {\mathbfcal{M}}]\nonumber\\
\Rightarrow N^\prime &=& \left [\frac{2 \pi } {\tilde {\mathcal{M}}_k} \right ]^{-\frac{N_p}{2}}
=\left \{\frac{(2 \pi )^{N_p}} {\text{Det}[\tilde {\mathbfcal{M}}]} \right \}^{-\frac{1}{2}}\ \ .
\end{eqnarray}
Now, $A({\bf p})$ can be expanded around the optimal parameter set ${\bf p_0}$ as 
in Eq. (\ref{chiquad}) by keeping only upto quadratic terms as,
\begin{eqnarray}
\label{expansion_a}
A({\bf p})&\approx&A({\bf p_0}) + \sum_{k=1}^{N_p}  ({\bf p}-{\bf p_0})_k  
\left (\frac{\partial A}{\partial {p}_k}\right)_{\bf p_0} + \frac{1}{2} \sum_{k,l=1}^{N_p}  ({\bf p}-{\bf p_0})_k 
\left (\frac{\partial^2 A}{\partial p_k \partial p_l}\right)_{\bf p_0}
({\bf p}-{\bf p_0})_l\nonumber \\
&=&A_{\bf 0} +  ({\bf p}-{\bf p_0})^T A^\prime_{\bf 0} + 
\frac{1}{2}  ({\bf p}-{\bf p_0})^T A^{\prime\prime}_{\bf 0} ({\bf p}-{\bf p_0}) \ \ .
\end{eqnarray}
Upon using the expansion of $A({\bf p})$ around ${\bf p_0}$ in Eq. (\ref{expansion_a}) 
and taking help from Eq. (\ref{multiellipse}), the average $\overline{A}$ can be calculated as,
\begin{eqnarray}
\overline{A}&=& \int_{- \infty}^{+ \infty} {L} ({\bf p})\cdot A({\bf p})d{\bf p}\nonumber\\
&=& N^\prime  \int_{- \infty}^{+ \infty} A({\bf p})\ \text{exp} \left [-\frac{1}{2} ({\bf p}-{\bf p_0})^T \mathbfcal{M} 
({\bf p}-{\bf p_0})  \right ] d{\bf p} \nonumber\\
&=& N^\prime  \int_{- \infty}^{+ \infty} \left\{A_{\bf 0} +  ({\bf p}-{\bf p_0})^T A^\prime_{\bf 0} + 
\frac{1}{2}  ({\bf p}-{\bf p_0})^T A^{\prime\prime}_{\bf 0} ({\bf p}-{\bf p_0}) \right\}\nonumber\\
&&\qquad\qquad\qquad\qquad \text{exp} \left [-\frac{1}{2} ({\bf p}-{\bf p_0})^T \mathbfcal{M} 
({\bf p}-{\bf p_0})  \right ] d{\bf p} \nonumber
\end{eqnarray}
\begin{eqnarray}
&=&A_{\bf 0}+0+\frac{1}{2} N^\prime  \int_{- \infty}^{+ \infty}
\left\{({\bf p}-{\bf p_0})^T A^{\prime\prime}_{\bf 0} ({\bf p}-{\bf p_0})\right\}\nonumber\\
&&\qquad\qquad\qquad\qquad \text{exp} \left [-\frac{1}{2} ({\bf p}-{\bf p_0})^T \mathbfcal{M} 
({\bf p}-{\bf p_0})  \right ] d{\bf p} \nonumber\\
&=&A_{\bf 0}+\frac{1}{2} N^\prime  \int_{- \infty}^{+ \infty}
\left\{({\bf p}-{\bf p_0})^T {\bf O} {\bf O}^T A^{\prime\prime}_{\bf 0} 
{\bf O} {\bf O}^T ({\bf p}-{\bf p_0})\right\}
\ \text{exp} \left [-\frac{1}{2} {\tilde{\bf p}}^T \tilde{\mathbfcal{M}}
{\tilde{\bf p}}  \right ] d{\tilde{\bf p}} \nonumber\\
&=&A_{\bf 0}+\frac{1}{2} N^\prime  \int_{- \infty}^{+ \infty}
\left\{{\tilde{\bf p}}^T \tilde{A^{\prime\prime}_{\bf 0}} {\tilde{\bf p}}\right\}
\ \text{exp} \left [-\frac{1}{2} {\tilde{\bf p}}^T \tilde{\mathbfcal{M}}
{\tilde{\bf p}}  \right ] d{\tilde{\bf p}}.
\end{eqnarray}
In the last step $\tilde{A^{\prime\prime}_{\bf 0}}={\bf O}A^{\prime\prime}_{\bf 0}{\bf O}^T$ 
was used. So the expression for $\overline{A}$ can be further simplified as, 
\begin{eqnarray}
\overline{A}&=&A_{\bf 0}+\frac{1}{2} N^\prime  \int_{- \infty}^{+ \infty}
\left\{\sum_{k^\prime,l^\prime=1}^{N_p}{\tilde p}_{k^\prime} 
\left(\tilde{A^{\prime\prime}_{\bf 0}}\right)_{k^\prime l^\prime} {\tilde p}_{l^\prime}\right\}
\ \text{exp} \left [-\frac{1}{2}\sum_{k=1}^{N_p} {\tilde{\mathcal{M}}}_k 
{\tilde p}_k^2  \right ] d{\tilde{\bf p}}\nonumber\\
&=&A_{\bf 0}+\frac{1}{2} N^\prime  \int_{- \infty}^{+ \infty}
\left\{\sum_{k^\prime,l^\prime=1}^{N_p}{\tilde p}_{k^\prime} 
\left(\tilde{A^{\prime\prime}_{\bf 0}}\right)_{k^\prime l^\prime} {\tilde p}_{l^\prime}\right\}
\ \prod_{k=1}^{N_p}\text{exp} \left [-\frac{1}{2} {\tilde{\mathcal{M}}}_k 
{\tilde p}_k^2  \right ] d{\tilde p}_k \nonumber\\
&=&A_{\bf 0}+\frac{1}{2} N^\prime  \int_{- \infty}^{+ \infty}
\left\{\sum_{k^\prime=1}^{N_p}{\tilde p}_{k^\prime} 
\left(\tilde{A^{\prime\prime}_{\bf 0}}\right)_{k^\prime k^\prime} {\tilde p}_{k^\prime}\right\}
\ \prod_{k=1}^{N_p}\text{exp} \left [-\frac{1}{2} {\tilde{\mathcal{M}}}_k 
{\tilde p}_k^2  \right ] d{\tilde p}_k\ \ .
\end{eqnarray}
For the last step the fact was used that any Gaussian integral with an odd 
power of variable multiplied to it vanishes. So the expression 
for $\overline{A}$ can be further simplified as, 
\begin{eqnarray}
\label{average_a}
\overline{A}&=&A_{\bf 0}+\frac{1}{2} N^\prime  \int_{- \infty}^{+ \infty}
\left\{\sum_{k^\prime=1}^{N_p} 
\left(\tilde{A^{\prime\prime}_{\bf 0}}\right)_{k^\prime k^\prime} {\tilde p}_{k^\prime}^2\right\}
\ \text{exp} \left [-\frac{1}{2} {\tilde{\mathcal{M}}}_{k^\prime} 
{\tilde p}_{k^\prime}^2  \right ] d{\tilde p}_{k^\prime} \nonumber\\
&&\qquad\qquad\int_{- \infty}^{+ \infty} \prod_{k\neq k^\prime ,1}^{N_p}
\text{exp} \left [-\frac{1}{2} {\tilde{\mathcal{M}}}_k 
{\tilde p}_k^2  \right ] d{\tilde p}_k\nonumber\\
&=&A_{\bf 0}+\frac{1}{2}\sum_{k^\prime=1}^{N_p}\left\{\frac{\tilde {\mathcal{M}}_{k^\prime}}{2\pi} \right\}^{\frac{N_p}{2}}
\left\{\frac{2\pi} {\tilde {\mathcal{M}}_{k^\prime}} \right\}^{\frac{N_p-1}{2}}
\left(\tilde{A^{\prime\prime}_{\bf 0}}\right)_{k^\prime k^\prime}\int_{- \infty}^{+ \infty}
{\tilde p}_{k^\prime}^2 \text{exp} \left [-\frac{1}{2} {\tilde{\mathcal{M}}}_{k^\prime} 
{\tilde p}_{k^\prime}^2  \right ] d{\tilde p}_{k^\prime}\nonumber\\
&=&A_{\bf 0}+\frac{1}{2}\sum_{k^\prime=1}^{N_p}
\left\{\frac{\tilde {\mathcal{M}}_{k^\prime}}{2\pi} \right\}^{\frac{1}{2}}
\left(\tilde{A^{\prime\prime}_{\bf 0}}\right)_{k^\prime k^\prime}
\left\{\frac{2\pi} {\tilde {\mathcal{M}}_{k^\prime}^3} \right\}^{\frac{1}{2}}\nonumber
\end{eqnarray}
\begin{eqnarray}
&=&A_{\bf 0}+\frac{1}{2}\sum_{k^\prime=1}^{N_p}
{\tilde {\mathcal{M}}_{k^\prime}}^{-1}
\left(\tilde{A^{\prime\prime}_{\bf 0}}\right)_{k^\prime k^\prime}\nonumber\\
&=&A_{\bf 0}+\frac{1}{2}\text{Tr}
\left[{\tilde{\mathbfcal{M}}}^{-1} \tilde{A_{\bf 0}^{\prime\prime}}\right]\nonumber\\
\Rightarrow \overline{A}&=&A_{\bf 0}+\frac{1}{2}\text{Tr}
\left[{\mathbfcal{M}}^{-1} A_{\bf 0}^{\prime\prime}\right]\ \ .
\end{eqnarray}
In the last step, simply the cyclic property during the matrix multiplication was used 
under the orthogonal transformation. The inverse of the Hessian matrix 
$\mathbfcal{M}^{-1}$ is commonly known as the "curvature" matrix. 
Now the deviation in the observable $A({\bf p})$ 
from its average $\overline{A}$ is given by,
\begin{eqnarray}
\label{deviation_a}
\Delta A&=& A({\bf p})-\overline{A}\nonumber\\
&=&A_{\bf 0} +  ({\bf p}-{\bf p_0})^T A^\prime_{\bf 0} + 
\frac{1}{2}  ({\bf p}-{\bf p_0})^T A^{\prime\prime}_{\bf 0} ({\bf p}-{\bf p_0})
-A_{\bf 0}-\frac{1}{2}\text{Tr}
\left[{\mathbfcal{M}}^{-1} A_{\bf 0}^{\prime\prime}\right]\nonumber\\
&=&({\bf p}-{\bf p_0})^T A^\prime_{\bf 0} + 
\frac{1}{2}  ({\bf p}-{\bf p_0})^T A^{\prime\prime}_{\bf 0} ({\bf p}-{\bf p_0})
-\frac{1}{2}\text{Tr} \left[{\mathbfcal{M}}^{-1} A_{\bf 0}^{\prime\prime}\right]\ \ .
\end{eqnarray}
Similarly, the deviation in a quantity $B$ can calculated following Eq. (\ref{deviation_a}). 
So one can write 
now,
\begin{eqnarray}
\Delta A({\bf p})\Delta B({\bf p})&\approx&\left\{({\bf p}-{\bf p_0})^T A^\prime_{\bf 0} + 
\frac{1}{2}  ({\bf p}-{\bf p_0})^T A^{\prime\prime}_{\bf 0} ({\bf p}-{\bf p_0})
-\frac{1}{2}\text{Tr} \left[{\mathbfcal{M}}^{-1} A_{\bf 0}^{\prime\prime}\right]\right\}\nonumber\\
&&\left\{{B^\prime_{\bf 0}}^T ({\bf p}-{\bf p_0}) + 
\frac{1}{2}  ({\bf p}-{\bf p_0})^T B^{\prime\prime}_{\bf 0} ({\bf p}-{\bf p_0})
-\frac{1}{2}\text{Tr} \left[{\mathbfcal{M}}^{-1} B_{\bf 0}^{\prime\prime}\right]\right\}\nonumber\\
&\approx&({\bf p}-{\bf p_0})^T A^\prime_{\bf 0}{B^\prime_{\bf 0}}^T ({\bf p}-{\bf p_0})+
\mathcal{O}\{\Delta {\bf p}^3\}\ \ .
\end{eqnarray}
Neglecting the contribution from the terms containing beyond $\mathcal{O}\{\Delta {\bf p}^2\}$ 
and calling $A^\prime_{\bf 0}{B^\prime_{\bf 0}}^T$ as $(AB)_{\bf 0}^{\prime}$, 
covariance between $A({\bf p})$ and $B({\bf p})$ can be obtained following the definition 
in Eq. (\ref{var_covar}) as,
\begin{eqnarray}
\overline{(\Delta A \Delta B)}&=&N^\prime  \int_{- \infty}^{+ \infty}
\left\{({\bf p}-{\bf p_0})^T (AB)_{\bf 0}^{\prime} ({\bf p}-{\bf p_0})\right\}
\ \text{exp} \left [-\frac{1}{2} ({\bf p}-{\bf p_0})^T \mathbfcal{M} 
({\bf p}-{\bf p_0})  \right ] d{\bf p} \nonumber\\
&=&N^\prime  \int_{- \infty}^{+ \infty}
\left\{({\bf p}-{\bf p_0})^T{\bf O}^T{\bf O} (AB)_{\bf 0}^{\prime}
{\bf O}{\bf O}^T ({\bf p}-{\bf p_0})\right\}\nonumber\\
&&\qquad\qquad\qquad\qquad \text{exp} \left [-\frac{1}{2} ({\bf p}-{\bf p_0})^T \mathbfcal{M} 
({\bf p}-{\bf p_0})  \right ] d{\bf p} \nonumber\\
&=&N^\prime  \int_{- \infty}^{+ \infty}
\left\{{\tilde{\bf p}}^T {\tilde{(AB)}}_{\bf 0}^{\prime} \tilde{\bf p}\right\}
\ \text{exp} \left [-\frac{1}{2} {\tilde{\bf p}}^T \tilde{\mathbfcal{M}}
{\tilde{\bf p}}  \right ] d{\tilde{\bf p}}\nonumber\\
&=&N^\prime  \int_{- \infty}^{+ \infty}
\left\{\sum_{k^\prime,l^\prime=1}^{N_p}{\tilde p}_{k^\prime} 
\left({\tilde{(AB)}}_{\bf 0}^{\prime}\right)_{k^\prime l^\prime} {\tilde p}_{l^\prime}\right\}
\ \prod_{k=1}^{N_p}\text{exp} \left [-\frac{1}{2} {\tilde{\mathcal{M}}}_k 
{\tilde p}_k^2  \right ] d{\tilde p}_k\ \ .
\end{eqnarray}
Now, following the same steps as $\overline {A}$ in Eq. (\ref{average_a}), the expression for 
covariance between $A$ and $B$ can be written as, 
\begin{eqnarray}
\label{covariance}
\overline{(\Delta A \Delta B)}&=&\text{Tr}
\left[{\mathbfcal{M}}^{-1} {(AB)}_{\bf 0}^{\prime}\right]\nonumber\\
&=&\sum_{k,l=1}^{N_p}{\mathcal{M}}^{-1}_{k,l}\left(A_{\bf 0}^{\prime}
{B_{\bf 0}^{\prime}}^T\right)_{l,k}\nonumber\\
&=&\sum_{k,l=1}^{N_p}{\mathcal{M}}^{-1}_{k,l}\left({A_{\bf 0}^{\prime}}^T
B_{\bf 0}^{\prime}\right)_{k,l}\nonumber\\
&=&\sum_{k,l=1}^{N_p}\left({A_{\bf 0}^{\prime}}\right)_k {\mathcal{M}}^{-1}_{k,l}
\left(B_{\bf 0}^{\prime}\right)_{l}\nonumber\\
\overline{(\Delta A \Delta B)}&=&\sum_{k,l=1}^{N_p}
\left (\frac{\partial A}{\partial {p}_k}\right)_{\bf p_0}{\mathcal{M}}^{-1}_{k,l}
\left (\frac{\partial B}{\partial {p}_l}\right)_{\bf p_0}.
\end{eqnarray}
Putting $A=B$, variance on any quantity $A$ can be calculated as,
\begin{eqnarray}
\label{variance}
\overline{(\Delta A)^2}&=&\sum_{k,l=1}^{N_p}
\left (\frac{\partial A}{\partial {p}_k}\right)_{\bf p_0}{\mathcal{M}}^{-1}_{k,l}
\left (\frac{\partial A}{\partial {p}_l}\right)_{\bf p_0}.
\end{eqnarray}
As, the parameters of a model are considered to be independent to one another, 
one can write, 
\begin{eqnarray}
\frac{\partial p_k}{\partial p_{k^\prime}}&=&\delta_{k k^\prime}\nonumber\\
\Rightarrow \overline{(\Delta p_k \Delta p_{k^\prime})}={\mathcal{M}}^{-1}_{k,k^\prime}\ &\& &\ 
 \overline{(\Delta p_k )^2}={\mathcal{M}}^{-1}_{k,k}.
\end{eqnarray}
So, the diagonal elements of the curvature matrix quantify the variance on 
the model parameters. Upon taking the square root of the diagonal elements, 
one can obtain the error on the parameters.

\section{Minimization and Sensitivity analysis}
In the present thesis work, parameters of a relativistic mean-field
(RMF) model are optimized. A close variant of Levenberg-Marquardt
\cite{Press92} was employed to minimize the $\chi^2$ function as defined
in Eq. (\ref{chidef}). The algorithm typically uses the inverse-Hessian
or curvature matrix method. Using the definition of $\chi^2$ in
Eq. (\ref{chidef}), Hessian matrix can be calculated as,
\begin{eqnarray}
\label{Hessian}
{\mathcal{M}}_{k,l}&=&\frac{1}{2}\left(\frac{\partial^2 \chi^2({\bf p_0})}{\partial 
{{p}_k}\partial {{p}_l}}\right)
=\frac{1}{2}\frac{\partial^2 }{\partial {{p}_k}\partial {{p}_l}}\left\{
\sum\limits_{i=1}^{N_d}\left (\frac{\mathcal{O}_i^{th}
-\mathcal{O}_i^{exp}}{\Delta\mathcal{O}_i}\right)^2\right\}\nonumber\\
&=&\sum\limits_{i=1}^{N_d}\frac{1}{\Delta \mathcal{O}_i^2}\left[ 
\left(\frac{\partial \mathcal{O}_i^{th}}{\partial p_k}\right)
\left(\frac{\partial \mathcal{O}_i^{th}}{\partial p_l}\right)
+\left(\mathcal{O}_i^{th}-\mathcal{O}_i^{exp}\right)
\left(\frac{\partial^2 \mathcal{O}_i^{th}}{\partial p_k \partial p_l}\right) \right]\ \ .
\end{eqnarray}
Here, $\left(\mathcal{O}_i^{th}-\mathcal{O}_i^{exp}\right)$ is the residual 
error in $\mathcal {O}_i$. If the model space is reasonable, residual errors 
are small. Moreover, one can expect that they are random in sign. Then 
the Hessian matrix is approximately obtained as \cite{Brandt97}
\begin{eqnarray}
\label{Jacobian}
{\mathcal{M}}_{k,l}&\approx&\sum\limits_{i=1}^{N_d}\frac{1}{\Delta \mathcal{O}_i^2}
\left(\frac{\partial \mathcal{O}_i^{th}}{\partial p_k}\right)
\left(\frac{\partial \mathcal{O}_i^{th}}{\partial p_l}\right)\nonumber\\
&=&\sum\limits_{i=1}^{N_d}\left(\frac{1}{\Delta \mathcal{O}_i}\frac{\partial \mathcal{O}_i^{th}}{\partial p_k}\right)
\left(\frac{1}{\Delta \mathcal{O}_i}\frac{\partial \mathcal{O}_i^{th}}{\partial p_l}\right)\nonumber\\
&=&\sum\limits_{i=1}^{N_d}J_{i,k}J_{i,l}\ \ .
\end{eqnarray}
So, the Hessian matrix and the corresponding curvature matrix can be obtained 
without ever calculating any double derivative of the observables with respect 
to the parameters. The ${\bf J}$ matrix is called the "Jacobian" matrix 
which is nothing but the derivative of an observable with respect to parameters 
weighted by the corresponding adopted error. This approximation along with the 
Levenberg-Marquardt method provides a very efficient and stable minimization 
procedure. 

\begin{figure}[]{}
\centering
\includegraphics[width=0.7\textwidth]{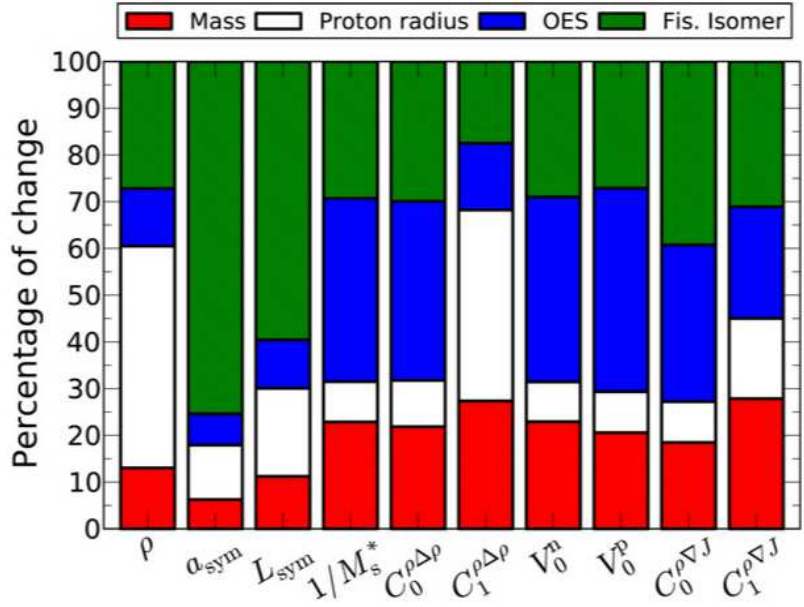}
\caption{\label{ch3_fig2}
Relative Sensitivity of different parameters of UNEDF1 \cite{Kortelainen12} to different 
type of data used in the fit, as given in Ref. \cite{Dobaczewski14}.} 
\end{figure}
It was mentioned in the previous section that proper estimation of the adopted 
errors i.e. $\Delta \mathcal{O}_i$ are very important. As there is no proper 
prescription to estimate $\Delta \mathcal{O}_i$, some arbitrariness is 
inevitable. However, the arbitrariness can be reduced significantly by introducing 
a global '$s$' factor \cite{Dobaczewski14,Birge32},
\begin{eqnarray}
\label{s_factor}
s=\frac{\chi^2({\bf p_0})}{N_d-N_p}.
\end{eqnarray}
While performing the covariance analysis as described in the previous section, 
ultimately the curvature matrix $\mathbfcal{M}^{-1}$ is replaced by $(s
\mathbfcal{M}^{-1})$ and rest of the procedure is followed. 

To study the overall impact of each type of data on the optimized parameters 
a "sensitivity matrix" ${\bf S}$ is defined as \cite{Kortelainen10, Kortelainen12, 
Dobaczewski14}, 
\begin{eqnarray}
\label{sen_mat}
{\bf S}({\bf p})&=&\left[{\bf J}({\bf p}){\bf J}^T({\bf p})\right]^{-1}{\bf J}({\bf p})\ .
\end{eqnarray}
For {\it k}-th row in the sensitivity matrix $S_{i,k}$ corresponding to a single parameter, 
one can compute the partial sums over different type of data $i_1,i_2\cdots$ where 
$i=i_1+i_2+\cdots$. Consequently, a percentage contribution from each type of data $i_1,i_2$ 
etc can be obtained by normalizing to the summation over all type of data as 100$\%$ i.e. 
$\sum_i S_{i,k}=100\%$ for the $k$-th parameter. A representative example is depicted in 
Fig. \ref{ch3_fig2} as given in Ref. \cite{Dobaczewski14}.

All the techniques described in Chapter \ref{ch3} have been extensively used 
in Chapter \ref{ch4} to test the merit of a relativistic mean field model. 
The numerical codes to perform the covariance and sensitivity analysis are 
also developed during the present thesis work.

%% file: chap4.tex
\chapter{Constraining the symmetry energy parameters using a relativistic mean field model}\label{ch4}
\section{Introduction}
\begin{figure}[]{}
\centering
\includegraphics[width=0.6\textwidth]{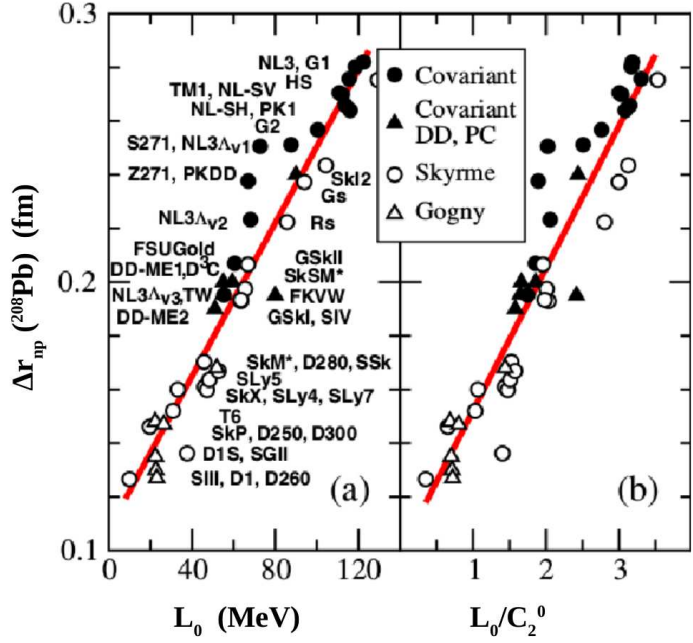} 
\caption{\label{prl102}
$\Delta r_{np}$ of $^{208}$Pb plotted as a function of $L_0$ (left panel) 
and $\frac{L_0}{C_2^0}$ (right panel), for $\sim40$ mean-field models as 
given in Ref \cite{Centelles09}.}
\end{figure}

\begin{figure}[]{}
\centering
\includegraphics[width=0.6\textwidth]{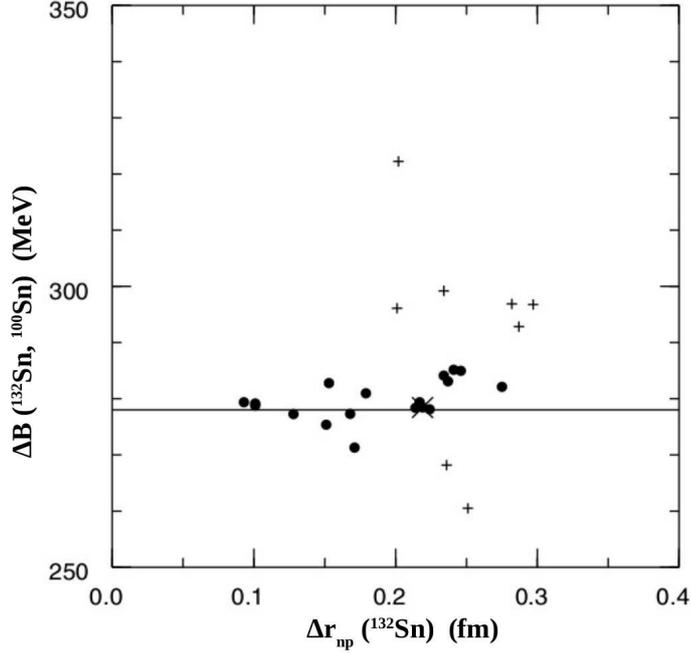} 
\caption{\label{ch4a_fig0}
The binding energy difference $\Delta B$ ($^{132}$Sn,  $^{100}$Sn) $=BE(^{132}\text{Sn})-BE(^{100}\text{Sn})$ 
plotted against neutron-skin thickness
$\Delta r_{np}$ of the $^{132}$Sn nucleus for different mean-filed models as 
shown in Ref. \cite{Brown00a}.} 
\end{figure}

\begin{figure}[]{}
\centering
\includegraphics[width=0.7\textwidth]{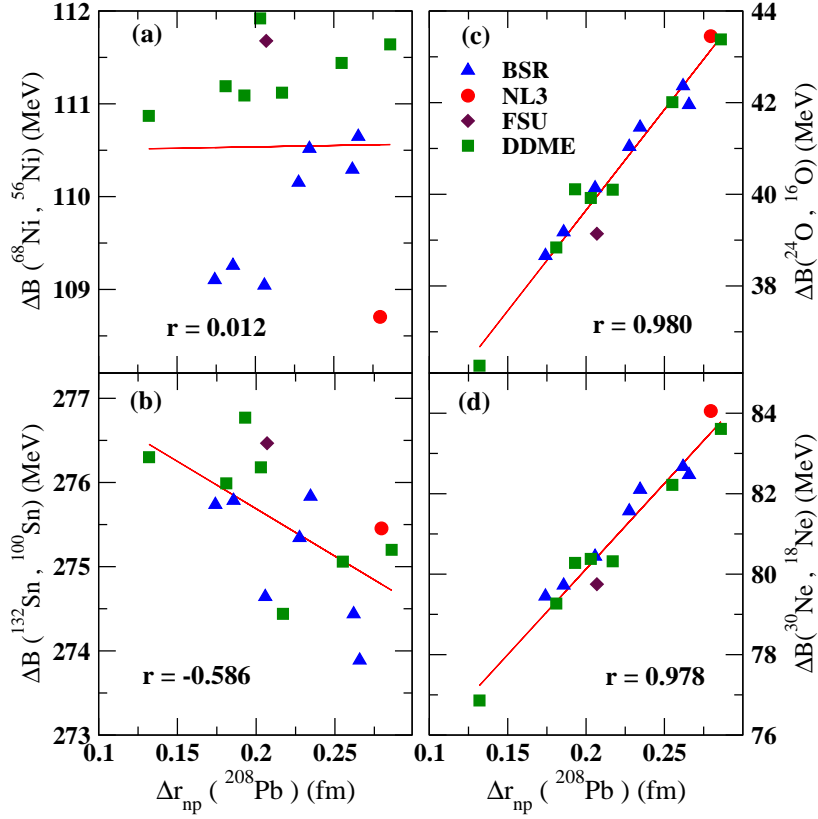} 
\caption{\label{ch4a_fig1}
The binding energy difference $\Delta BE(X,Y)=BE(X)-BE(Y)$ for four
different pairs of isotopes are plotted against neutron-skin thickness
$\Delta r_{np}$ in the $^{208}$Pb nucleus for 16 different RMF models
(See text for details). The values of correlation coefficients $r$
are also displayed. }
\end{figure}

The symmetry energy coefficient $C_2^0$ (Eq. (\ref{sym_expand})) is well constrained from binding
energies of finite nuclei with its mean value $\sim32$ MeV \cite{
Myers66,Moller95,Pomorski03,Xu10,Moller12,Jiang12}. However, symmetry
energy slope $L_0$, shows a wide variation $\sim20$ - 120 MeV \cite{
Centelles09,Steiner12,Shetty07,Famiano06,Trippa08, Roca-Maza13a,
Roca-Maza13,Carbone10,Chen11,Tsang12,Niksic08,Zhao10,Chabanat98,
Roca-Maza12,Roca-Maza11,Dutra12,Dutra14}. It can be realized by looking at the 
abscissa of Fig. (\ref{prl102}), where neutron-skin thickness $\Delta r_{np}$ of $^{208}$Pb 
is plotted as a function of $L_0$ for a set of $\sim$ 40 mean-filed models, 
which is taken from Ref. \cite{Centelles09}. One can also observe a linear 
correlation between $\Delta r_{np}$ of $^{208}$Pb and $L_0$. Thus, precise 
information of $\Delta r_{np}$ of $^{208}$Pb can constrain the value of $L_0$ tightly. 
From the analysis of data on the precisely known nuclear masses in macroscopic nuclear models,
$L_0$ is known with a fair amount of accuracy, $L_0=60\pm20$ MeV \cite{Moller12,Jiang12,
Agrawal12,Agrawal13}.  Energy density functionals (EDF) in microscopic
mean field models, parametrized to reproduce the binding energies
of nuclei along with some other specific nuclear observables do not,
however, display such constraints on $L_0$.  Questions then arise
how the information content of symmetry energy gets blurred in the
exploration of nuclear masses in microscopic models. For example in
Ref. \cite{Brown00a}, correlation between the binding energy difference $\Delta B$
of $^{132}$Sn and $^{100}$Sn and the $\Delta r_{np}$ of $^{132}$Sn for different
sets of Skyrme EDFs were studied. In principle, $\Delta B$ ($^{132}$Sn,  $^{100}$Sn) 
should contain a major contribution coming from the asymmetric part of the $^{132}$Sn, which 
in return should be related to symmetry energy and consequently to $\Delta r_{np}$. 
However, no noticeable correlation between $\Delta B$ ($^{132}$Sn,  $^{100}$Sn) and
$\Delta r_{np}$ was found, which is depicted in Fig. (\ref{ch4a_fig0}).
In the present thesis work, this lack of correlation was reconciled by
studying the binding energy difference between four pairs of nuclei
in different RMF models with increasing asymmetry effects namely,
($^{68}\text{Ni}- ^{56}\text{Ni}$), ($^{132}\text{Sn}-^{100}\text{Sn}$),
($^{24}\text{O}-^{16}\text{O}$) and ($^{30}\text{Ne}-^{18}\text{Ne}$).
The neutron rich $^{68}$Ni and $^{132}$Sn nuclei have asymmetries
$\delta=0.176$ and 0.242 respectively ($\delta$ is the isospin asymmetry
parameter $(N-Z)/A$); $^{24}$O and $^{30}$Ne have $\delta\approx0.33$
i.e. $N/Z\approx 2$.  The Ni and Sn isotopes are doubly closed shell
nuclei. So also the O-nuclei, $^{24}$O is recently seen to be an
unexpectedly stable doubly magic nucleus \cite{Kanungo09, Janssens09}.
The Ne-nuclei have their neutron shells closed but have valence
protons. The binding energy difference between the two Ne-nuclei is
expected to cancel the pairing and the possible core-polarization effects
arising from the two valence protons partially.  In Fig.\ref{ch4a_fig1}
the binding energy difference between the four pairs of nuclei are plotted
against $\Delta r_{np}$ of $^{208}$Pb, the $\Delta r_{np}$ and the binding
energies being calculated for seven models of BSR family \cite{Dhiman07,
Agrawal10}, NL3 \cite{Lalazissis97}, FSU \cite{Todd-Rutel05} and
for seven models of Density Dependent Meson Exchange (DDME) family
\cite{Vretenar03}.  The correlation coefficient for the Ni-pair is seen
to be only 0.012, for the Sn-pair, it has increased to 0.586. For the O
and Ne pairs, they are quite high, 0.980 and 0.978, respectively. One can
not fail to notice the increasingly high correlation with increasing
asymmetry, particularly for the latter two cases.

\section{A Covariance analysis}
The method of Covariance analysis as described in Chapter \ref{ch3} 
provides the perfect tool to unveil the effect of extremely asymmetric 
nuclei to construct the model parameters of a RMF model.
\subsection{Fit data and model parameters}
The occurrence of strong correlation for the case of O and Ne pairs
probably suggest that selective combination of suitable binding energies
of nuclei of low and high isospin may be ideally suited to better
constrain the isovector part of the nuclear interaction.  To explore
this idea, two RMF models (model-I and model-II) corresponding to
different sets of fit-data are constructed.  The observables  explored
are the symmetry energy $C_2^0$, symmetry energy slope $L_0$ along with
the $\Delta r_{np}$ of $^{208}$Pb.  The effective Lagrangian density
for the RMF model employed in the present work is similar to that
of the FSU one \cite{Todd-Rutel05, Furnstahl97, Boguta77, Boguta83}
(see Chapter \ref{ch2}).  The values of the parameters entering the
EDF of the RMF model are obtained from an optimal $\chi^2$ fit of the
experimental observables with the theoretically calculated values,
as described in Chapter \ref{ch3}.  In model-I the binding energies
and charge radii of some standard set of nuclei ($^{16}$O, $^{40}$Ca,
$^{48}$Ca, $^{56}$Ni, $^{68}$Ni, $^{90}$Zr, $^{100}$Sn, $^{116}$Sn,
$^{132}$Sn, $^{144}$Sm and $^{208}$Pb) spanning the entire periodic
table are taken as fit-data.  In model-II, we have the same set
of experimental observables, but with the addition of the binding
energy difference $\Delta B$ of ($^{24}\text{O},^{16}\text{O}$) and
of ($^{30}\text{Ne},^{18}\text{Ne}$).  The parameters of model-I and
model-II are obtained by optimizing \cite{Bevington69} the objective
function $\chi^2$({\bf p}) as described in Chapter \ref{ch3}.

Once the optimized parameter set is obtained the correlation coefficient
between two quantities $\mathcal{A}$ and $\mathcal{B}$, which may
be a parameter as well as an observable, can be evaluated within the
covariance analysis as described in Chapter \ref{ch3}.  The parameters
for model-I and II corresponding to minimum value of the objective
function $\chi^2$({\bf p}) (=$\chi^2(\text{\bf p}_0)$) along with their
statistical errors are listed in Table \ref{ch4a_tab1}.
\begin{table*}[] 
\centering
\caption{\label{ch4a_tab1} The best fit values for the parameters of
model-I and model-II.  $m_\sigma$ is the mass of $\sigma$ meson given in
units of MeV. The masses of $\omega$ and $\rho$ mesons are kept fixed to
$m_\omega$= 782.5 MeV and $m_\rho$= 763 MeV and nucleon mass is taken
to be $M$= 939 MeV.  Statistical errors on the fitted parameters  are
also given for both the models.}
\begin{tabular}{ccccccccc}
\hhline{=========}
Name  & $g_{\sigma}$ & $g_{\omega}$ & $g_{\rho}$ & ${\kappa_3}$ &
${\kappa_4}$ & ${\eta_{2\rho}}$ & $\zeta_0$ & $m_{\sigma}$ \\
\hline
model-I & -10.6246 & 13.8585 & 12.077 & 1.46285 &
-0.9673 & 28.33 & 5.2056 & 496.007 \\
(Error)& 0.246 & 0.662 & 2.60 & 0.275 & 3.66 & 29.9 & 3.21 &
 12.2\\
model-II& -10.6212 & 13.8599 & 12.436 & 1.46223 &
-0.8566 & 32.50 & 5.3220 & 495.815 \\
(Error)& 0.149 & 0.262 & 1.54 & 0.290 & 1.53 & 18.1 & 0.099 &
 8.23\\
\hhline{=========}
\end{tabular}
\end{table*}
Overall, the errors on the parameters  for the case of model-II are
smaller than those obtained for the model-I indicating that the inclusion
of the fit data on the binding energy differences constrain the model
parameters better. In particular, the errors on the parameters $g_\rho$
and $\eta_{2\rho}$, which govern the isovector part of the effective
Lagrangian, are smaller for the model-II. The large error on the
parameters $\kappa_3$ and $\kappa_4$ for both the models may be due to
the fact that the fit data does not include any observable which could
constrain the value of the nuclear matter incompressibility coefficient
\cite{Boguta83}.  In Table \ref{ch4a_tab2} different observables
$\mathcal{O}_i$, adopted errors on them $\Delta\mathcal{O}_i$,
their experimental values along with the results obtained for
model-I and model-II using the corresponding best fit parameters are
listed. The values of $\mathcal{O}_i$ and $\Delta \mathcal{O}_i$,
except for the $\Delta B$ of ($^{24}\text{O},^{16}\text{O}$) and
($^{30}\text{Ne},^{18}\text{Ne}$) and $r_{ch}$ of $^{132}$Sn, are
exactly same as used in Ref. \cite{Klupfel09}. The experimental
data for $\Delta B$ of ($^{24}\text{O},^{16}\text{O}$) and
($^{30}\text{Ne},^{18}\text{Ne}$) are taken from \cite{Audi12} and that
for the $r_{ch}$ of $^{132}$Sn from \cite{Angeli13}.
\begin{table*}[]
\centering
\caption{\label{ch4a_tab2}
Observables $\mathcal{O}$ of different nuclei, adopted errors on them
$\Delta\mathcal{O}$, their experimental values and the ones obtained for
model-I and II. $BE$ and $r_{ch}$ refers to binding energy and charge
radius of a nucleus respectively, and $\Delta B$ is binding energy
difference of two isotopes of a nucleus as indicated. $BE$ and $\Delta B$
are in units of MeV and $r_{ch}$ in fm.  }
\begin{tabular}{cccccc}
\hhline{======}
Nucleus & $\mathcal{O}$ &$\Delta\mathcal{O}$ & Expt. &  
model-I & model-II \\
\hline
$^{16}$O & $BE$ &4.0 & 127.62 &  127.781$\pm$0.990 & 127.783$\pm$0.576 \\
& $r_{ch}$ &0.04 & 2.701 &  2.700$\pm$0.017 & 2.699$\pm$0.013 \\
$^{16}$O, $^{24}$O & $\Delta B$ &2.0 & 41.34 &  - & 40.995$\pm$1.046 \\
$^{18}$Ne, $^{30}$Ne & $\Delta B$ &2.0 & 79.147 &  - & 79.149$\pm$1.296 \\
$^{40}$Ca & $BE$ &3.0 & 342.051 &  342.929$\pm$1.064 & 342.927$\pm$0.927 \\
& $r_{ch}$ &0.02 & 3.478 &  3.457$\pm$0.013 & 3.455$\pm$0.010 \\
$^{48}$Ca & $BE$ &1.0 & 415.99 &  414.883$\pm$0.720 & 414.751$\pm$0.541 \\
& $r_{ch}$ &0.04 & 3.479 &  3.439$\pm$0.007 & 3.439$\pm$0.006 \\
$^{56}$Ni & $BE$ &5.0 & 483.99 &  483.752$\pm$2.495 & 483.619$\pm$1.646 \\
& $r_{ch}$ &0.18 & 3.750 &  3.695$\pm$0.025 & 3.693$\pm$0.020 \\
$^{68}$Ni & $BE$ &1.0 & 590.43 &  592.294$\pm$0.784 & 592.162$\pm$0.736 \\ 
$^{90}$Zr & $BE$ &1.0 & 783.893 &  782.855$\pm$4.833 & 782.776$\pm$1.621 \\
& $r_{ch}$ &0.02 & 4.269 &  4.267$\pm$0.009 & 4.267$\pm$0.034 \\
$^{100}$Sn & $BE$ &2.0 & 825.8 &  827.987$\pm$1.753 & 827.757$\pm$1.534 \\
$^{116}$Sn & $BE$ &2.0 & 988.32 &  987.169$\pm$0.946 & 987.072$\pm$0.760 \\
& $r_{ch}$ &0.18 & 4.626 &  4.623$\pm$0.009 & 4.623$\pm$0.008 \\
$^{132}$Sn & $BE$ &1.0 & 1102.9 &  1102.851$\pm$1.146 & 1102.631$\pm$0.856 \\
& $r_{ch}$ &0.02 & 4.71 &  4.711$\pm$0.011 & 4.712$\pm$0.010 \\
$^{144}$Sm & $BE$ &2.0 & 1195.74 &  1195.834$\pm$1.240 & 1195.736$\pm$1.287 \\
& $r_{ch}$ &0.02 & 4.96 &  4.956$\pm$0.009 & 4.956$\pm$0.009 \\
$^{208}$Pb & $BE$ &1.0 & 1636.446 &  1636.457$\pm$4.301 & 1636.383$\pm$0.917 \\
& $r_{ch}$ &0.02 & 5.504 &  5.530$\pm$0.012 & 5.531$\pm$0.010 \\
\hhline{======}
\end{tabular}
\end{table*}

\subsection{Results}
\begin{figure}[]{}
\centering
\includegraphics[width=0.5\textwidth]{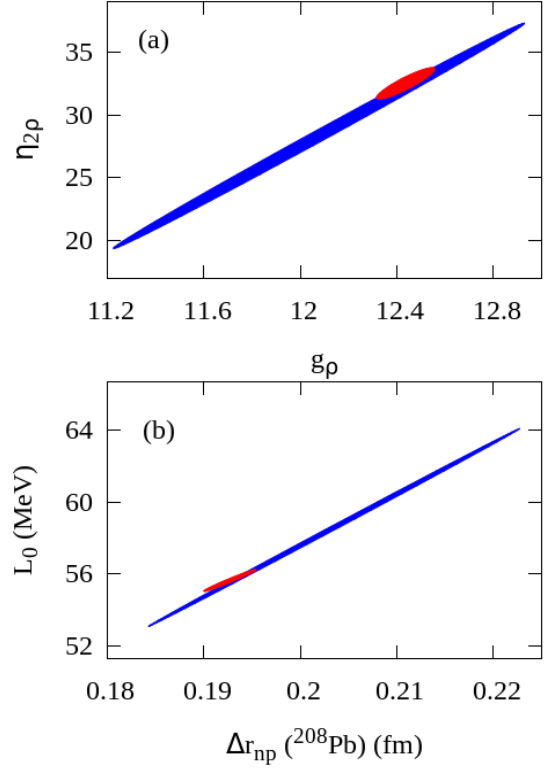} 
\caption{\label{ch4a_fig2} 
The covariance ellipsoids for the parameters $g_\rho \text{ - }
\eta_{2\rho}$ (upper panel) and the corresponding $L_0 \text{ - }
\Delta r_{np}$ (lower panel) for the model I (blue) and model II
(red).  The area inside the ellipsoids indicate the reasonable
domain of the parameters.}
\end{figure}

The results obtained for the model-I and model-II are compared to see
up to what extent the inclusion of the experimental data on the binding
energy differences between the pair of  O and Ne nuclei can constrain the
iso-vector part of the effective Lagrangian.  In Fig. \ref{ch4a_fig2}(a)
the covariance ellipsoids for the parameters $g_\rho$ and $\eta_{2\rho}$
are displayed (see Chapter \ref{ch3}).
\begin{table*} 
\centering
\caption{\label{ch4a_tab3}
The values for the binding energy per nucleon $\mathcal{E}_0$,
incompressibility coefficient $K_0$, Dirac effective mass of nucleon
$M^*/M$, symmetry energy coefficient $C_2^0$ and density slope parameter
of symmetry energy $L_0$ for the nuclear matter evaluated at saturation
density $\rho_0$ along with the correlated errors on them obtained
within the covariance analysis for the models I and II. The results for
neutron-skin thickness $\Delta r_{np}$ in  $^{48}$Ca, $^{132}$Sn and
$^{208}$Pb are also presented.  }
\begin{tabular}{ccc} 
\hhline{===}
Observable  & model-I & model-II\\ \hline $\mathcal{E}_0$
(MeV)           &  $-16.036\pm0.070$  & $-16.036\pm0.051$   \\ $K_0$
(MeV)             &  $210.12\pm27.87$   & $209.64\pm28.52$    \\
$\rho_0$ (fm$^{-3}$)  &  $0.150\pm0.003$    & $0.150\pm0.003$     \\
$M^*/M$               &  $0.585\pm0.012$    & $0.585\pm0.010$     \\
$C_2^0$ (MeV)         &  $32.03\pm3.08$     & $31.69\pm1.51$      \\
$L_0$ (MeV)           &  $57.62\pm17.08$    & $55.63\pm7.00$      \\
$\Delta r_{np}$ ($^{48}$Ca) (fm) & $0.191\pm0.036$ & $0.187\pm0.016$\\ 
$\Delta r_{np}$ ($^{132}$Sn) (fm) & $0.266\pm0.070$ & $0.257\pm0.031$\\ 
$\Delta r_{np}$ ($^{208}$Pb) (fm) & $0.201\pm0.065$ & $0.193\pm0.030$\\ 
\hhline{===}
\end{tabular} 
\end{table*}
For these sets of parameters, the values of the symmetry energy slope
parameter $L_0$ and the $\Delta r_{ np}$  in the $^{208}$Pb nucleus are
displayed in Fig. \ref{ch4a_fig2}(b).  The inclined and elongated shapes
of the ellipsoids indicate that the correlations amongst $g_\rho \text{
- } \eta_{2\rho}$ and $L_0 \text{ - } \Delta r_{np}$ are strong. In
fact, the values of the correlation coefficients (Eq.  (\ref{corr})) for
these pairs of quantities for both the models turn out to be $\sim$0.95.
It is evident that the ellipsoids depicting the results for the model-II
(red) are narrower in comparison  to those for the model-I (blue). 
This is suggestive of the fact that the inclusion of the binding
energies for the $^{24}$O and $^{30}$Ne put tighter constraints on the
isovector part of the effective Lagrangian density.

Nuclear matter properties for model-I and model-II are compared in
Table \ref{ch4a_tab3}. Errors on the entities describing the isoscalar
behavior of nuclear matter ($\mathcal{E}_0$, $K_0$, $\rho_0$ and $M^*/M$) are pretty
much the same for both the models concerned.  For model-II, however, a
significant improvement (by a factor $\sim2$) on the spread of parameters
like $C_2^0$ and $L_0$, which describe the symmetry behavior of nuclear
matter, is achieved over model-I.  Strikingly, the errors on $C_2^0$
and $L_0$ for the model-II agree very well with the ones obtained for
the SAMi Skyrme force \cite{Roca-Maza12} which includes the variational
EoS for the pure neutron matter as pseudo data  in the fitting protocol.
We also provide the values of $\Delta r_{np}$ for $^{48}$Ca, $^{132}$Sn
and $^{208}$Pb nuclei in Table \ref{ch4a_tab3}.  The reduction in the
errors on $\Delta r_{np}$ for the model-II in comparison to those for the
model-I are in harmony with the results depicted in Fig.  \ref{ch4a_fig2}.

\begin{figure}[]{}
\centering
\includegraphics[width=0.7\textwidth]{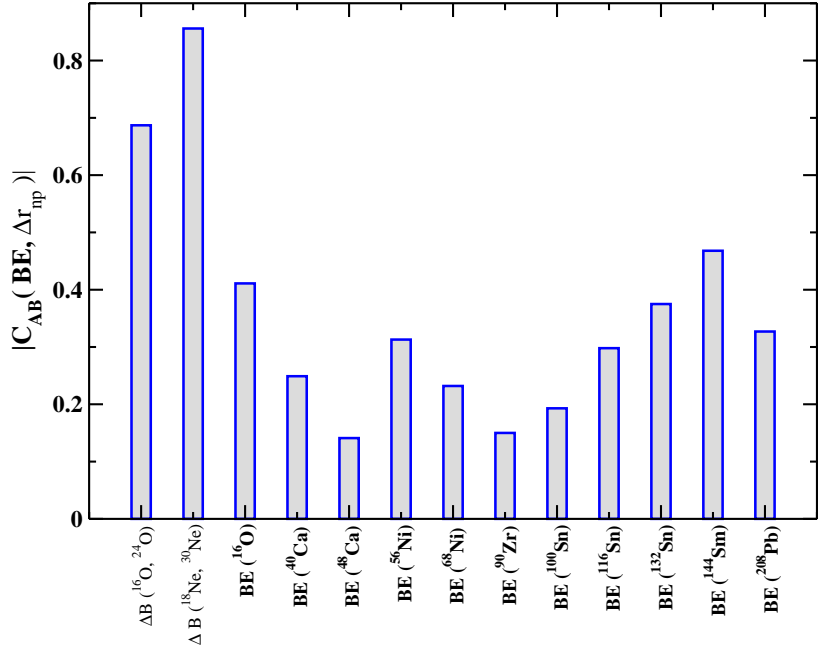} 
\caption{\label{ch4a_fig3} 
The correlation between $\Delta r_{np}$ of $^{208}$Pb and different
fit-data on binding energies used to obtain the model parameters of
model-II.}
\end{figure}
In Fig. \ref{ch4a_fig3} correlation between $\Delta r_{np}$ of $^{208}$Pb
and fit-data on binding energies for model-II are plotted. The stronger
correlation for the data on $\Delta B$ involving highly asymmetric nuclei
($\delta\sim0.33$) compared to others reproduces the similar features
of Fig. \ref{ch4a_fig1}, however, analysis being done within a single
model using covariance analysis. One can notice that these correlation
coefficients between $\Delta r_{np}$ of $^{208}$Pb and data on $\Delta B$
are not as high as compared to those obtained from the analysis involving
several models in Fig. \ref{ch4a_fig1}. Incidentally, these correlations
obtained from a single model using covariance analysis and using several
models should not be compared directly. The correlation in the covariance
analysis may depend on the set of data and the model parameters chosen
for an analysis. The correlations depicted in Fig. \ref{ch4a_fig3} is
more of an indicator that $\Delta B$ of Oxygen and Neon pair are more
sensitive to the $\Delta r_{np}$ of $^{208}$Pb compared to other binding
energy data used to optimize the parameters of model-II.

\section{A sensitivity analysis}
To analyze the sensitivity of the symmetry energy elements of nuclear
matter to highly neutron rich systems as obtained in the previous section,
a systematic analysis is essential. For this purpose Sensitivity analysis 
was employed as described in Ref. \cite{Dobaczewski14} (see also Chapter \ref{ch3}).
\subsection{The RMF models SINPB and SINPA}
Two different RMF models are constructed namely, SINPB and SINPA with an expanded data
set compared to those in model-I and model-II in the previous section. A
comparative study on the nuclear matter properties of these two models
is executed in detail.  In SINPB binding energies ($BE$) and charge radii
($r_{ch}$) of some standard set of nuclei across the whole nuclear chart
are taken as fit-data (see Tab. \ref{ch4b_tab2}).  The binding energies
of $^{54}$Ca, $^{78}$Ni and $^{138}$Sn nuclei having somewhat larger
asymmetry ($\delta\sim$ 0.26 - 0.28) are also included in the fitting
protocol. The model SINPA includes some highly asymmetric nuclei, namely,
$^{24}$O, $^{30}$Ne, $^{36}$Mg and $^{58}$Ca ($\delta>$ 0.3) in addition
to the data set used in the base model SINPB. SINPA also contains the
symmetric $^{20}$Ne and $^{24}$Mg nuclei and the observed maximum mass
of neutron star $M_{max}^{NS}$ as fit-data.

\begin{table*}[]
\centering
\caption{\label{ch4b_tab1}
Optimum values of the parameters for the models SINPB and SINPA,
statistical errors on them are given. Mass of the $\sigma$ meson
($m_\sigma$) is given in units of MeV.  The masses of $\omega$ and $\rho$
mesons are kept fixed to $m_\omega$= 782.5 MeV and $m_{\boldsymbol\rho}$=
763 MeV and nucleon mass is taken to be $M$= 939 MeV.  }
\begin{tabular}{ccccccccc}
\hhline{=========}
Name  & $g_{\sigma}$ & $g_{\omega}$ & $g_{\boldsymbol\rho}$ & ${\kappa_3}$
& ${\kappa_4}$ & ${\eta_{2\boldsymbol\rho}}$ & $\zeta_0$ & $m_{\sigma}$ \\
\hline
SINPB &-10.6007 & 13.8767 & 10.613 & 1.4868 & -0.802 & 13.487 & 5.467 & 
493.850\\
(Error)& 0.14 & 0.24 & 1.29 & 0.19 & 1.15 & 12.26 & 0.45 & 4.98\\
SINPA &-10.6292 & 13.8532 & 12.831 & 1.5375 & -1.190 & 38.179 & 5.363 & 
495.394 \\
(Error)& 0.16 & 0.33 & 0.82 & 0.06 & 0.47 & 11.92 & 0.45 & 3.86\\
\hhline{=========}
\end{tabular}
\end{table*}
In Table \ref{ch4b_tab1}, the optimal values of the parameters {\bf
p$_0$} for SINPB and SINPA are given along with the errors on them.
respectively. One can observe that the errors on the parameters
$g_{\boldsymbol\rho}$, ${\kappa_3}$, ${\kappa_4}$ and $m_{\sigma}$
decreased by a noticeable amount in SINPA in comparison to SINPB. For
the parameters $g_{\sigma}$, $\zeta_0$ and ${\eta_{2\boldsymbol\rho}}$
the errors are almost the same for both the models and for the case of
$g_{\omega}$ its value is slightly higher in SINPA than in SINPB. The
pairing is treated within the BCS approximation with cut-off energy in
pairing space taken as $\hbar\omega_0 = 41A^{-1/3}$ MeV. The BCS pairing
strengths for neutron and proton for the models SINPB and SINPA were kept
fixed to $G_n=20/A$ and $G_p=25/A$.  The neutron and proton pairing gaps
($\Delta_n, \Delta_p$) in MeV for the neutron rich nuclei  are $^{30}$Ne
(0.0, 2.3), $^{36}$Mg (2.5, 2.0), $^{54}$Ca (1.1, 0.0), $^{58}$Ca (1.0,
0.0), $^{138}$Sn (1.3, 0.0). The pairing gaps for other non-magic nuclei
are close to $12/\sqrt{A}$ MeV. The neutron pairing gap for   $^{24}$O
practically vanishes, since, the first unoccupied $1d_{3/2}$ orbit is
about 4.5 MeV above the completely filled $2s_{1/2}$ orbit \cite{Chen15}.

\begin{table*}[]
\centering
\caption{\label{ch4b_tab2}
Various observables $\mathcal{O}$, adopted errors on them
$\Delta\mathcal{O}$, corresponding experimental data (Expt.) and their
best-fit values for SINPB and SINPA. $BE$ and $r_{ch}$ corresponds
to binding energy and charge radius of a nucleus, respectively and
$M_{max}^{NS}$ is the maximum mass of neutron star (NS). Values of $BE$
are given in units of MeV and $r_{ch}$ in fm.  $M_{max}^{NS}$ is in
units of Solar Mass ($M_{\odot}$).}
\begin{tabular}{cccccc}
\hhline{======}
& $\mathcal{O}$ &$\Delta\mathcal{O}$ & Expt. &  
SINPB & SINPA \\
\hline
$^{16}$O & $BE$ &4.0 & 127.62 &  127.78 & 128.35\\
& $r_{ch}$ &0.04 & 2.699 &  2.704 & 2.696\\
$^{24}$O & $BE$ &2.0 & 168.96 & - & 169.28\\
$^{20}$Ne& $BE$ &4.0 & 160.64 & - & 155.89\\
$^{30}$Ne& $BE$ &3.0 & 211.29 & - & 214.37\\
$^{24}$Mg& $BE$ &3.0 & 198.26 & - & 195.87\\
$^{36}$Mg& $BE$ &2.0 & 260.78 & - & 261.68\\
$^{40}$Ca& $BE$ &3.0 & 342.05 &  343.19 & 343.66\\
& $r_{ch}$ &0.02 & 3.478 &  3.460 & 3.452\\
$^{48}$Ca& $BE$ &1.0 & 416.00 &  415.27 & 415.47\\
& $r_{ch}$ &0.04 & 3.477 &  3.437 & 3.437\\
$^{54}$Ca& $BE$ &2.0 & 445.37 &  445.63 & 443.79\\
$^{58}$Ca& $BE$ &2.0 & 454.43 & - & 456.33\\
$^{56}$Ni& $BE$ &5.0 & 483.99 &  483.38 & 484.34\\
& $r_{ch}$ &0.18 & 3.750 &  3.700 & 3.686\\
$^{68}$Ni& $BE$ &2.0 & 590.41 &  592.86 & 592.97\\ 
$^{78}$Ni& $BE$ &2.0 & 641.78 &  642.10 & 641.59\\ 
$^{90}$Zr& $BE$ &1.0 & 783.90 &  783.02 & 783.20\\
& $r_{ch}$ &0.02 & 4.269 &  4.266 & 4.264\\
$^{100}$Sn& $BE$ &2.0 & 825.30 & 828.11 & 827.93\\
$^{116}$Sn& $BE$ &2.0 & 988.68 & 987.45 & 987.32\\
& $r_{ch}$&0.18 & 4.625 &  4.620 & 4.622\\
$^{132}$Sn& $BE$ &1.0 & 1102.84 & 1103.28 & 1103.40\\
& $r_{ch}$&0.02 & 4.709 &  4.706 & 4.710\\
$^{138}$Sn& $BE$ &2.0 & 1119.59 &  1118.65 & 1117.05\\
$^{144}$Sm& $BE$ &2.0 & 1195.73 &  1196.00 & 1195.67\\
& $r_{ch}$&0.02 & 4.952 &  4.955 & 4.955\\
$^{208}$Pb& $BE$ &1.0 & 1636.43 &  1636.38 & 1636.57\\
& $r_{ch}$&0.02 & 5.501 &  5.528 & 5.530\\
\hline
NS& $M_{max}^{NS}$ &0.04 & 2.01 & - & 1.98\\
\hhline{======}
\end{tabular}
\end{table*}
In Table \ref{ch4b_tab2} different observables $\mathcal{O}$ pertaining to
finite nuclei and neutron star, their experimental values, their obtained
values from SINPB and SINPA along with $\Delta\mathcal{O}$, the adopted
errors on them are listed. The experimental values of binding energies
of all the nuclei except for $^{54}$Ca used in the fit are taken from
the latest compilation AME-2012 \cite{Wang12}. Recently, binding energy
of $^{54}$Ca was measured very accurately at TRIUMF \cite{Gallant12}
and CERN \cite{Wienholtz13}.  For this present calculation, the
experimental value of the binding energy for $^{54}$Ca is taken from
Ref. \cite{Wienholtz13}.  Experimental values for the charge radii used
in the fit are obtained from the compilation by Angeli and Marinova
\cite{Angeli13}.  For the optimization of SINPA, observed maximum mass
of neutron star $M_{max}^{NS}$ is taken from Ref. \cite{Demorest10,
Antoniadis13}.  It may be pointed out that, experimental value for some of
the fit data are little different in the present calculation in comparison
to the previous section.  Except for $^{68}$Ni, $\Delta\mathcal{O}$ for
all the fit-data common to both the models SINPB and SINPA are taken
from Ref. \cite{Klupfel09}.  As the obtained value of binding energy
of $^{68}$Ni from both the models SINPB and SINPA deviate by more than
2 MeV from its experimental value, demanding too much accuracy on that
particular datum costs a larger amount in total $\chi^2$ compared to other
data points.  For this reason $\Delta\mathcal{O} = 2$ was taken MeV for
the binding energy of $^{68}$Ni unlike in Ref.\cite{Klupfel09}, where
$\Delta\mathcal{O} = 1$ MeV.  Calculated errors on the binding energies
and charge radii due to uncertainties in the model parameters for the
fitted nuclei for both the models SINPB and SINPA lie within the range
from 0.51 - 1.89 MeV and 0.005 - 0.016 fm, respectively. In model SINPA
the obtained maximum neutron star mass $M_{max}^{NS}$ (1.98$\pm$0.03
$M_{\odot}$) compares well with the observed value.  It should be
noted that the two isotopes of Mg nuclei used in the optimization of
SINPA are deformed. The numerical computation is done with 20 oscillator
shells being taken as the basis states for the nucleons.  The quadrupole
deformation parameter $\beta_2$ calculated from SINPA for $^{24}$Mg and
$^{36}$Mg nuclei are found to be 0.47 and 0.37, respectively.

\subsection{Results for SINPB and SINPA}
Energy per nucleon $\mathcal{E} (\rho,0)$ for symmetric nuclear matter
(SNM) can be expressed in terms of model parameters as (see Chapter
\ref{ch2}),
\begin{eqnarray}
\label{erho}
\mathcal{E} (\rho,0)=&& \frac{2}{\pi^2}\int_0^{k_F} dk\
k^2 \sqrt{k^2+{M^*}^2}\nonumber \\ &&+\frac{1}{2}m_{\sigma}^2
\sigma^2 + \frac{{\kappa_3}}{6M} g_{\sigma}m_{\sigma}^2\sigma^3 +
\frac{{\kappa_4}}{24M^2}g_{\sigma}^2 m_{\sigma}^2\sigma^4 \nonumber \\
&&-\frac{1}{2}m_{\omega}^2 \omega^2 - \frac{1}{24}\zeta_0 g_{\omega}^{2}
\omega^4,
\end{eqnarray}
and, $C_2(\rho)$ is expressed as,
\begin{equation}
C_2(\rho)=\frac{k_F^2}{6(k_F^2+{M^*}^2)^{1/2}}+\frac{g_{\boldsymbol
\rho}^2}{12\pi^2}\frac{k_F^3}{{m_{\boldsymbol\rho}^*}^2}.
\label{csymrho}
\end{equation}
Here, $k_F$ is the nucleon Fermi momentum in symmetric nuclear matter
at density $\rho$ (=$\frac{2k_F^3}{3\pi^2}$). The Dirac effective mass
of nucleon $M^*$ is given by $M^*=M-g_{\sigma}\sigma$ 
and, the effective mass of $\boldsymbol\rho$ meson,
$m_{\boldsymbol\rho}^*$ is expressed as \cite{Horowitz01a},
\begin{equation}
{m_{\boldsymbol\rho}^*}^2 = m_{\boldsymbol\rho}^2\left(1+\frac{1}{2M^2}
\eta_{2\boldsymbol\rho}g_{\omega}^2\omega^2\right).
\label{mrhostar}
\end{equation}
From Eq. (\ref{csymrho}) one can see that, the kinetic part of $C_2(\rho)$
depends on the effective mass of nucleon $M^*$, which has dependence on
the parameter $g_{\sigma}$ and the field value of $\sigma$. However,
the interaction part of $C_2(\rho)$ mainly depends on the isovector
parameters $g_{\boldsymbol \rho}$ and $\eta_{2\boldsymbol\rho}$.

\begin{table}[]
\centering
\caption{\label{ch4b_tab3}
Different nuclear matter properties: the binding energy per nucleon for
symmetric matter $\mathcal E_0$, incompressibility coefficient $K_0$,
Dirac effective mass of nucleon $M^*_0$ (scaled by nucleon mass $M$),
symmetry energy coefficient $C_2^0$ and density slope parameter of
symmetry energy $L_0$ for the nuclear matter evaluated at saturation
density $\rho_0$  along with the correlated errors on them for the models
SINPB and SINPA. The values of $C_2(\rho_{c})$ and $L(\rho_{c})$
calculated at crossing density $\rho_c$ along with the neutron skin
$\Delta r_{np}$ in  $^{208}$Pb are also presented for these two models.  }
\begin{tabular}{ccc} 
\hhline{===}
Observable  & SINPB & SINPA\\ \hline 
$\mathcal E_0$ (MeV)           &  $-16.04\pm0.06$  & $-16.00\pm0.05$   \\ 
$K_0$ (MeV)             &  $206\pm20$   & $203\pm6$    \\
$\rho_0$ (fm$^{-3}$)  &  $0.150\pm0.002$    & $0.151\pm0.001$     \\
$M^*_0/M$               &  $0.59\pm0.01$    & $0.58\pm0.01$     \\
$C_2^0$ (MeV)         &  $33.95\pm2.41$     & $31.20\pm1.11$      \\
$C_2(\rho_{c})$ (MeV)         & $26.08\pm0.41$     & $25.60\pm0.51$     \\
$L_0$ (MeV)           &  $71.55\pm18.89$    & $53.86\pm4.66$      \\
$L(\rho_{c})$ (MeV)           &  $55.98\pm13.78$   & $38.47\pm5.43$     \\
$\Delta r_{np}$ ($^{208}$Pb) (fm) & $0.241\pm0.040$ & $0.183\pm0.022$\\
\hhline{===}
\end{tabular}
\end{table}
Once the objective functions for the models SINPB and SINPA are
optimized, different nuclear matter properties can be extracted from
them and compared.  In Table \ref{ch4b_tab3} values of different nuclear
matter parameters along with the corresponding errors evaluated within
the covariance analysis are listed for SINPB and SINPA.  The properties
associated with symmetric nuclear matter are evaluated at the saturation
density $\rho_0$, while, those characterizing the asymmetric nuclear
matter are evaluated at $\rho_0$ and  the crossing-density $\rho_c$ which
is taken as $\frac{0.11}{0.16}\times\rho_0$ \cite{Wang15}.  Errors on
binding energy per nucleon $\mathcal E_0$ ($=\mathcal{E} (\rho_0,0)$),
saturation density $\rho_0$ and Dirac effective mass of nucleon $M^*_0/M$
(=$M^*(\rho_0)/M$) are pretty much the same for both the models concerned.
However, a noticeable improvement is observed for the model SINPA over
SINPB for the calculated errors on the symmetry energy parameters $C^0_2$
(=$C_2(\rho_0)$), $L_0$ (=$L(\rho_0)$) and $L(\rho_c)$.  The refinement
in the error in SINPA in comparison to SINPB is also to be noted for the
incompressibility coefficient at saturation density, $K_0$.  Error on
the neutron-skin $\Delta r_{np}$ in $^{208}$Pb also reduces by almost
a factor of $2$ in SINPA in comparison to SINPB. The central values of
$L_0$ and $\Delta r_{np}$ of $^{208}$Pb obtained for the model SINPB
are seen to differ from  those obtained from the  model-I of previous
section; this can be attributed to the differences in the adopted error
on the binding energy of $^{68}$Ni and to the differences in some of
the experimental fit data.

The observation of improved constraint in the symmetry elements calculated
from model SINPA over those from SINPB  clearly indicates that the
additional data of four highly asymmetric nuclei ($^{24}$O, $^{30}$Ne,
$^{36}$Mg and $^{58}$Ca) with $\delta>0.3$ and the observed maximum mass
of neutron star $M_{max}^{NS}$ contain more distilled information on
isovector elements in the nuclear interaction.  It is striking to note
that the addition of the binding energies of $^{54}$Ca, $^{78}$Ni and
$^{138}$Sn ($\delta \sim 0.26$-$0.28$) as fit data in the optimization
of the model SINPB did  not improve the uncertainties in the  symmetry
energy parameters as compared to those for the model-I in the previous
section. On the other hand, inclusion of  highly asymmetric ($\delta
>0.3$) $^{36}$Mg and $^{58}$Ca nuclei in the fitting protocol of the model
SINPA yields smaller uncertainties in the symmetry energy parameters in
comparison to the model-II of previous section which does not include
these nuclei.  This clearly emphasizes that the binding energies of
nuclei with $\delta > 0.3$ play a crucial role in constraining the
symmetry energy parameters and is thus a pointer to the necessity of
taking data for very asymmetric nuclei in the optimization of the RMF
model. In the next section we are going to analyze this more critically.
\begin{figure}[]{}
\centering
\includegraphics[width=0.7\textwidth]{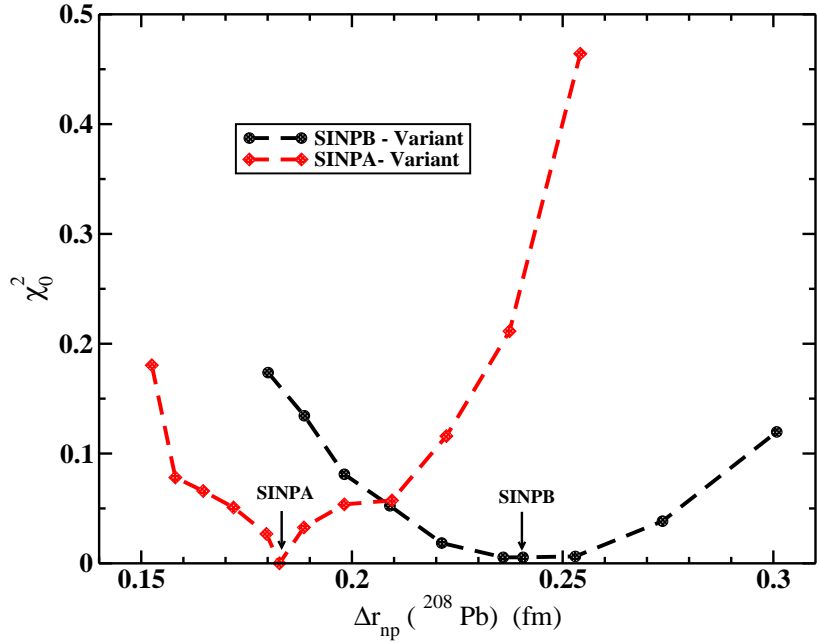} 
\caption{\label{ch4b_fig1}
Optimum values of the objective function ($\chi_0^2$)
are plotted as a function of $\Delta r_{np}$ (neutron skin of $^{208}$Pb)
for two families of models, namely, SINPB-Variant and SINPA-Variant
(see text for details).}
\end{figure}

Now the sensitivity of symmetry energy parameters to the properties of
the neutron rich systems are discussed in detail.  Before embarking on
the analysis in terms of sensitivity matrix (see Chapter \ref{ch3}), we
make a simple examination of the results.  We look into the dependence
of the optimal value of the objective function   on the neutron
skin of $^{208}$Pb.  Fixing $\eta_{2\boldsymbol\rho}$ to a preset
value  and optimizing the $\chi^2$ function by adjusting the rest
of the model parameters,  one can get a particular value of $\Delta
r_{np}$ of $^{208}$Pb for the models SINPB and SINPA \cite{Sil05}.
Two families of RMF models so constructed are called SINPB-Variant and
SINPA-Variant. Different input values of $\eta_{2\boldsymbol\rho}$
would yield different $\Delta r_{np}$ in both these models. In
Fig. \ref{ch4b_fig1} optimal values of the objective function $\chi^2$
(i.e. $\chi_0^2$) for these two models are displayed as a function of
$\Delta r_{np}$ of $^{208}$Pb; the values of $\chi_0^2$ are so adjusted
that their minimum value within a family vanishes. Visual comparison of
results from the two families of models shows that there is a stronger
\begin{figure}[]{}
\centering
\includegraphics[width=0.7\textwidth]{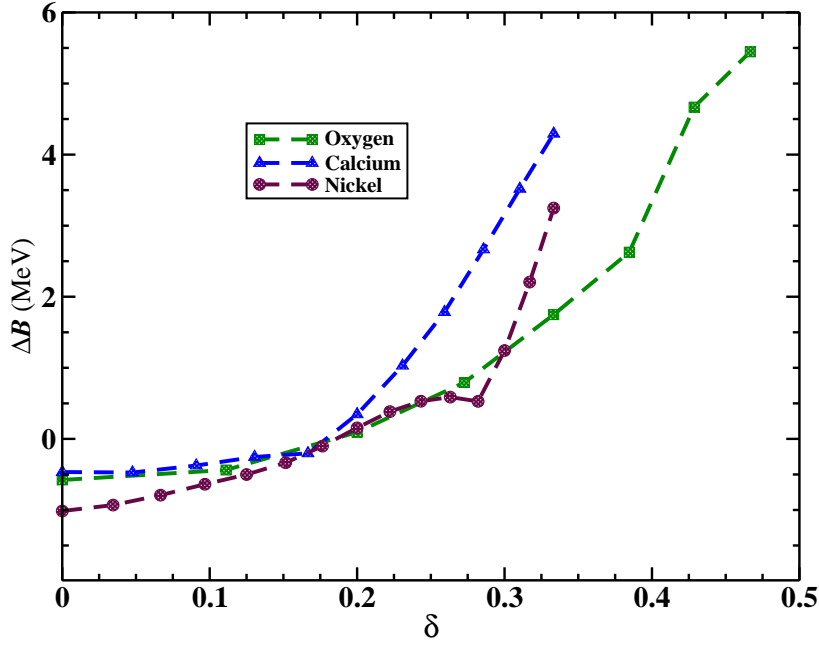} 
\caption{\label{ch4b_fig2}
Binding energy differences $\Delta B$ (=$BE({\rm
SINPB})-BE({\rm SINPA})$) extracted using models SINPB and SINPA for
even isotopes of O, Ca and Ni nuclei plotted as a function of asymmetry
$\delta$.}
\end{figure}
preference to a particular value of $\Delta r_{np}$ of $^{208}$Pb
in the SINPA-Variant family.  It is worthwhile to mention that,
SINPB-Variant family has $^{54}$Ca, $^{78}$Ni and $^{138}$Sn in the
fitted data set where asymmetry $\delta \sim 0.26$ - $0.28$. The
$\chi_0^2$ function is still rather flat, making it tenuous to give a
reasonable bound on the value of $\Delta r_{np}$ of $^{208}$Pb. The role
of ultra neutron-rich nuclei in the SINPA-Variant family where nuclei
with $\delta > 0.3$ (e.g. $^{24}$O, $^{30}$Ne, $^{36}$Mg, $^{58}$Ca)
are further included in the fitting protocol are eminently evident in
Fig \ref{ch4b_fig1}.  As $\Delta r_{np}$ of $^{208}$Pb is correlated to
$L_0$ \cite{Centelles09,Chen05}, one finds a tighter constraint on $L_0$
as well from SINPA as compared to SINPB (see Tab. \ref{ch4b_tab3}).

The two Variant families so constructed from selective optimization of
the parameter set ${\bf p_0}$ keeping $\Delta r_{np}$ of $^{208}$Pb fixed
should affect the calculated binding energies. In Fig. \ref{ch4b_fig2}
binding energy differences of three isotopic chains of O, Ca and Ni
extracted from models SINPB  and SINPA ($\Delta r_{np}$ ($^{208}$Pb)
= 0.241 fm and 0.183 fm, respectively at absolute minima of $\chi_0^2$,
see Fig. \ref{ch4b_fig1}) are plotted as a function of asymmetry $\delta$.
The differences in the binding energies so calculated for all the isotopic
chains show significant enhancement when one goes from $\delta$ just below
0.3 to higher values \cite{Chen15}. Nuclei beyond $\delta$ = 0.3 thus
show a high sensitivity towards $\Delta r_{np}$ of $^{208}$Pb. Several
experimental efforts are being made to accurately measure binding
energies of these exotic nuclei.  These measurements may impose very
tight constraint on the value of $\Delta r_{np}$ of $^{208}$Pb.
\begin{figure}[]{}
\centering
\includegraphics[width=0.7\textwidth]{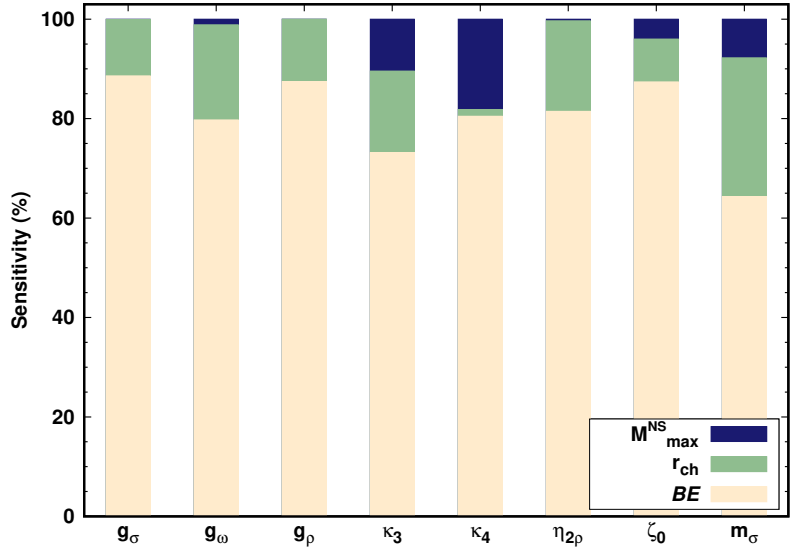} 
\caption{\label{ch4b_fig3}
Relative sensitivity of different parameters of the
effective Lagrangian density to three groups of fit data used in
optimization of SINPA. These groups are nuclear binding energies
($BE$), charge radii ($r_{ch}$) and maximum mass of neutron star
($M_{max}^{NS}$).}
\end{figure}

\begin{figure}[]{}
\centering
\includegraphics[width=0.7\textwidth]{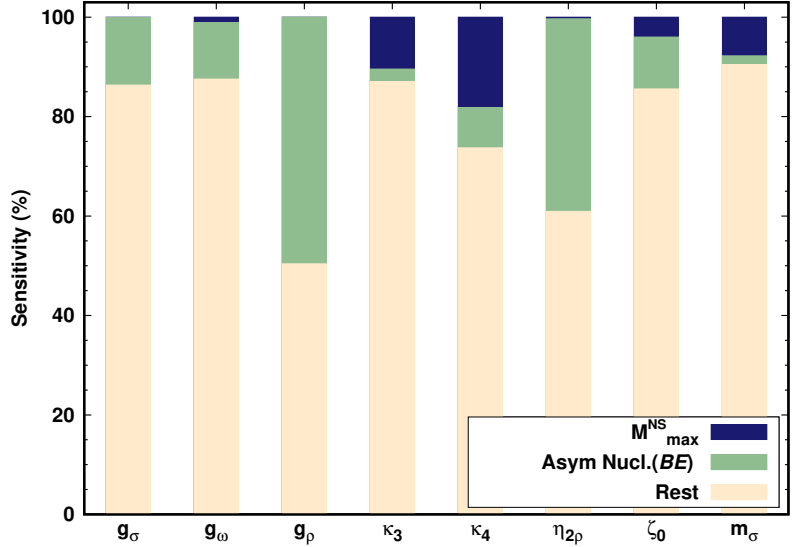} 
\caption{\label{ch4b_fig4}
Same as Fig. \ref{ch4b_fig3}, but, with different grouping
of the fit data of finite nuclei. One group contains binding energies of
highly asymmetric nuclei ($^{24}$O, $^{30}$Ne, $^{36}$Mg and $^{58}$Ca)
and another contains remaining fit data on the finite nuclei.}
\end{figure}
Further, the sensitivity analysis based on a sensitivity matrix
was employed in model SINPA to understand the impact of the
new fit-data considered to optimize it (see Chapter \ref{ch3}).
In Fig. \ref{ch4b_fig3} the relative sensitivity of different parameters
of the effective Lagrangian density to three broad data-types (binding
energies $BE$, charge radii $r_{ch}$ of finite nuclei and maximum mass
of neutron star $M_{max}^{NS}$) are displayed.  It is evident that
all the parameters are maximally sensitive ($>$65$\%$) to the binding
energies of nuclei.  The higher relative sensitivity of the parameters
to the binding energies of nuclei can be attributed partly to their
large number used in the fit.  The parameter $\kappa_4$ shows almost
no sensitivity towards the charge radii.  The parameters $\kappa_3$,
$\kappa_4$ and $m_{\sigma}$ are seen to be appreciably sensitive to the
single data of neutron star $M_{max}^{NS}$ as they have a crucial role
in the determination of the high density behavior of the nuclear EoS
which in turn governs the value of $M_{max}^{NS}$.

In Fig. \ref{ch4b_fig4} the analysis was performed by regrouping the data
on binding energies and charge radii so that the sensitivity of the RMF
model parameters to the binding energies of highly asymmetric nuclei can
be assessed.  One of the group consists of only the binding energies of
$^{24}$O, $^{30}$Ne, $^{36}$Mg and $^{58}$Ca nuclei, while the other
group contains the remaining data on the finite nuclei.  One can not
fail to notice that, the parameters $g_{\mathbf{\boldsymbol\rho}}$
and ${\eta_{2\boldsymbol\rho}}$, which control the isovector part of
the effective Lagrangian, are relatively more sensitive ($\sim 40\%$)
to the binding energies of highly asymmetric nuclei.  The sensitivity
of $g_{\boldsymbol\rho}$ and ${\eta_{2\boldsymbol\rho}}$ to the
value of $M_{max}^{NS}$ is not observed in Figs. \ref{ch4b_fig3} and
\ref{ch4b_fig4} partly because $M_{max}^{NS}$ is a single datum, but
mainly because it is overshadowed by the relative contributions to the
sensitivity from the binding energies of asymmetric nuclei.
\begin{figure}[]{}
\centering
\includegraphics[width=0.7\textwidth]{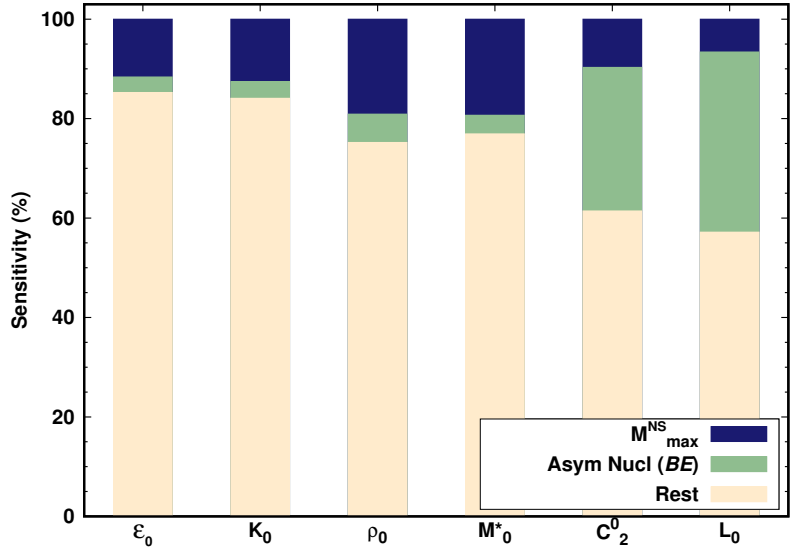} 
\caption{\label{ch4b_fig5}
Relative sensitivity of the nuclear matter properties at
saturation density to the fit data of SINPA with the same grouping as
in Fig. \ref{ch4b_fig4}.}
\end{figure}

In Fig. \ref{ch4b_fig5} the sensitivity of different empirical data
pertaining to the saturation density $\rho_0$ of nuclear matter are
displayed to the data-set used in the optimization of the model SINPA. To
do so, the same grouping of data was used as in Fig. \ref{ch4b_fig4}. Since
the parameters $g_{\sigma}$, $g_{\omega}$ etc. of the effective Lagrangian
are optimally determined from the full data set, it is no wonder that the
empirical nuclear matter data obtained from the energy density functional
are maximally sensitive to the group of fit data "Rest", as it contains
the largest number of data elements.  The high sensitivity of $C_2^0$
($\sim 30\%$) and $L_0$ ($\sim 40\%$) to the binding energies of the
highly asymmetric $^{24}$O, $^{30}$Ne, $^{36}$Mg and $^{58}$Ca nuclei,
which form a very small subset of the data-set used in the optimization
of SINPA (4 out of 30) is a reflection of the high sensitivity of the
model parameters $g_{\boldsymbol\rho}$ and ${\eta_{2\boldsymbol\rho}}$
to the masses of these highly asymmetric nuclei as seen earlier in
Fig. \ref{ch4b_fig4}.  Appreciable sensitivity of all the nuclear
matter properties to the single data on neutron star $M_{max}^{NS}$
can not also be missed either.  Accurate knowledge of $M_{max}^{NS}$
is required for the precision determination of the EDF involving high
densities beyond saturation, any small change in it thus may result in
large change in the value of the nuclear matter properties ($\mathcal E_0,
K_0, \rho_0, M^*_0$) calculated from the EDF. This can be appreciated
from the sensitivity of $\kappa_3, \kappa_4$ and partly $\zeta_0$
(governing the scalar mass and the number density) on $M_{max}^{NS}$
displayed in Fig. \ref{ch4b_fig4}.  The not-too-insignificant sensitivity
of $C_2^0$ and $L_0$ to $M_{max}^{NS}$ demands attention.  It stems
from the dependence of the kinetic part of $C_2(\rho)$ on $M^*$
(Eq. (\ref{csymrho})) whose value at saturation density is found
appreciably sensitive to the maximum mass of neutron star.  The value
of $\sigma$-field  determining the effective mass of nucleon depends
on the coupling constants $g_{\sigma}$, $\kappa_3$, $\kappa_4$ and the
value of $m_{\sigma}$.  High sensitivity of these coupling constants
to $M_{max}^{NS}$ (see Figs. \ref{ch4b_fig3} and \ref{ch4b_fig4}) gets
reflected in the sensitivity analysis of the symmetry energy parameters
to $M_{max}^{NS}$.


\subsection{Nuclear Matter properties at high density}
\begin{figure}[]{}
\centering
\includegraphics[width=0.7\textwidth]{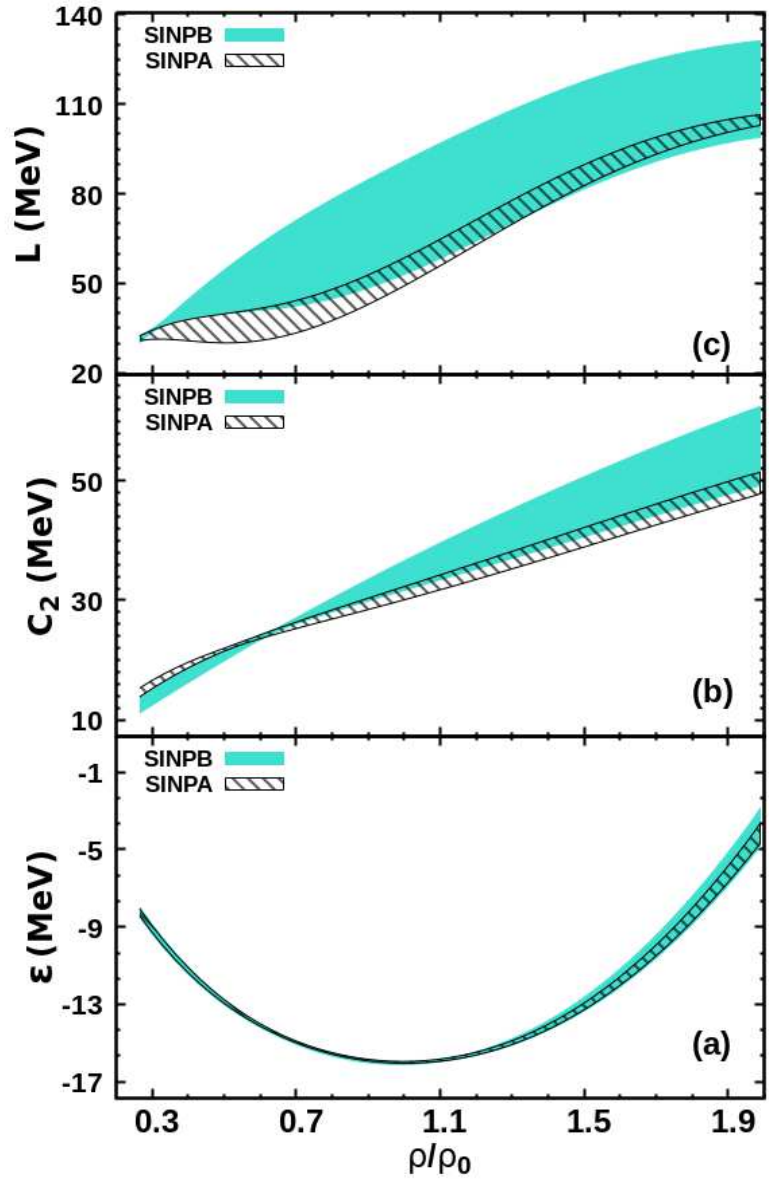} 
\caption{\label{ch4b_fig6}
Binding energy per nucleon for symmetric matter $\mathcal E$,
symmetry energy parameter $C_2$ and its density derivative $L$ along
with their errors as a function of density $\rho/\rho_0$ for SINPB and
SINPA. }
\end{figure}
The calculation of nuclear matter properties with both the models
SINPB and SINPA were extended for densities beyond saturation. This
provides valuable informations to construct theories for dense nuclear
systems viz. neutron star and several other astrophysical objects
from EoS so constrained at saturation density. In Fig. \ref{ch4b_fig6}
different nuclear matter properties, e.g. binding energy per nucleon
for symmetric matter $\mathcal E$ (Fig. \ref{ch4b_fig6}(a)), symmetry
energy coefficient $C_2$ (Fig \ref{ch4b_fig6}(b)) and its density
derivative $L$ (Fig. \ref{ch4b_fig6}(c)) were plotted as a function
of density $\rho/\rho_0$ for the models SINPB (turquoise) and SINPA
(black-pattern) along with their associated errors.  The errors are
calculated within the covariance analysis.  The energy per nucleon
$\mathcal E$ in the explored density region for SINPB and SINPA are
almost identical as seen from Fig. \ref{ch4b_fig6}(a). Most stringent
constraint on the values of $\mathcal E$ appear at $\rho \sim \rho_0$
for both the models and they grow as one moves away from $\rho_0$
\cite{De15}. In Fig.  \ref{ch4b_fig6}(b) allowed regions of $C_2$
show similar trend for SINPB and SINPA, both of them having their
minimum variance at $\rho \sim 0.7\rho_0$ \cite{Zhang13}. However, a
significant improvement is observed over the errors on $C_2$ for SINPA
in comparison to SINPB at higher densities.  Comparison of calculated
electric dipole polarizability of $^{208}$Pb from several Skyrme and RMF
interactions with the corresponding experimental data recently yielded
a very tightly constrained value of $C_2$ at density $\rho_0/3$, $C_2
(\rho_0/3)= 15.91\pm0.99$ MeV \cite{Zhang15}. It is interesting to note
that the model SINPB has overlap with this constraint at the lower end,
$C_2 (\rho_0/3) = 13.69$ - $16.31$ MeV, whereas SINPA agrees with this
result at the higher end, $C_2 (\rho_0/3)= 16.41$ - $17.67$ MeV.

\begin{figure}[]{}
\centering
\includegraphics[width=0.7\textwidth]{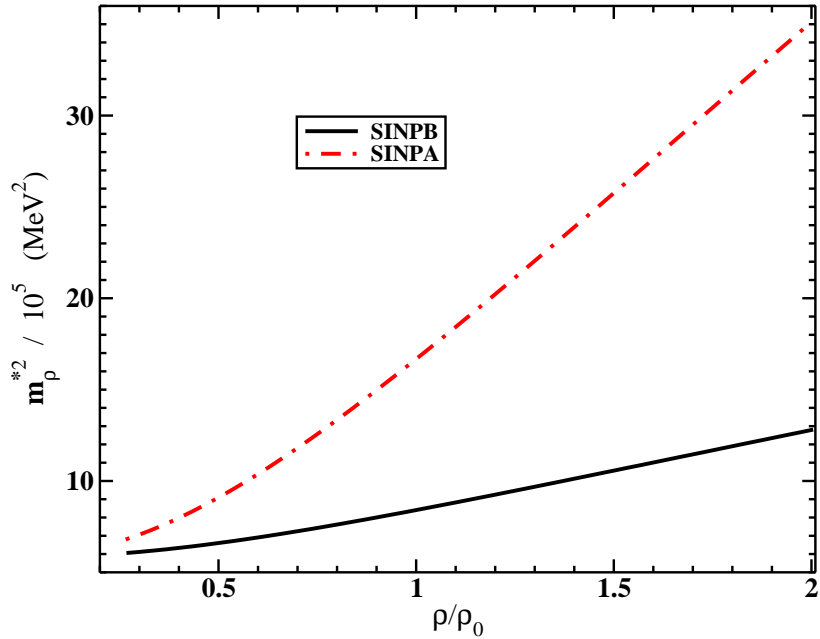} 
\caption{\label{ch4b_fig7}
Square of the effective mass of $\boldsymbol\rho$ meson
(scaled by $10^5$) as a function of density $\rho/\rho_0$ plotted for
SINPB and SINPA. }
\end{figure}

In Fig. \ref{ch4b_fig6}(c) a curious behavior in the variance of $L$
with density was observed.  For the model SINPB, the variance in $L$
grows up to a certain density $\sim \rho_0$ and from there onwards it
remained almost constant all the way up to $2\rho_0$. In contrast, in
SINPA error on $L$ grows only up to $\rho \sim 0.7\rho_0$ and shows a
monotonically decreasing trend afterwards.  This particular result may
appear intriguing.  A model primarily obtained by fitting some ground
state properties of finite nuclei, where concerned central density is
$\sim \rho_0$ and average density is $\sim 0.7\rho_0$ is not normally
expected to show better constraint on nuclear matter properties
at ultra-saturation densities. To investigate this, the expression
of $C_2$ as a function of density given in Eq. (\ref{csymrho}) was
recalled. $C_2$ has a dependence on ${m_{\boldsymbol\rho}^*}^2$, the
square of the effective mass of ${\boldsymbol\rho}$ meson.  The density
variation of ${m_{\boldsymbol\rho}^*}^2$ for both the models are
displayed in Fig. \ref{ch4b_fig7}.  A rapid difference in the value
of ${m_{\boldsymbol\rho}^*}^2$ (scaled by $10^5$) calculated in models
SINPA and SINPB builds up with increasing density. As the value of the
parameter $\eta_{2\boldsymbol\rho}$ is much larger in SINPA (38.18)
compared to that in SINPB (13.49) [see Table \ref{ch4b_tab1}], at high
densities the second term in the expression of $C_2$ (Eq. (\ref{csymrho}))
gets diluted due to ${m_{\boldsymbol\rho}^*} ^2$ (Eq. (\ref{mrhostar}))
by a much greater rate for the model SINPA in comparison to SINPB.
This explains why the error on $C_2$ grows at much faster rate in
SINPB than in SINPA. Now, if one takes density derivative of $C_2$,
the second term in Eq. (\ref{csymrho}) gives rise to two terms with
$\eta_{2 \boldsymbol\rho}$ in the denominator for the expression of $L$
as a function of density due to varying $\omega$ field value.  That is why
$\eta_{2\boldsymbol\rho}$ becomes a very crucial factor for the values
of $L$ at higher densities. This fact explains why in SINPA error on
$L$ decreases at higher densities, whereas in SINPB it remains almost
constant as shown in Fig. \ref{ch4b_fig6}(c).

\section {Summary}
To sum up, an investigation is made on the extraction of the precision
information from experimental data on the isovector content of the nuclear
interaction and their observable derivatives like the symmetry energy of
nuclear matter and its density slope $L_0$ at saturation density. The
relativistic mean field model is chosen as the framework for the
realization of this goal. A comparative study of the covariance analysis
of the interaction strengths and the symmetry observables ($C_2^0$,
$L_0$, $\Delta r_{np}$ of $^{208}$Pb) made with two sets of models: (I)
with model-I and SINPB (these included in the fit data observables from
nearly symmetric and few asymmetric nuclei); (II) with model-II and SINPA
(which included further data from extremely asymmetric nuclei right at
the edge of neutron drip line with neutron to proton ratio $\sim$ 2 and
the observed maximum mass $M_{max}^{NS}$ of neutron star for SINPA) shows
that the nuclear symmetry energy properties and the neutron skin thickness $\Delta
r_{np}$ of $^{208}$Pb are determined in much narrower constraints from the
latter set of models.  This is a pointer to the necessity of inclusion
of extremely neutron-rich systems in any data analysis for filtering
out information on isovector entities in the nuclear interaction. The
conclusion is further reinforced  from the sensitivity analysis of
the different model parameters of SINPA entering the nuclear effective
interaction to the experimental data set taken for such an analysis.

%% file: chap5.tex
\chapter{Model dependence in the symmetry energy parameters}\label{ch5}
\section{Introduction}
The macroscopic nuclear droplet model (DM) given by Myers and Swiatecki
\cite{Myers69, Myers80} suggests that various symmetry energy parameters
and the neutron-skin thickness in a heavy nucleus are related to one
another. The neutron skin thickness is defined as the difference between
the rms radii for the density distributions of the neutrons and protons
in the nucleus:
\begin{equation}
\Delta r_{\rm np} \equiv \langle r^2 \rangle_n^{1/2} 
- \langle r^2 \rangle_p^{1/2}.
\label{skin}
\end{equation} 
Nuclear mean-field models predict a nearly linear correlation of $\Delta
r_{\rm np}$ of a heavy nucleus such as $^{208}$Pb with the slope of the
equation of state of neutron matter at a subsaturation density around
0.1 fm$^{-3}$ \cite{Brown00a,Brown01}, with the density derivative
of the symmetry energy $L_0$ \cite{Furnstahl02,Centelles09,Warda09,
Danielewicz03,Avancini07,Vidana09, Chen05}, and with the surface symmetry
energy in a finite nucleus \cite{Danielewicz03,Centelles09,Satula06}.
The correlation of a finite nucleus property such as $\Delta r_{\rm
np}$ with a bulk property of infinite nuclear matter such as $L_0$ can be
interpreted as basically due to the dependence of $\Delta r_{\rm np}$
on the surface symmetry energy.  In a local density approximation
the surface symmetry energy can be correlated with $L_0$, and this
fact therefore implies the correlation between $\Delta r_{\rm np}$
and $L_0$. Macroscopic approaches such as the DM \cite{Myers69,Myers80}
often provide insightful guidance into the global features of many of
these correlations \cite{Centelles09,Warda09,Centelles10}.

Hadronic probes based on strong interaction, measure the $\Delta 
r_{\rm np}$ of $^{208}$Pb with fair accuracy \cite{Hoffmann80,Zenihiro10,
Krasznahorkay04,Klos07,Friedman09,Tarbert14}. However, being immensely 
dependent on the formulation of the strong interaction, these measurements 
are heavily model dependent. The Lead Radius Experiment (PREX) 
\cite{Horowitz01, Abrahamyan12} based on parity violating electron 
scattering \cite{Dubach89}, provides the most model independent 
measurement of $\Delta r_{\rm np}$ of $^{208}$Pb with fair accuracy. 
Ongoing efforts are underway to perform an accurate and model independent
measurement of the neutron-skin thickness in the $^{208}$Pb nucleus 
\cite{Jlab}. At the same time, it may not be straightforward for theory to extract
various symmetry energy parameters from the neutron-skin thickness in a
model-independent fashion. $\Delta r_{\rm np}$ of $^{208}$Pb was extracted 
recently from comparison of theory with the measured electric dipole polarizability in
$^{208}$Pb \cite{Roca-Maza13, Roca-Maza15, Tamii11, Hashimoto15, Rossi13}.
However, the focus has mainly been on the linear correlation between the neutron-skin thickness
and the slope parameter $L_0$ of the symmetry energy \cite{Brown00a,Brown00,Brown01,Furnstahl02}. The correlation
is satisfied to a large degree in the microscopic calculations
with mean field models but it is not perfect and a certain model
dependence appears in the results \cite{Brown00,Brown01,Furnstahl02,
Danielewicz03,Centelles09,Warda09,Centelles10}.  
In the present thesis work the correlations of $\Delta r_{\rm
np}$ with various symmetry energy parameters were revisited to look for the plausible causes
for the existence of a model dependence in these correlations. 
This complements the calculations done in Chapter \ref{ch4}, 
where the correlation between $\Delta r_{\rm np}$ of $^{208}$Pb 
and $L_0$ was studied within a relativistic mean field model using 
covariance analysis.

\section{Neutron-skin thickness and symmetry energy \\parameters in Droplet Model}
\label{theo}
From a geometrical point of view, the neutron skin thickness in a nucleus
may be thought as originated by two different effects. One effect is
due to the separation between the mean sharp surfaces of the neutron
and proton density distributions. Since this effect corresponds to a
different extent of the bulk region of the neutron and proton densities,
it is referred as the bulk contribution to the neutron skin thickness. The
other effect is due to the different surface widths of the neutron and
proton densities, which is called the surface contribution to the neutron
skin thickness. To compute the bulk and surface contributions to the
neutron skin thickness in a nucleus requires a proper definition of
these quantities based on the nuclear densities. In this respect 
the method described by Hasse and Myers \cite{Hasse88} is closely followed.

In order to determine the position of the neutron and proton effective
surfaces one can define different radii. In particular, one can define
the central radius $C$ as
\begin{equation}
\label{c}
C= \frac{1}{\rho(0)} 
\int_0^{\infty} \rho(r) dr .
\end{equation}
Another option for the mean position of the surface is the equivalent
radius $R$, which is the radius of a uniform sharp distribution whose
density equals the bulk value of the actual density and has the same
number of particles:
\begin{equation}
\label{r}
\frac 43 \pi R^3 \rho({\rm bulk}) =
4\pi \int_0^{\infty} \rho(r) r^2 dr \,.
\end{equation}
Finally, one can also define the  equivalent rms radius $Q$ that
describes a uniform sharp distribution with the same rms radius as the
given density:
\begin{equation}
\label{q}
\frac 35 \, Q^2= \langle r^2 \rangle \,.
\end{equation}

The radii $C$, $R$, and $Q$ are related by the expressions \cite{Hasse88}
\begin{equation}
\label{qr}
Q= R\left(1+\frac{5}{2}\frac {b^2}{R^2}+ ...\right) \quad
C= R\left(1-\frac {b^2}{R^2}+ ...\right),
\end{equation}
where $b$ is the surface width of the density profile defined as
\begin{equation}
\label{b}
b^2= - \frac{1}{\rho(0)}
\int_0^{\infty} (r-C)^2 \frac{d \rho(r)}{dr} dr,
\end{equation}
which provides a measure of the extent of the surface of the nucleus.
The neutron skin thickness, which is defined through the rms radii, can 
be expressed by
\begin{equation}
\label{r0}
\Delta r_{np}=\sqrt{\frac{3}{5}} \left(Q_n-Q_p\right) ,
\end{equation}
and using Eq.(\ref{qr}) reads:
\begin{equation}
\Delta r_{\rm np} = \sqrt{\frac{3}{5}} \left[\left(R_n - R_p\right)
+ \frac{5}{2}\left(\frac{b_n^2}{R_n}-\frac{b_p^2}{R_p}\right)\right],
\label{rnptotal}
\end{equation}
which clearly separates the bulk and surface contributions as
 \begin{equation}
\Delta r_{\rm np}^{\rm bulk} \equiv \sqrt{\frac{3}{5}} \left(R_n - R_p\right),
\label{rnpbulk}
\end{equation}
and 
\begin{equation}
\Delta r_{\rm np}^{\rm surf} \equiv\sqrt{\frac{3}{5}} \frac{5}{2}
\left(\frac{b_n^2}{R_n}-\frac{b_p^2}{R_p}\right).
\label{rnpsurf}
\end{equation}
In Eqs. (\ref{rnptotal}) and (\ref{rnpsurf}), 
$\mathcal{O}\left[b^4/R^3\right]$ and higher-order terms are neglected since they
represent a small correction \cite{Centelles10} to $\Delta r_{\rm
np}$ -- of less or around a 1-2\% -- that will leave the conclusions
unchanged.

The quantal proton and neutron densities obtained
within the Skyrme Hartree-Fock or the relativistic mean-field models are 
described in Chapter \ref{ch2}.
In order to extract the bulk and surface contributions to the neutron
skin thickness from these distributions of neutrons and protons, the 
method which was followed closely resembles Refs. \cite{Centelles10, 
Warda10}. The self-consistent quantal proton and neutron densities were 
fitted by two-parameter Fermi (2pF) distributions
\begin{equation}
\rho_q(r)=\frac{\rho_{0,q}}{1+{\rm exp}[(r-C_q)/a_q]},
\label{2pF}
\end{equation}
where $q=n,p$. The parameters $\rho_{0,q}$,  $C_q$ and $a_q$ are adjusted
to reproduce the nucleon numbers as well as the values for the second
and fourth moments of the actual density distributions, i.e.,  $\langle
r^2_q\rangle$ and $\langle r^4_q\rangle$.  Once this fit is done,
one can express Eqs. (\ref{rnptotal})--(\ref{rnpsurf}) for the neutron
skin thickness in terms of the parameters $C_q$ and $a_q$ taking into
account Eq.(\ref{qr}) and the fact that for a 2pF distribution $b=\pi
a/\sqrt{3}$. Therefore, the bulk and surface contributions to the neutron
skin thickness can be written as
\begin{equation}
\Delta r_{\rm np}^{\rm bulk}= \sqrt{\frac{3}{5}}
\left[(C_n-C_p) +\frac{\pi^2}{3}\left(\frac{a_n^2}{C_n}-
\frac{a_p^2}{C_p}\right)\right ], 
\label{blk}
\end{equation}
\begin{equation}
\Delta r_{\rm np}^{\rm surf}=  \sqrt{\frac{3}{5}}\frac {5 \pi^2}{6}
\left(\frac{a_n^2}{C_n}-\frac{a_p^2}{C_p}\right),
\label{srf}
\end{equation}
up to terms of order $\mathcal{O}\left[a^4/C^3\right]$. It should
be mentioned that, the $\Delta r_{\rm np}$ values calculated from the
actual densities obtained self consistently match very well with the ones
calculated by summing Eqs. (\ref{blk}) and (\ref{srf}) after applying
our prescription to determine the parameters of the Fermi function.

Some insight about possible correlations between the neutron skin
thickness and different observables related to the symmetry energy is
provided by the DM \cite{Myers80}. Within this model, which neglects
shell correction effects, the neutron skin thickness is expressed by
\begin{equation}
\Delta r_{\rm np} = \sqrt{\frac{3}{5}} \left[t - \frac{e^2Z}{70 C_2^0}
+ \frac{5}{2R}\left(b_n^2 - b_p^2\right)\right],
\label{rnpdm}
\end{equation}
where $e^2Z/70C_2^0$ is a correction due to the Coulomb interaction, $R=r_0
A^{1/3}$ is the nuclear radius, and $b_n$ and $b_p$ are the surface
widths of the neutron and proton density profiles.  The quantity $t$
in (\ref{rnpdm}) represents the distance between the location of the
neutron and proton mean surfaces and therefore is proportional to the
bulk contribution to the neutron skin thickness. In the DM its value is
given by
\begin{equation}
t = \frac{3}{2}r_0 \frac{C_2^0}{Q_{\rm stiff}} \frac{I-I_C}{1 + x_A}, 
\label{t}
\end{equation}
with 
\begin{equation}
I_C = \frac{3e^2}{5r_0}\frac{Z}{12C_2^0}A^{-1/3} \quad  \text{and} \quad 
x_A = \frac{9C_2^0}{4Q_{\rm stiff}}A^{-1/3},
\label{ICXA}
\end{equation}
where $I=(N-Z)/A$, $C_2^0$ is the bulk symmetry energy at saturation, and
$Q_{\rm stiff}$ is the surface stiffness. For each mean field model,
the parameters $r_0$ and $C_2^0$ can be obtained from calculations in
infinite nuclear matter and $Q_{\rm stiff}$ from calculations performed
in semi-infinite nuclear matter \cite{Warda09,Centelles98,Estal99}.

Within the DM, the symmetry energy coefficient of a finite nucleus of
mass number $A$ is given by
\begin{equation}
a_{sym}(A) = \frac{C_2^0}{1 + x_A}.
\label{symdm}
\end{equation}
Replacing $a_{sym}(A)$ in Eq. (\ref{t}), the separation distance between
the mean surfaces of neutrons and protons can be recast as
\begin{equation}
t = \frac{2r_0}{3C_2^0}[C_2^0 - a_{sym}(A)]A^{1/3} (I -I_C). 
\label{t1}
\end{equation}
The link between a property in finite nuclei such as $a_{sym}(A)$ and some
symmetry energy parameters in infinite nuclear matter may be obtained from
the observation \cite{Centelles09} that for a heavy nucleus there is a
subsaturation density, which for $^{208}$Pb is around 0.1 fm$^{-3}$, such
that the symmetry energy coefficient in the finite nucleus $a_{sym}(A)$
equals the symmetry energy in nuclear matter $C_2(\rho)$ computed at that
density. This relation is roughly independent of the mean field model
used to compute it. Around the saturation density $\rho_0$ the symmetry
energy can be expanded as
\begin{equation}
C_2(\rho) \simeq C_2^0 -  L_0\Big(\frac{\rho_0-\rho}{3\rho_0}\Big) 
+ \frac{1}{2}K_{sym}\Big(\frac{\rho_0-\rho}{3\rho_0}\Big)^2.
\label{s_rho}
\end{equation}
Consequently, the distance $t$ can be finally expressed approximately as
\cite{Centelles09}
\begin{equation}
t = \frac{2r_0}{3C_2^0}L_0\Big(\frac{\rho-\rho_0}{3\rho_0}\Big)
\Big[1 - \frac{K_{sym}}{2L_0} \Big(\frac{\rho-\rho_0}{3\rho_0}\Big) \Big] A^{1/3} (I -I_C).
\label{t2}
\end{equation}
Equations (\ref{t1}) and (\ref{t2}) suggest correlations between
the bulk neutron skin thickness in finite nuclei and some isovector
indicators such as $C_2^0-a_{sym}(A)$, $a_{sym}(A)/C_2^0$ and $L_0$. 
To compute the average symmetry
energy of a finite nucleus with the DM (Eq. (\ref{symdm})) requires
the knowledge of the surface stiffness $Q_{\rm stiff}$, which in turn
requires semi-infinite nuclear matter calculations \cite{Warda09}. An
efficient procedure to circumvent this, is to evaluate $a_{sym}(A)$
within a local density approximation as \cite{Agrawal12}
\begin{eqnarray}
a_{\rm sym}(A)=\frac{4 \pi}{AI^2}\int \ [r^2 \rho(r)I^2(r)] C_2(\rho(r))dr,
\label{eq:Asym}
\end{eqnarray}
where $I(r) = \frac{\rho_n(r) -\rho_p(r)}{\rho(r)}$ is the local isospin
asymmetry and $\rho (r)$ is the sum of the neutron and proton densities.
This approximation works very well for medium heavy $^{132}$Sn or heavy
$^{208}$Pb nuclei \cite{Liu13}.

\section{Results and discussions}
The neutron-skin thickness and several symmetry energy parameters are
calculated using five different families of systematically varied
models, namely, the \mbox{SAMi-J} \cite{Roca-Maza12, Roca-Maza13},
DDME \cite{Vretenar03}, FSV, TSV and KDE0-J models. The energy density
functional associated with DDME, FSV, and TSV corresponds to an effective
Lagrangian density typical of the relativistic mean-field models, whereas
SAMi-J and KDE0-J are based on the standard form of the Skyrme force 
(see Chapter \ref{ch2}).

The different families of systematically varied
parameter sets were obtained so that they explore different values of the symmetry
energy parameters around an optimal value, while reasonably keeping the
quality of the best fit.  The values of the neutron-skin thickness in a
heavy nucleus like $^{208}$Pb vary over a wide range within the families
due to the variations of the symmetry energy parameters. The parameter
sets for the FSV, TSV and KDE0-J families were obtained in the present
thesis work. The effective Lagrangian density employed for the FSV family is
similar to that for the FSU model \cite{Todd-Rutel05}.  In addition to
the coupling of $\rho$ meson to the nucleons as conventionally employed,
the presence of a cross-coupling between the $\omega$ and $\rho$ mesons in
the FSU model enables one to vary the symmetry energy, and accordingly
the symmetry energy slope parameter $L_0$, over a wide range without
significantly affecting the quality of the fit to the bulk properties
of the finite nuclei. The TSV family is obtained using the effective
Lagrangian density as introduced in Ref. \cite{Estal01} in which the
$\rho-$meson and its coupling to the $\sigma-$meson govern the isovector
part of the interactions between the nucleons. The $\omega-\rho$ cross
coupling in the FSV family and the $\sigma-\rho$ cross coupling in the
TSV family produce different behaviors in the density dependence of
the symmetry energy, because the source term for the $\omega$-field
is governed by the baryon density and that for the $\sigma$-field
is governed by the scalar density. The experimental data employed to
determine the TSV and FSV families are the total binding energies for
the $^{16}{\rm O}, ^{40,48}{\rm Ca}, ^{68}{\rm Ni}, ^{90}{\rm Zr},
^{100,132}{\rm Sn}, ^{208}{\rm Pb}$ nuclei, and the root mean square
charge radii for the $^{16}{\rm O}, ^{40,48}{\rm Ca}, ^{90}{\rm Zr},
^{208}{\rm Pb}$ nuclei. The energy density functional for the KDE0-J
family calculated within the Skyrme ansatz is taken from the KDE0 force
of Ref. \cite{Agrawal05}.  The model parameters are constrained to
yield the nuclear matter incompressibility coefficient in the range of
225--250 MeV. The calculated values of the total binding energy and the
charge radius for the $^{208}$Pb nucleus obtained for all the models
considered deviate from the experimental data only within $0.25\%$
and $0.8\%$, respectively.

\subsection{Correlation plots associated with isovector \\indicators}
\begin{figure}
\centering
\includegraphics[width=0.7\textwidth]{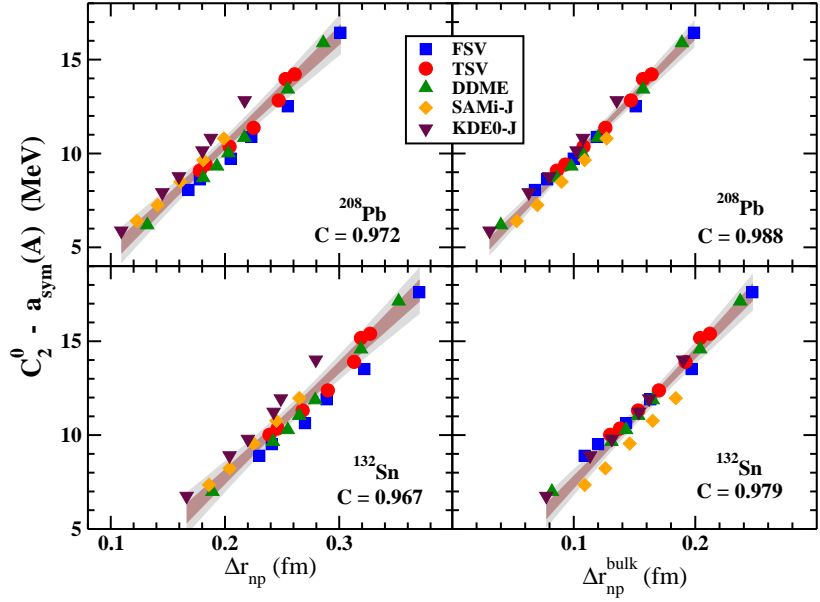}
\caption{\label{ch5_fig1} Plots for the difference between the
symmetry energy coefficient for  infinite nuclear matter $C_2^0$ and that
for finite nuclei $a_{\rm sym}(A)$  as a function of the neutron-skin
thickness (left panels) and of the bulk part of the neutron-skin thickness
(right panels).  The results are obtained using five different families
of mean-field models, namely, FSV (blue squares), TSV (red circles), DDME
(green triangles), SAMi-J (orange diamonds) and KDE0-J (maroon inverted
triangles). The correlation coefficients are: $C(C_2^0 - a_{\rm sym}(A),
\Delta r_{\rm np} )$ = 0.972 (0.967) and  $C(C_2^0- a_{\rm sym}(A), \Delta
r_{\rm np}^{\rm bulk})$ = 0.988 (0.979)  for  $^{208}$Pb ($^{132}$Sn)
nuclei.  The inner (outer) colored regions depict the loci of the 95\%
confidence (prediction) bands of the regression \cite{Draper81}.}
\end{figure}
The DM provides a useful guideline to
suggest the kind of correlations that one can expect between the neutron
skin thickness and the symmetry energy parameters. As shown in Ref.
\cite{Centelles10}, these correlations are mainly due to the bulk term
of Eq.(\ref{rnpdm}) rather than to the surface contribution to $\Delta
r_{\rm np}$. In the bulk part of $\Delta r_{\rm np}$, the quantity $\left(
C_2^0-a_{\rm sym}(A) \right)/C_2^0$ determines the ratio of the surface symmetry
to volume symmetry energies, see Eq.(\ref{t1}); the close relation
of different isovector observables in finite nuclei with the ratio of
the surface and volume symmetry energies has been observed in several
studies \cite{Satula06,Roca-Maza13a}. The values of $r_0$ for the 
various models considered in the
present thesis work display only a small variation indicating that the total
neutron-skin thickness $\Delta r_{\rm np}$ of a given heavy nucleus may
be correlated to the ratio $\left( C_2^0-a_{\rm sym}(A) \right)/C_2^0$, or also
to the difference $(C_2^0-a_{\rm sym}(A))$ provided the value of $C_2^0$ does not
show a large variation as compared to $\left( C_2^0-a_{\rm sym}(A) \right)$.

In Fig. \ref{ch5_fig1}, the values of $C_2^0 - a_{\rm sym}(A)$ are plotted 
as a function of $\Delta r_{\rm np}$
in the left panel, and as a function of the bulk part of the neutron-skin
thickness $\Delta r_{\rm np}^{\rm bulk}$ in the right panel, for $^{208}$Pb 
and $^{132}$Sn nuclei. The results
are reported for the five different families of systematically varied
models, namely, FSV, TSV, SAMi-J, DDME and KDE0-J as indicated in the
figure. Fairly evident linear correlations are observed between $C_2^0 -
a_{\rm sym}(A)$ and both $\Delta r_{\rm np}$ and $\Delta r_{\rm np}^{\rm
bulk}$. More quantitatively, the Pearson's correlation
coefficients $C(X,Y)$ \cite{Brandt97} are calculated, their values are $C(C_2^0-a_{\rm
sym}(A),\Delta r_{\rm np})$ = 0.972 (0.967) and $C(C_2^0-a_{\rm sym}(A),
\Delta r_{\rm np}^{\rm bulk})$ = 0.988 (0.979) for the $^{208}$Pb
($^{132}$Sn) nuclei, respectively.  Thus, the correlation of $C_2^0-a_{\rm
sym}(A)$ with $\Delta r_{\rm np}^{\rm bulk}$ is a little higher than
with $\Delta r_{\rm np}$ for both $^{208}$Pb and $^{132}$Sn nuclei.

\begin{figure}{}
\centering
\includegraphics[width=0.7\textwidth]{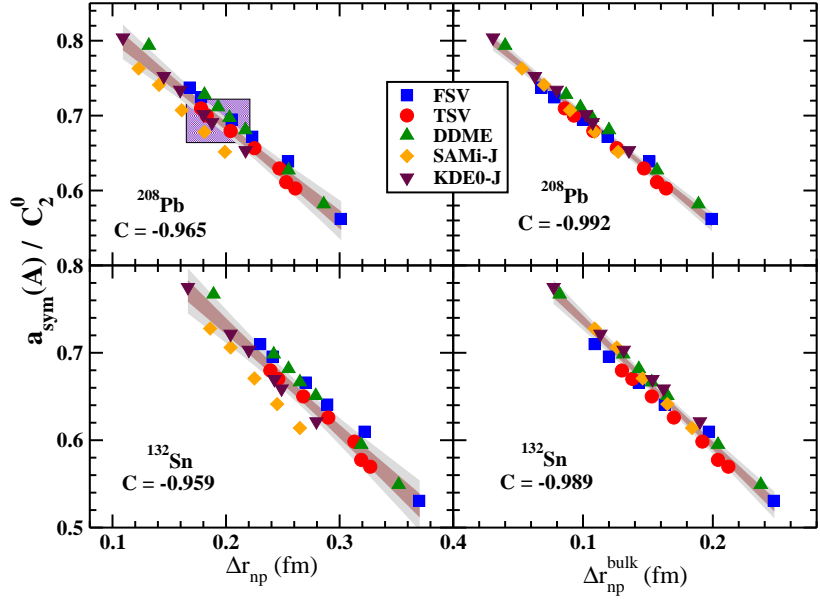}
\caption{\label{ch5_fig2} Plots for the ratio of the  nuclear
symmetry energy coefficient for finite nuclei $a_{\rm sym}(A)$ to that for
infinite nuclear matter $C_2^0$, as a function of the neutron-skin thickness
(left panels) and of the bulk part of the neutron-skin thickness (right
panels).  The square shaded region in the upper-left panel corresponds to
$ a_{\rm sym}(A) = 22.4\pm 0.3$ \cite{Fan14} MeV and $C_2^0= 32.3 \pm 1.3$ MeV
\cite{Carbone10}. The correlation coefficients are $|C(a_{\rm sym}(A)/C_2^0,
\Delta r_{\rm np} )|$ = 0.965 (0.959) and  $|C(a_{\rm sym}(A)/C_2^0, \Delta
r_{\rm np}^{\rm bulk} )|$ = 0.992 (0.989)  for $^{208}$Pb ($^{132}$Sn)
nuclei.  The inner (outer) colored regions depict the loci of the 95\%
confidence (prediction) bands of the regression \cite{Draper81}.}
\end{figure}
Following Eq. (\ref{t1}) one can directly correlate $\left( C_2^0-a_{\rm
sym}(A) \right)/C_2^0$ (or equivalently $a_{\rm sym}(A)/C_2^0$) with $\Delta
r_{\rm np}$ of a heavy nucleus.  In Fig. \ref{ch5_fig2} the
ratio $a_{\rm sym}(A)/C_2^0$ as a function of $\Delta r_{\rm np}$ and of
$\Delta r_{np}^{\rm bulk}$ are displayed for the $^{208}$Pb and $^{132}$Sn nuclei. The
correlations of $a_{\rm sym}(A)/C_2^0$ with $\Delta r_{\rm np}$ are relatively
weaker in comparison to those with $\Delta r_{\rm np}^{\rm bulk}$. In
the case of $a_{\rm sym}(A)/C_2^0$ and $\Delta r_{\rm np}$ the correlation
coefficient is $|C(a_{\rm sym}(A)/C_2^0, \Delta r_{\rm np} )| = 0.965$ (0.959)
for $^{208}$Pb ($^{132}$Sn), whereas in the case of $a_{\rm sym}(A)/C_2^0$
and $\Delta r_{\rm np}^{\rm bulk}$ the correlation coefficient increases
up to high values $|C(a_{\rm sym}(A)/C_2^0, \Delta r_{\rm np}^{\rm bulk} )|
= 0.992$ (0.989) for $^{208}$Pb ($^{132}$Sn).

It is interesting to address the constraints on the
neutron-skin thickness that may be deduced from the present study. The
rectangular shaded region in the upper-left panel of Fig. \ref{ch5_fig2}
corresponds to $ a_{\rm sym}(A) = 22.4\pm 0.3$ MeV for $^{208}$Pb
\cite{Fan14} and $C_2^0= 32.3 \pm 1.3$ MeV \cite{Carbone10}, which yields
$\Delta r_{\rm np}= 0.193 \pm 0.028$ fm in the $^{208}$Pb nucleus. This
value is compatible with the recent constraints on the the neutron
skin thickness of $^{208}$Pb derived from the measured electric dipole
polarizability in $^{68}$Ni, $^{120}$Sn and $^{208}$Pb \cite{Roca-Maza15}.
The constraint $a_{\rm sym}(A)= 22.4\pm 0.3$ MeV was evaluated in
Ref. \cite{Fan14} using the experimental binding energy differences.
Furthermore, the effect of the Coulomb interaction on the surface
asymmetry and the effect of the surface diffuseness on the Coulomb
energy were taken into account. The value of $C_2^0 = 32.3\pm 1.3$ MeV
\cite{Carbone10}, as obtained by analyzing the experimental data on the
pygmy dipole resonance combined with the correlation between $L_0$ and $C_2^0$,
has a quite reasonable overlap with the values of $C_2^0$ that have been
extracted either from a version of the finite-range droplet model (FRDM)
that performs very well in reproducing the experimental mass systematics
\cite{Moller12}, or from specific manipulation of the semi-empirical
mass formula \cite{Jiang12}, or through analysis of the properties of
semi-infinite nuclear matter \cite{Danielewicz09}. This value of $C_2^0$ also
overlaps with the conclusions provided in recent papers \cite{Tsang12,
Lattimer13, Roca-Maza15}.

\begin{figure}
\centering
\includegraphics[width=0.7\textwidth]{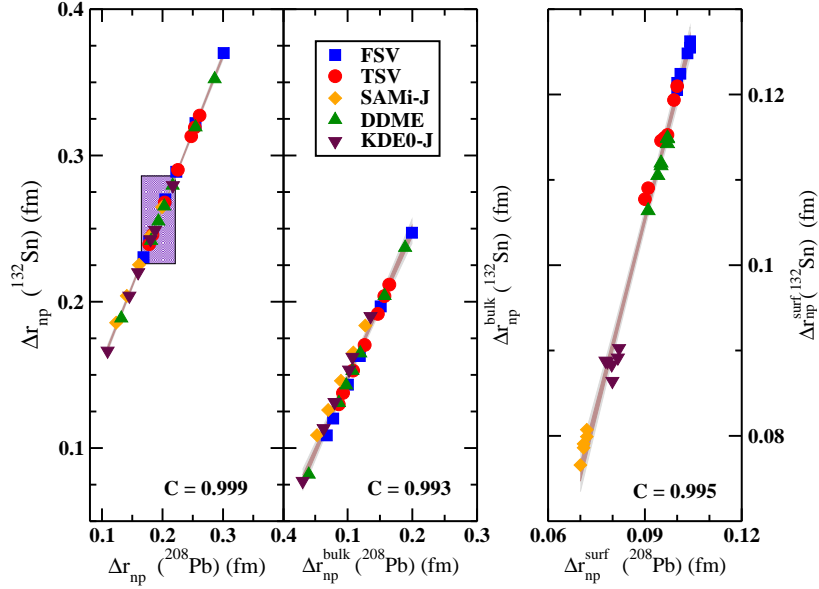}
\caption{\label{ch5_fig3} 
Neutron-skin thickness (left) and its bulk (middle) and surface
(right) contributions for the $^{132}$Sn nucleus plotted against
the same quantities for the $^{208}$Pb nucleus.  The shaded region
corresponds to the values of the neutron-skin thickness in $^{132}$Sn
determined from the ones estimated for the $^{208}$Pb nucleus (see also
Fig. \ref{ch5_fig2}).  The correlation coefficients obtained for the results
presented in the left, middle  and right panels are 0.999, 0.993 and
0.995, respectively.  The inner (outer) colored regions depict the loci
of the 95\% confidence (prediction) bands of the regression \cite{Draper81}.}
\end{figure}
It is desirable to check the degree of consistency between the results
for different heavy nuclei, in particular between $^{208}$Pb and
$^{132}$Sn which would allow to predict the neutron skin thickness of
the nucleus $^{132}$Sn assumed that the one of $^{208}$Pb is known. In
the left panel of Fig. \ref{ch5_fig3}, we plot $\Delta r_{\rm np}$ for the
$^{132}$Sn nucleus against that for the $^{208}$Pb nucleus. Similarly,
the results for $\Delta r_{\rm np}^{\rm bulk}$ and $\Delta r_{\rm np}^{\rm
surf}$ are plotted in the middle and right panels of Fig. \ref{ch5_fig3},
respectively. It is observed that the values of $\Delta r_{\rm np}$,
$\Delta r_{\rm np}^{\rm bulk}$ and $\Delta r_{\rm np}^{\rm surf}$ for
the $^{132}$Sn nucleus are very well correlated with the corresponding
values in the $^{208}$Pb nucleus. This is in harmony with earlier work
\cite{Piekarewicz12}. Hence, the information provided by the neutron skin
of two heavy nuclei on the isovector channel of the nuclear effective
interaction is mutually inclusive. Such an observation allows one to
predict $\Delta r_{\rm np}= 0.256 \pm 0.030\,{\rm fm}$ for $^{132}$Sn
nucleus by using the above estimated value for ${}^{208}$Pb of $\Delta
r_{\rm np}= 0.193 \pm 0.028$ fm.

\begin{figure}
\centering
\includegraphics[width=0.7\textwidth]{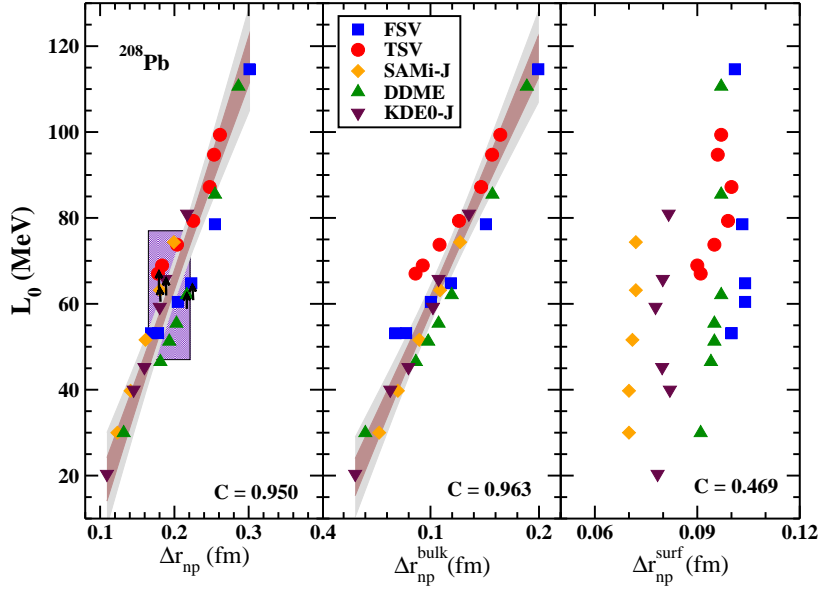}
\caption{\label{ch5_fig4} Plots for the symmetry energy slope
parameter $L_0$ as a function of the neutron-skin thickness (left), its bulk
part (middle) and its surface part (right) for the $^{208}$Pb nucleus.
The shaded region in the left panel projects out the values of $L_0=
62 \pm 15$ MeV obtained from $\Delta r_{\rm np} = 0.193 \pm 0.028$
fm which, in turn, is obtained  by using the empirical values of $C_2^0$
and $a_{\rm sym}(A)$ (see also Fig. \ref{ch5_fig2}).  The arrow marks in the
left panel indicate the points with the slope parameter $L_0 \sim$ 65 MeV.
 The  values of the correlation coefficients are
$C(L_0,\Delta r_{\rm np})$ = 0.950, $C(L_0,\Delta r_{\rm np}^{\rm bulk})$ =
0.963 and $C(L_0,\Delta r_{\rm np}^{\rm surf})$ = 0.469.  The inner (outer)
colored regions depict the loci of the 95\% confidence (prediction)
bands of the regression \cite{Draper81}.}
\end{figure}
As discussed in the literature \cite{Centelles10}, 
the correlation between the neutron-skin thickness and $\left(
C_2^0 - a_{\rm sym}(A) \right) / C_2^0$ leads to a correlation between the
neutron-skin thickness and the symmetry energy slope parameter $L_0$. In
Fig. \ref{ch5_fig4}, the variation of $L_0$ as a function of $\Delta
r_{\rm np}$ (left), $\Delta r_{\rm np}^{\rm bulk}$ (middle) and $\Delta
r_{\rm np}^{\rm surf}$ (right panel) are depicted for the $^{208}$Pb nucleus 
for the four families of models obtained in this present thesis work.  
Using the constraint on $\Delta r_{\rm np}$
($^{208}$Pb) obtained in Fig.  \ref{ch5_fig2}, the bound on the value of $L_0$
comes out to be $L_0=\ 62\pm15$ MeV; displayed as the shaded region of
left panel in Fig. \ref{ch5_fig4}.  The correlation coefficients of $L_0$ with
$\Delta r_{\rm np}$ and with $\Delta r_{\rm np}^{\rm bulk}$ are lower
than in the case of the correlations displayed in Figs. \ref{ch5_fig1} and
\ref{ch5_fig2}, suggesting that the neutron-skin thickness is slightly better
correlated with $C_2^0 - a_{\rm sym}(A)$ or the ratio $a_{\rm sym}(A)/C_2^0$ than
with the slope parameter $L_0$.  This might be a feature of the families 
chosen chosen in the present thesis and does not necessarily apply to the situation in which one
employs a large set of unbiasedly selected models \cite{Centelles10}. 

The ``arrow''marks in Fig. \ref{ch5_fig4} indicate the five models, each
from a different family, with $L_0$ varying in a narrow range of 62.1
MeV to 67.0 MeV. For these five models, there happens to be a spread in
$\Delta r_{\rm np}$ of almost $0.05$ fm which is larger than expected.
In comparison, the equation of the linear fit of the results of all
models in the left panel of Fig. \ref{ch5_fig4} gives a variation in the
value of $\Delta r_{\rm np}$ ($^{208}$Pb) with the change of $L_0$ as,
$\delta (\Delta r_{\rm np})\simeq 0.002\,\delta L_0$, so that a change in
$L_0$ of 5 MeV implies an average change in $\Delta r_{\rm np}$ of about
0.01 fm only, which is smaller than the observed spread of 0.05 fm in
the five models mentioned above. The DM supports a similar conclusion,
as it can be seen from Eq. (\ref{t2}) that the DM predicts an average
variation of $\Delta r_{\rm np}$ ($^{208}$Pb) with $L_0$ approximately as,
$\delta (\Delta r_{\rm np})\simeq 0.003\,\delta L_0$.  The two mentioned
models from the TSV and SAMi-J families have $L_0=67$ MeV and $L_0=63.2$ MeV,
respectively, and yield in $^{208}$Pb smaller values of $\Delta r_{\rm np}
\simeq 0.18$ fm, whereas the two models from the FSV and DDME families
have $L_0=64.8$ MeV and $L_0=62.1$ MeV, respectively, and give rise to larger
values of $\Delta r_{\rm np} \simeq 0.22$ fm. The model from KDE0-J family
with $L_0=65.7$ MeV yields an intermediate value of $\Delta r_{\rm np}$
($^{208}$Pb) $\simeq 0.19$ fm.  Actually, it comes as an intriguing fact
that the extracted values of $\Delta r_{\rm np}$ differ by $\sim0.05$
fm for the two models of the FSV and TSV families with similar $L_0$,
although the parameters for these two families are obtained by using
exactly the same kind of fitting protocol. In the next subsection, 
the plausible interpretations for such differences in the
neutron skin thickness corresponding to models with similar $L_0$ values 
are investigated.

\subsection{Systematic differences between the families of \\functionals}
\begin{figure}[]
\centering
{\includegraphics[width=0.7\textwidth]{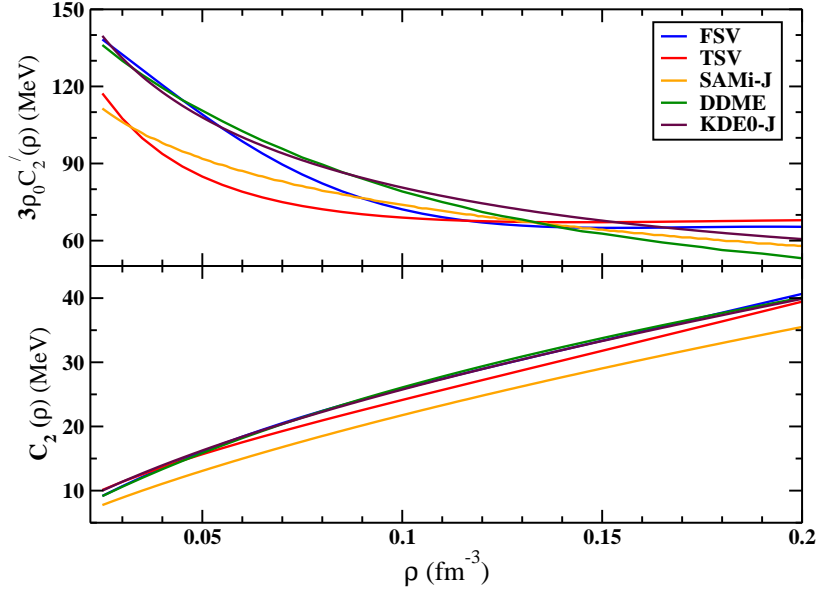}}
\caption{\label{ch5_fig5} 
The nuclear symmetry energy $C_2$ (lower panel) and its density derivative
$C_2'$ multiplied by $3\rho_0$ (upper panel)  as a function of density
for the five different models associated with the slope parameter for
nuclear matter $L_0$ $\sim 65$ MeV.  Each of these models belongs to a
different family (see also Table \ref{ch5_tab1}).}
\end{figure}

In an attempt to understand the issues raised at the end of the previous
subsection, a detailed comparison is made between the results for the
five models belonging to different families but yielding almost the
same values for $L_0$. First a closer look is given in Fig. \ref{ch5_fig5}
into the values of the symmetry energy $C_2(\rho)$ (lower panel) and its
density derivative $3\rho_0C_2^\prime(\rho)$ (upper panel) as a function
of density for these models. The behavior of $C_2(\rho)$ as a function of
density seemingly appears to be similar for the five models. But the
values of $3\rho_0C_2^\prime(\rho)$ show significant differences in the
low density region ($\rho<0.10\ \text{fm}^{-3}$). Furthermore, one may
note that the TSV and SAMi-J models corresponding to $\Delta r_{\rm
np}(^{208}\text{Pb}) \sim 0.18$ fm and the KDE0-J model with $\Delta
r_{\rm np}(^{208}\text{Pb}) \sim 0.19$ fm display a relatively similar
behavior in the density dependence of $C_2^\prime(\rho)$.  The same
is true for the FSV and DDME models corresponding to $\Delta r_{\rm
np}(^{208}\text{Pb}) \sim 0.22$ fm.

To investigate whether such differences in the values of the density
derivative of the symmetry energy at lower densities have an influence
in the finite nuclei calculations, and motivated by Eq. (\ref{eq:Asym}),
an effective value of the slope parameter $L_{\rm eff}$ is defined,
which might be more sensitive to the relative distributions of neutrons
with respect to protons in finite nuclei, as follows:
\begin{equation}
 L_{\rm eff} = \frac{3\rho_0\int \left[r^2\rho(r)I^2(r)\right]C_2^\prime(\rho(r))dr} {\int \left[r^2\rho(r)I^2(r)\right]dr}.
\label{leff}
\end{equation}
Here, $I(r)$ is the local asymmetry parameter defined as, $I(r)
\equiv (\rho_n(r)-\rho_p(r))/\rho(r)$. If one assumes $C_2(\rho)$ to
be linear in density, the $L_{\rm eff}$ parameter coincides with $L_0$
(see Eq. (\ref{s_rho})). However, one can see in Fig. \ref{ch5_fig5}
that $C_2(\rho)$ can depart significantly from linearity at low
densities. Therefore, the $L_{\rm eff}$ parameter as defined in
Eq. (\ref{leff}) tries to take into account this effect. At very low
densities ($\rho < 0.01\ {\rm fm}^{-3}$) $C_2(\rho)$ deviates largely
from linearity. The integrals in the numerator and denominator of
Eq. (\ref{leff}) are thus evaluated by integrating from the center of the
nucleus, where the density $\rho(r)$ is of the order of $\rho_0$, up to
the point where the density of the nucleus falls to 0.01 fm$^{-3}$, which
corresponds to a radial coordinate $r$ of about 9 fm. It is worthwhile
to mention that here the goal was to study the effect of $C_2^\prime(\rho)$
but not the quantity $L(\rho)\ (\equiv 3\rho C_2^\prime(\rho))$ on the
$\Delta r_{\rm np}$ of a heavy nucleus. That is why $\rho_0$ was kept 
outside the integral of the numerator in Eq. (\ref{leff}). The values of
$L_{\rm eff}$ along with various other properties evaluated for the five
models corresponding to $L_0\sim 65$ MeV are compared in Table \ref{ch5_tab1}.
\begin{table}[]
\centering
\caption{\label{ch5_tab1} Comparison of the properties of infinite
nuclear matter (NM) and of the $^{208}$Pb and $^{132}$Sn nuclei for the
five different models that yield a value of $L_0$ around 65 MeV.}
\begin{tabular}{ccccccc}
\hhline{=======} 
&	& SAMi-J	&	TSV	&	FSV	&	DDME & KDE0-J\\		
\hline
NM&$\rho_0$(fm$^{-3}$)&0.157&0.147	&0.149	&0.152& 0.162\\
&$L_0$(MeV)	& 63.2& 	67.0		&	64.8	& 	62.1& 65.7\\
&$C_2^0$(MeV)	&	30.00&	31.29		&33.16	& 	34.00& 35.00\\
$^{208}$Pb&$a_{\rm sym}(A)$(MeV)	& 20.35	&	22.20	&22.28& 	23.15& 24.18\\
&$\Delta r_{\rm np}$(fm)	& 	0.181	&	0.178&		0.223	& 	0.217& 0.188\\
&$\Delta r_{\rm np}^{\rm bulk}$(fm)& 0.109	&	0.086	&	0.119	& 	0.120& 0.108\\
&$L_{\rm eff}$(MeV)	& 81.2& 	82.7		&	95.7	& 	96.5& 90.8\\
$^{132}$Sn&$a_{\rm sym}(A)$(MeV)	& 19.24	&	21.27	&21.25& 	22.13& 23.06\\
&$\Delta r_{\rm np}$(fm)	& 0.245	&		0.239&		0.289	& 	0.279& 0.249\\
&$\Delta r_{\rm np}^{\rm bulk}$(fm)& 0.165	&	0.130	&	0.163	& 	0.165& 0.163\\
&$L_{\rm eff}$(MeV)	& 84.3& 	85.7		&  101.2	& 	98.0& 97.8\\
\hhline{=======} 
\end{tabular}
\end{table}

It can be easily observed in Table \ref{ch5_tab1} that though the values
of $L_0$ for these models vary only by $\sim$\,5 MeV, the values of
$\Delta r_{\rm np}$ of heavy nuclei calculated from the same models
can differ by $\sim$\,0.05 fm, which is larger than the average spread
of the correlation between $\Delta r_{\rm np}$ and $L_0$. Interestingly,
when one looks at the extracted $L_{\rm eff}$ parameter, the models from
SAMi-J and TSV families those predict $\Delta r_{\rm np}(^{208}\text{Pb})
\sim 0.18$ fm give similar $L_{\rm eff}\sim 82$ MeV, and the models from
FSV and DDME families those predict $\Delta r_{\rm np}(^{208}\text{Pb})
\sim 0.22$ fm give similar $L_{\rm eff}\sim 96$ MeV. The model from
the KDE0-J family with $\Delta r_{\rm np}(^{208}\text{Pb}) \sim 0.19$
fm predicts $L_{\rm eff}\sim 91$ MeV.  That is, the models with larger
$L_{\rm eff}$ give larger $\Delta r_{\rm np}$ and vice versa. In fact,
further inspection of Fig. \ref{ch5_fig4} reveals that two members of
the FSV and DDME families with $\Delta r_{\rm np}(^{208}\text{Pb})
\sim 0.18$ fm, same as the SAMi-J and TSV models in Table \ref{ch5_tab1},
predict departing $L_0$ values ($L_0=53.2$ MeV in the FSV model and $L_0=46.5$
MeV in the DDME model). It turns out that these FSV and DDME models
also explore similar values of $L_{\rm eff}$ (83.9 MeV in FSV and 86.6
MeV in DDME) as done by the models from the SAMi-J and TSV families
displayed in Table \ref{ch5_tab1} with $\Delta r_{\rm np} \sim 0.18$ fm.
In principle, one can also define $L_{\rm eff}$ without the $I^2(r)$
terms in Eq. (\ref{leff}). That is why, the calculations
of $L_{\rm eff}$ were repeated by taking $I^2(r)$ to be unity in Eq.  (\ref{leff})
and similar trends were found as explained above. In Table \ref{ch5_tab1},
concerning the properties of uniform matter, it is also noticeable that
the models do not display the same value of the saturation density. For
the non-relativistic functionals belonging to the SAMi-J and KDE0-J
family this value is about 5--10$\%$ larger than the values explored by
the relativistic functionals. This fact has some impact on the extracted
values of $L_{\rm eff}$ for these models (see Eq. (\ref{leff})).

\begin{figure}[]
\centering
\includegraphics[width=0.7\textwidth]{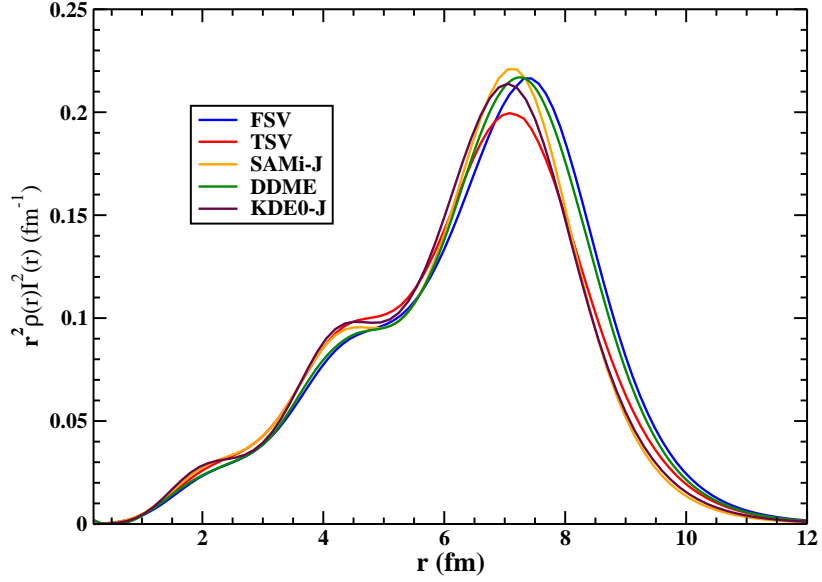}
\caption{\label{ch5_fig6} The variation of $r^2\rho(r)I^2(r)$
as a function of the radial coordinate $r$ in $^{208}$Pb for the five
models that yield a symmetry energy slope parameter $L_0 \sim 65$ MeV.}
\end{figure}
To have a better insight into the source of the differences between
the values of $L_{\rm eff}$ for the models with similar values of $L_0$
at $\rho_0$, in Fig. \ref{ch5_fig6} the total density distribution
$\rho(r)$ of $^{208}$Pb multiplied by $r^2I^2(r)$ is plotted for the models with
$L_0\sim 65$ MeV. The values of $r^2\rho(r)I^2(r)$ for all the different
cases are close to each other up to $r \sim 6$ fm, in this region
$\rho(r) \geqslant 0.1 \text{ fm}^{-3}$. With further increase in $r$,
the differences in the values of $r^2\rho(r)I^2(r)$ gradually become
noticeable.  One can argue that different behaviors in the surface region
may be responsible for different values of $L_{\rm eff}$ and consequently
lead to different values of $\Delta r_{\rm np}$ in heavy nuclei like
$^{208}$Pb or $^{132}$Sn. The question still remains whether $L_{\rm eff}$
is more sensitive to the density dependence of $C_2^\prime(\rho)$ (upper
panel of Fig. \ref{ch5_fig5}) or to the density distributions of nucleons
inside the nucleus (Fig. \ref{ch5_fig6}). To unmask this, the
values of $L_{\rm eff}$ were calculated using $C_2^\prime(\rho)$ of a given model, but with
the density distributions of nucleons from the five models that have
$L_0\sim$ 65 MeV. This calculation was repeated for the different choices
of $C_2^\prime(\rho)$ of these five models. The values of $L_{\rm eff}$
so obtained did not show the trend as observed in Table \ref{ch5_tab1},
where $C_2^\prime(\rho)$ and the density distributions of nucleons used
correspond to the same model consistently.  Thus, the values of $L_{\rm
eff}$ are sensitive to both the density dependence of the symmetry
energy and the density distributions of nucleons inside the nucleus. 
It should be pointed out that the differences in the values
of $L_{\rm eff}$ for the models with similar $L_0$ parameter are mainly
due to the differences in the low density behavior of $C_2^\prime(\rho)$
and the distributions of nucleons in the surface region of the nucleus.

\section{Summary}
To summarize, the correlations of the neutron-skin thickness
in finite nuclei with various symmetry energy parameters pertaining
to infinite nuclear matter were revisited. Particular attention is paid to the model
dependence in such correlations that can play a role in understanding
the density dependence of the nuclear symmetry energy. The finite
nuclei analyzed are $^{208}$Pb and $^{132}$Sn. The symmetry energy
parameters considered are $C_2^0 - a_{\rm sym}(A)$, $a_{\rm sym}(A)/C_2^0$ and
$L_0$, where $C_2^0$ and $L_0$ are the symmetry energy and the symmetry energy
slope associated with infinite nuclear matter at the saturation density,
and $ a_{\rm sym}(A)$ corresponds to the symmetry energy parameter
in finite nuclei. Five different families of systematically varied
mean-field models corresponding to different energy density functionals
are employed to calculate the relevant quantities for the finite nuclei
and those for the infinite nuclear matter.

In general, the correlations of the neutron-skin thickness with the
different symmetry energy parameters are strong within the individual
families of the models. Once the results for all the different families
are combined, the correlation coefficients become smaller, indicating
a model dependence. The correlations of the symmetry energy parameters with the
bulk part $\Delta r_{\rm np}^{\rm bulk}$ of the neutron-skin thickness
are less model dependent than with the total neutron-skin thickness
$\Delta r_{\rm np}$.  Exceptionally, the bulk part of the neutron-skin
thickness is found to be correlated with $C_2^0 - a_{\rm sym}(A)$ and 
$a_{\rm sym}(A)/C_2^0$ in an almost model independent manner.

To understand better the model dependence in the various correlations
considered, the results are compared for the models belonging to different
families, but yielding similar values of $L_0$. An
effective value of the symmetry energy slope parameter $L_{\rm eff}$ was determined 
using the density distributions of nucleons and the density derivative
of the symmetry energy for these models. It is found that the values of
$\Delta r_{\rm np}$, which differ for the models with the same $L_0 \sim$
65 MeV, are in harmony with the values of $L_{\rm eff}$. 
Differences in the values of $L_{\rm eff}$ are caused by differences
in the density distributions of nucleons in the surface region and the
derivative of the symmetry energy at subsaturation densities. 

%% file: chap6.tex
\chapter{Interdependence among the symmetry energy parameters}\label{ch6}
\section{Introduction}
The symmetry energy coefficient $C_2(\rho)$ is now  known in tighter
bounds at the saturation  density $\rho_0$ \cite{Moller12,Jiang12}
of symmetric nuclear matter (SNM).  From analysis of the giant dipole
resonance (GDR) of $^{208}$Pb nucleus, a well-constrained estimate
of $C_2(\rho)$ at a somewhat lesser density ($\rho =0.1$ fm$^{-3}$)
\cite{Trippa08} is also known. The value of the density slope of the
symmetry energy $L_0$ is less certain \cite{Centelles09,Agrawal12,Li15a}.
\begin{figure}[]{}
\centering
\includegraphics[width=0.7\textwidth]{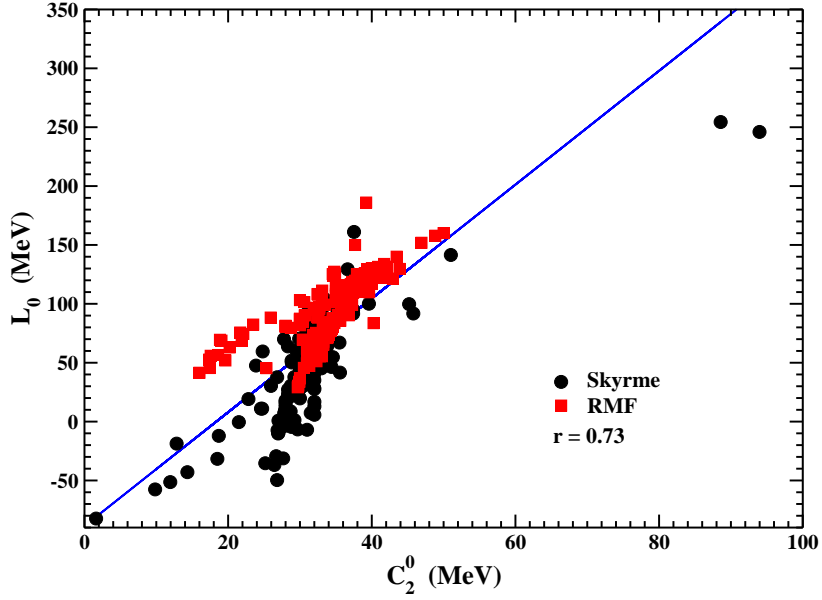}
\caption{\label{ch6_fig1}
Values of $L_0$ plotted against $C_2^0$ as obtained from 500 EDFs based
on both relativistic and non-relativistic mean-field \cite{Dutra12,
Dutra14}. The black circles correspond to the non-relativistic
Skyrme-inspired EDFs and the red squares refer to those obtained from
relativistic mean field (RMF) models. }
\end{figure} 
Tremendous amount of efforts are being made over last decade or so to
constrain the value of $L_0$.  In chapter (\ref{ch4}) a stringent
constraint on the value of $L_0$ is obtained in a relativistic
mean-field (RMF) framework by incorporating binding energies of
highly asymmetric nuclei (number of neutrons is twice to that of
protons i.e. $N\approx2Z$) in the fitting protocol to obtain the
parameters of the model. The currently accepted value of $L_0$ is
lying between 50 and 60 MeV. However, this is not the case for even
higher order derivatives of the symmetry energy \Big[e.g.  $K_{sym}^0$
\Big($=K_{sym}(\rho_0)=9\rho_0^2 \left(\frac{\partial^2C_2}{\partial
\rho^2}\right)_{\rho_0}$\Big) or $Q_{sym}^0$
\Big($=Q_{sym}(\rho_0)=27\rho_0^3\left(\frac{\partial^3C_2}{\partial
\rho^3}\right)_{\rho_0}$\Big)\Big] and on the difference between the
neutron and proton  effective masses $\Delta m^*_0$ [=$(m_n^*-m_p^*)/m$]
in neutron-rich matter at $\rho_0$.  The values of $K_{sym}^0$ and
$Q_{sym}^0$, in different parametrizations of the Skyrme energy
density functional (EDF) lie in very wide ranges [$-700$ MeV $<
K_{sym}^0 < $ 400 MeV; $-800$ MeV $ < Q_{sym}^0 < 1500 $ MeV ]
\cite{Dutra12, Dutra14} whereas there are divergent predictions on the
value of $\Delta m^*_0$ from theoretical studies based on microscopic
many-body theories \cite{Zuo05,Dalen05} or phenomenological approaches
\cite{Ou11,Sellahewa14,Chen09,Kong17}.  Such large uncertainties belie
a satisfactory understanding of the isovector part of the nuclear
interaction.

There is a sliver of expectation that the entities $C_2^0$
$(=C_2(\rho_0))$, $L_0$, $K_{sym}^0$, etc. may have an intrinsic
correlation among them.  Finding a correlated structure for these symmetry
energy elements helps in making a somewhat more precise statement on
an otherwise uncertain isovector indicator as it may be tied up to
other quantities known with more certainty. The correlations need to be
really strong so that one can extract meaningful constraints regarding
the uncertain symmetry energy parameters. Moreover, the correlations
should not depend on the choice of models.  In Fig. (\ref{ch6_fig1})
values of $C_2^0$ are plotted against $L_0$ using 500 mean-field models
from the literature both relativistic and non-relativistic \cite{Dutra12,
Dutra14}.  Only a weak positive correlation was observed with correlation
coefficient $r=0.73$. With this degree of correlation, even with the
precise information on $C_2^0$, one can not infer about the value of $L_0$
with good precision.

\begin{figure}[]{}
\centering
\includegraphics[width=0.7\textwidth]{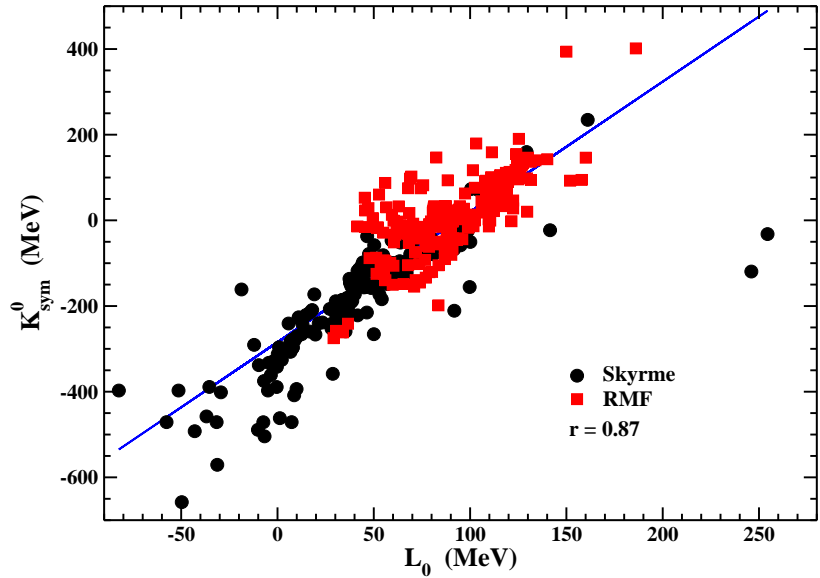}
\caption{\label{ch6_fig4}
The correlation between $L_0$ and $K_{sym}^0$ is plotted for 500
relativistic and non-relativistic EDFs \cite{Dutra12,Dutra14}. The
black circles correspond to the non-relativistic Skyrme-inspired EDFs,
the red squares refer to those obtained from relativistic mean field
(RMF) models.}
\end{figure}
From observation of the computed values of $L_0$ and $K_{sym}^0$ with
selected sets of non-relativistic and relativistic EDFs, an empirical
linear relationship between $K_{sym}^0$ and $L_0$ is also suggested
\cite{Yoshida06,Danielewicz09,Chen09,Vidana09,Ducoin11,Dong12,Santos14}.
For example, in Ref. \cite{Vidana09}, using a selective set of
mean-field models the correlation coefficient was found to be $r=0.87$.
In Ref. \cite{Dong12}, using a different set of mean-field models the
correlation coefficient between the same quantities was found to be
$r=0.97$. Clearly the correlation between $K_{sym}^0$ and $L_0$ has
dependence on the choice of set of models.  To understand the degree of
model dependence in the correlation between $K_{sym}^0$ and $L_0$, in
Fig. (\ref{ch6_fig4}) $K_{sym}^0$ versus $L_0$ is plotted for a diverse
set of 500 relativistic and non-relativistic mean-field models as compiled
by Dutra et al \cite{Dutra12,Dutra14}. The correlation coefficient was
found to be $r=0.87$, which is certainly not as high as it was found in
\cite{Dong12}. To constrain the value $K_{sym}^0$ from the better known
nuclear matter properties, search for a universal correlation is thus
called for.

\section{Theoretical Framework}
Using few basic equations of statistical mechanics, a theoretical
framework to calculate the properties of nuclear matter is given in
the following.
\subsection{Symmetric Nuclear Matter}
For symmetric nuclear matter at density $\rho$, with energy density
${\cal H}$, and at zero temperature ($T=0$), the chemical potential of
the nucleon is given by
\bea
\label{chem1}
\mu\;=\;\mathcal E_F\;=\;\frac{P_F^2}{2m^*}+V\;=\;\frac{P_F^2}{2m}+U,
\eea
where $\mathcal E_F$ is the Fermi energy, $P_F$ is the Fermi momentum,
the effective mass $m^*$ and the single-particle potential $V$ are
given by $\hbar^2/2m^* = \delta {\cal H} / \delta {\mathcal K}$ and $V=
\delta {\cal H} / \delta \rho$, where $\frac{\hbar^2}{2m}\mathcal K$
is the kinetic energy density. One also can redefine the single-particle
potential as $U$ by including within it the effective mass contribution,
as done in the r.h.s. of Eq.~(\ref{chem1}). No special assumption about
the nucleonic interaction is made except that it is density dependent
to simulate many-body forces and that it depends quadratically on the
momentum; thus, the single-particle potential $U$ separates into three
parts,
\bea
\label{sp1}
U=V_0+P_F^2 V_1+V_2.
\eea
The term $(V_0+P_F^2 V_1)$ on the right is the Hartree-Fock potential
and the last term $V_2$ is the rearrangement potential that arises from
the density dependence of the interaction. The term $V_1$ comes from
the momentum dependence:
\bea
\label{mstar5}
\frac{P_F^2}{2m^*}&=&\frac{P_F^2}{2m}+P_F^2 V_1\nonumber\\
\Rightarrow\frac{1}{m^*}&=&\frac{1}{m}+2 V_1
\eea
In general, $m^*$ is momentum and energy dependent, in the mean-field
level the energy dependence is ignored and the momentum dependence
is taken at the Fermi surface.  The rearrangement energy does not
enter explicitly in the energy expression when written in terms of the
mean-field potential \cite{Brueckner59,Bandyopadhyay90}, the energy per
nucleon for SNM at density $\rho$ is then given by,
\bea
\label{eners}
e&=&\frac{1}{2m}\left<p^2\right>+\frac{1}{2}\left<p^2\right>V_1+\frac{1}{2}V_0\nonumber\\
&=&\frac{\left<p^2\right>}{2m^*}\left(\frac{m^*}{m}+m^* V_1\right)+\frac{1}{2}V_0\nonumber\\
&=&\frac{\left<p^2\right>}{2m^*}\left(\frac{m^*}{m}+\frac{1}{2}-\frac{1}{2}\frac{m^*}{m}
\right)+\frac{1}{2}V_0\nonumber\\
&=&\frac{\left<p^2\right>}{2m^*}\frac{1}{2}\left(1+\frac{m^*}{m}\right)+\frac{1}{2}V_0\nonumber\\
&=&\frac{1}{4}\left(\frac{1}{m}+\frac{1}{m^*}\right)\left<p^2\right>+\frac{1}{2}V_0\nonumber\\
&=&\left(\frac{1}{m}+\frac{1}{m^*}\right)\frac{3 P_F^2}{20}+\frac{1}{2}V_0.
\eea
Here the average of the square of the momentum $\langle p^2 \rangle$
is calculated by using the Fermi distribution $\tilde n (p)$ as,
\begin{eqnarray}
\langle p^2 \rangle &=& \frac{\int_0^{P_F}\tilde n (p) p^2 d^3 p}
{\int_0^{P_F}\tilde n (p) d^3 p};\ \ \ \ \text{with}\ \ \ \tilde n (p)
=\frac{1}{e^{\frac{E-E_F}{pT}}+1}\nonumber\\
&=&\frac{4\pi\int_0^{P_F}\tilde n (p) p^2p^2 dp}{4\pi\int_0^{P_F}\tilde n (p)p^2 dp}\nonumber\\
\Rightarrow \langle p^2 \rangle &=&\frac{3 P_F^2}{5}.
\end{eqnarray}
To arrive at the last step, the fact was used that at $T=0$, below Fermi
energy $E_F$ (i.e. $E< E_F$), $\tilde n (p)=1$. At $T=0$, energy ($E$)
of a system is given by the Helmholtz free energy ($F$) i.e.
\bea
\label{helmholtz}
F=E=-PV+\mu N.
\eea
Here, $P,V,N$ are the pressure, volume and number of particles of the
system, respectively. Immediately, energy per particle (nucleon) $e$
can be connected to chemical potential $\mu$ as,
\bea
\label{gibbs}
\frac{E}{N}&=&-P\frac{V}{N}+\mu\nonumber\\
\Rightarrow e&=&-\frac{P}{\rho}+\mu\nonumber\\
\Rightarrow \mu &=&e+\frac{P}{\rho}.
\eea
This relation is known as the Gibbs-Duhem relation. At zero pressure
this leads to the Hugenholtz-Van Hove theorem \cite{Hugenholtz58} which
has recently been used to link nucleon single-particle characteristics
to macroscopic isovector properties in Ref. \cite{Chen12}. Keeping this
in mind, starting from equating Eqs. (\ref{chem1}) and (\ref{gibbs})
and invoking Eqs. (\ref{eners}) and (\ref{sp1}) therein one can write,
\bea
\frac{P_F^2}{2m}+U&=&e+\frac{P}{\rho}\nonumber\\
\Rightarrow \frac{P_F^2}{2m}+V_0+P_F^2 V_1+V_2&=&\left(\frac{1}{m}+\frac{1}{m^*}\right)
\frac{3 P_F^2}{20}+\frac{1}{2}V_0+\frac{P}{\rho}\nonumber\\
\Rightarrow \frac{P_F^2}{2m}+\frac{1}{2}V_0+P_F^2 V_1&=&\left(\frac{1}{m}+\frac{1}{m^*}\right)
\frac{3 P_F^2}{20}+\frac{P}{\rho}-V_2\nonumber\\
\Rightarrow \left(\frac{1}{m}+\frac{1}{m}+2 V_1\right)\frac{3 P_F^2}{20} +\frac{1}{2}V_0&=&\left(\frac{1}{m}+\frac{1}{m^*}\right)
\frac{3 P_F^2}{20}+\frac{P}{\rho}-V_2-\frac{P_F^2}{5m}-\frac{7}{10} P_F^2 V_1\nonumber\\
\Rightarrow \left(\frac{1}{m}+\frac{1}{m^*}\right)\frac{3 P_F^2}{20} +\frac{1}{2}V_0&=&\left(\frac{1}{m}+\frac{1}{m^*}\right)
\frac{3 P_F^2}{20}+\frac{P}{\rho}-V_2-\frac{P_F^2}{5m}\nonumber\\
&&\ \ \ \ \ \ \ \ -\frac{7}{20} P_F^2 \left(\frac{1}{m^*}-\frac{1}{m}\right)
\eea
Recognizing L.H.S of the above equation from Eq. (\ref{eners}), the
energy per nucleon for SNM can be written as \cite{De15},
\bea
\label{ener1}
e&=&\frac{3}{10}\frac{P_F^2}{m}-\frac{1}{5}\frac{P_F^2}{m^*}-V_2+\frac{P}{\rho}\nonumber\\
&=&\frac{P_F^2}{10m}\left(3-2\frac{m}{m^*}\right) -V_2+\frac{P}{\rho }.
\eea
The state dependence of single-particle effective potential can be
taken care in terms of an effective mass of the nucleon $m^*$. In a
non-relativistic prescription for SNM, $\frac{m}{m^*}$ can be expanded
as a function of $\rho$ \cite{Baldo14}. Keeping terms only upto linear
in $\rho$, the expansion is given by \cite{De15}
\bea
\label{mstarsnm}
\frac{m}{m^*(\rho)}=1+k\rho. 
\eea
The density dependence of the rearrangement potential of SNM can be
taken as \cite{De15},
\bea
\label{vreasnm}
V_2(\rho)=a\rho^{\tilde\alpha}. 
\eea
Then, energy per nucleon $e$ of SNM in Eq. (\ref{ener1}) takes the form
\bea
\label{rhoalpha}
e&=&\frac{P_F^2}{10m}\left(3-2\frac{m}{m^*}\right) -a\rho^{\tilde\alpha}+\frac{P}{\rho }\nonumber\\
&=&\frac{g^2\left(\frac{\rho}{2}\right)^{\frac{2}{3}}}{5\cdot2m}
\left(3-2-2k\rho \right) -a\rho^{\tilde\alpha}+\frac{P}{\rho }\nonumber\\
&=&\frac{\lambda}{5}\rho^{\frac{2}{3}}
\left(1-2k\rho \right) -a\rho^{\tilde\alpha}+\frac{P}{\rho }.
\eea
Here, $\lambda$ is given by, $\lambda=\frac{g^2}{2^{2/3}\cdot2m}$.
The pressure for SNM is then given by,
\bea
\label{presssnm}
P\ =\  \rho^2\frac{\partial e}{\partial \rho}\ =\ \frac{\lambda}{15}\rho^{\frac{5}{3}}
-\frac{1}{3}\lambda k \rho^{\frac{8}{3}}-\frac{1}{2}\tilde{\alpha}a\rho^{\tilde\alpha+1}
+\frac{1}{2}\rho\frac{\partial P}{\partial \rho}.
\eea
At $\rho=\rho_0$, the pressure vanishes ($P=0$) and the
incompressibility is given by $K_0=9\left.\frac{\partial P}{\partial
\rho}\right|_{\rho=\rho_0}$. Extracting the value of $a\rho^{\tilde\alpha}$
from Eq. (\ref{rhoalpha}) at $\rho_0$ ($P=0$) the above equation can give the
value of $\tilde\alpha$ as,
\bea
\label{deralp}
0&=&\frac{\lambda}{15}\rho_0^{\frac{5}{3}}
-\frac{1}{3}\lambda k \rho_0^{\frac{8}{3}}-\frac{1}{2}\tilde{\alpha}a\rho_0^{\tilde\alpha+1}
+\rho_0\frac{K_0}{18}\nonumber\\
\Rightarrow 0&=&\frac{\lambda}{15}\rho_0^{\frac{2}{3}}
-\frac{1}{3}\lambda k \rho_0^{\frac{5}{3}}-\frac{1}{2}\tilde{\alpha}\left(a\rho_0^{\tilde\alpha}\right)
+\frac{K_0}{18}\nonumber\\
\Rightarrow 0&=&\frac{\lambda}{15}\rho_0^{\frac{2}{3}}
-\frac{1}{3}\lambda k \rho_0^{\frac{5}{3}}-\frac{1}{2}\tilde{\alpha}\left[\frac{\lambda}{5}\rho_0^{\frac{2}{3}}
\left(1-2k\rho_0 \right)-e_0\right]
+\frac{K_0}{18}\nonumber\\
\Rightarrow K_0&=&18\left\{-\frac{\lambda}{15}\rho_0^{\frac{2}{3}}
+\frac{1}{3}\lambda k \rho_0^{\frac{5}{3}}+\frac{1}{2}\tilde{\alpha}\left[\frac{\lambda}{5}\rho_0^{\frac{2}{3}}
\left(1-2k\rho_0 \right)-e_0\right]\right\}\nonumber\\
\Rightarrow K_0&=&-\frac{6}{5}\lambda\rho_0^{\frac{2}{3}}
+6\lambda \rho_0^{\frac{2}{3}}\left(\frac{m}{m^*_0}-1\right)
+9\tilde{\alpha}\left[\frac{\lambda}{5}\rho_0^{\frac{2}{3}}
\left(3-2\frac{m}{m^*_0} \right)-e_0\right]\nonumber
\eea
\bea
\Rightarrow \tilde\alpha&=&\frac{K_0+\frac{6}{5}\lambda\rho_0^{\frac{2}{3}}
-6\lambda \rho_0^{\frac{2}{3}}\left(\frac{m}{m^*_0}-1\right)}
{9\left[\frac{\lambda}{5}\rho_0^{\frac{2}{3}}
\left(3-2\frac{m}{m^*_0} \right)-e_0\right]}.
\eea
Using the expression for Fermi energy at $\rho_0$ as
$E_F^0=\lambda\rho_0^{\frac{2}{3}}$ the value of $\tilde\alpha$ can be
written as
\bea
\label{alpha}
\tilde\alpha =\frac{\frac{K_0}{9}+\frac{E_F^0}{3}(\frac{12}{5}-2\frac{m}{m_0^*})}
{\frac{E_F^0}{5}(3-2\frac{m}{m_0^*})-e_0}.
\eea

\subsection{Asymmetric Nuclear Matter}
For asymmetric nuclear matter (ANM), the equation for the energy per nucleon
can be generalized  as
\bea
\label{ener4}
e(\rho,\delta)=\frac{1}{\rho}\left[\sum_{\tau}\frac{P_{F,\tau}^2}{10m}\rho_{\tau}\left(
3-2\frac{m}{m_{\tau}^*(\rho)}\right)\right]-V_2(\rho,\delta)
+\frac{P(\rho,\delta)}{\rho}.
\eea
In Eq.(\ref{ener4}), $\tau $ is the isospin index, $\rho_{\tau}=(1+\tau
\delta )\rho /2$; here, $\tau =1$ for neutrons and $\tau =-1$ for
protons.  The Fermi momentum for the individual species can be written as
$P_{F,\tau}=g\rho_{\tau}^{1/3}$ with $g=( 3\pi^2)^{1/3}\hbar$.  
Generalizing the expression in Eq. (\ref{mstarsnm}) density-dependent nucleon
effective mass for asymmetric matter is written as
\bea
\label{mstar6}
\frac{m}{m_{\tau}^*(\rho)}=1+\frac{k_+}{2}\rho+\frac{k_-}{2}\rho \tau \delta.
\eea
The constant $k_+$ for ANM in Eq. (\ref{mstar6}) is equivalent to $k$
in Eq. (\ref{mstarsnm}) with $k=\frac{k_+}{2}$. 
Following the expression of rearrangement potential for SNM in
Eq. (\ref{vreasnm}) the density dependence in the rearrangement potential
for asymmetric nuclear matter can be generalized as (keeping upto
$\sim\delta^2$)
\bea
\label{vrea}
V_2(\rho,\delta)=(a+b\delta^2)\rho^{\tilde\alpha} ,
\eea
which is independent of the isospin index $\tau$.  The constant $a$
weighs the rearrangement potential for SNM, whereas the constant $b$ is
a measure of the asymmetry dependence of the rearrangement potential.

The energy per nucleon $e(\rho,\delta)$ can also be written  in terms
of the symmetry energy coefficients as
\bea
\label{ener5}
e(\rho,\delta)&=&e(\rho,\delta=0)+\frac{1}{2!}\left(\frac{\partial^2 e(\rho,\delta)}
{\partial \delta^2}\right)_{\delta =0}\delta^2+\frac{1}{4!}\left(\frac{\partial^4 e(\rho,\delta)}
{\partial \delta^4}\right)_{\delta =0}\delta^4+\cdots\nonumber\\
&=& e(\rho,0)+C_2(\rho )\delta^2+C_4(\rho )\delta^4+\cdots
\eea
As the nuclear force is invariant under isospin exchange, only the
even powers of $\delta$ survive in the expansion of $e(\rho,\delta)$.
An expression for the  pressure $P(\rho,\delta )= \rho^2\frac{\partial
e}{\partial \rho}$ follows  from the above equation as,
\bea
\label{press}
\frac{P(\rho,\delta )}{\rho}=\rho\frac{\partial e(\rho,0)}{\partial \rho}+\rho\frac{\partial C_2(\rho )}
{\partial \rho}\delta^2+\rho\frac{\partial C_4(\rho )}{\partial \rho}\delta^4+\cdots.
\eea
The right hand side of Eq.(\ref{ener4}) can be expanded in powers of
$\delta $ using the expressions for $P(\rho,\delta )$ and $V_2(\rho,\delta
)$ and using Eq.(\ref{mstar6}), keeping only upto order of $\delta^2$
in $\frac{P(\rho,\delta)}{\rho}$ as,
\bea
\label{expand}
e(\rho,\delta)&=&\frac{1}{\rho}\left[\sum_{\tau}\frac{P_{F,\tau}^2}{10m}(1+\tau\delta)\frac{\rho}{2}
\left(3-2\frac{m}{m_{\tau}^*(\rho)}\right)\right]-V_2(\rho,\delta)
+\rho\frac{\partial e(\rho,0)}{\partial \rho}+\rho\frac{\partial C_2(\rho )}
{\partial \rho}\delta^2\nonumber
\eea
\bea
&=&\left[\sum_{\tau}\frac{\{(3\pi^2)^{\frac{1}{3}}\hbar\rho_{\tau}^{\frac{1}{3}}\}^2}{10m}
\frac{(1+\tau\delta)}{2}
\Big(3-2-k_+\rho-k_-\rho\tau\delta\Big)\right]-V_2(\rho,\delta)\nonumber\\
&&+\rho\frac{\partial e(\rho,0)}{\partial \rho}+\rho\frac{\partial C_2(\rho )}
{\partial \rho}\delta^2\nonumber\\
&=&\frac{g^2}{10m}\frac{1}{2^{2/3}\cdot 2}\left[\sum_{\tau}
\left\{\left(1+\tau\delta\right)\rho\right\}^{\frac{2}{3}}\left(1+\tau\delta\right)
\Big(1-k_+\rho-k_-\rho\tau\delta\Big)\right]-\left(a+b\delta^2\right)\rho^{\tilde\alpha}\nonumber\\
&&+\rho\frac{\partial e(\rho,0)}{\partial \rho}+\rho\frac{\partial C_2(\rho )}
{\partial \rho}\delta^2\nonumber\\
&=&\frac{g^2}{10m}\frac{\rho^{\frac{2}{3}}}{2^{2/3}\cdot 2}\left[\sum_{\tau=1,-1}
\left(1+\tau\delta\right)^{\frac{5}{3}}
\Big(1-k_+\rho-k_-\rho\tau\delta\Big)\right]-\left(a+b\delta^2\right)\rho^{\tilde\alpha}\nonumber\\
&&+\rho\frac{\partial e(\rho,0)}{\partial \rho}+\rho\frac{\partial C_2(\rho )}
{\partial \rho}\delta^2\nonumber\\
&=&\frac{g^2}{10m}\frac{\rho^{\frac{2}{3}}}{2^{2/3}\cdot 2}\left\{
\left[\left(1+\delta\right)^{\frac{5}{3}}\Big(1-k_+\rho-k_-\rho\delta\Big)\right]
+\left[\left(1-\delta\right)^{\frac{5}{3}}\Big(1-k_+\rho+k_-\rho\delta\Big)\right]
\right\}\nonumber\\&&-\left(a+b\delta^2\right)\rho^{\tilde\alpha}
+\rho\frac{\partial e(\rho,0)}{\partial \rho}+\rho\frac{\partial C_2(\rho )}
{\partial \rho}\delta^2\nonumber\\
\Rightarrow e(\rho,\delta)&=&\frac{g^2}{10m}\frac{\rho^{\frac{2}{3}}}{2^{2/3}\cdot 2}\left\{
\left[\left(1+\frac{5}{3}\delta+\frac{5}{9}\delta^2\right)\Big(1-k_+\rho-k_-\rho\delta\Big)\right]\right.\nonumber\\
&&\left.\ \ \ \ \ \ \ \ \ \ \ \ \ \ \ \ \ \ \ \ \ \ \ \  
+\left[\left(1-\frac{5}{3}\delta+\frac{5}{9}\delta^2\right)\Big(1-k_+\rho+k_-\rho\delta\Big)\right]
\right\}\nonumber\\
&&-\left(a+b\delta^2\right)\rho^{\tilde\alpha}
+\rho\frac{\partial e(\rho,0)}{\partial \rho}+\rho\frac{\partial C_2(\rho )}
{\partial \rho}\delta^2
\eea

\subsection{Symmetry energy parameters}
Comparing then  with Eq.(\ref{ener5}) and equating coefficients of
the same order in $\delta$, one gets the expression for $C_2(\rho)$
by putting $y=\frac{g^2}{10m}\frac{1}{2^{2/3}}$ as,
\bea
\label{c2r1}
C_2(\rho)&=&-b\rho^{\tilde\alpha}+\rho\frac{\partial C_2(\rho)}{\partial \rho}+
\frac{g^2}{10m}\frac{\rho^{\frac{2}{3}}}{2^{2/3}\cdot 2}\left[
-\frac{10}{3}k_-\rho+\frac{10}{9}(1-k_+\rho)\right]\nonumber
\eea
\bea
\Rightarrow C_2(\rho)&=&-b\rho^{\tilde\alpha}+\rho\frac{\partial C_2(\rho)}{\partial \rho}+y\rho^{\frac{2}{3}}\left[
-\frac{5}{3}k_-\rho+\frac{5}{9}(1-k_+\rho)\right].
\eea
The relation between $C_2(\rho)$ and its density derivative is
a direct consequence of the Gibbs-Duhem relation.  Using $k_+
\rho_0=2(\frac{m}{m^*_0}-1)$ [$\rho_0$ correspond to $\delta=0$; see
Eq. (\ref{mstar6})] at saturation, the symmetry energy coefficient $C_2$
reads as,
\bea
\label{c2r2}
C_2^0&=&-b\rho_0^{\tilde\alpha}+\frac{L_0}{3}+E_F^0\left[-\frac{1}{3}k_-\rho_0
+\frac{1}{9}(1-k_+\rho_0)\right]\nonumber\\
&=&-b\rho_0^{\tilde\alpha}+\frac{L_0}{3}+E_F^0\left[-\frac{1}{3}k_-\rho_0
+\frac{1}{9}\left(3-2\frac{m}{m^*_0}\right)\right],
\eea
where $E_F^0=5y\rho_0^{2/3}$ is the Fermi energy at $\rho_0$.  Similar
equations can be obtained for higher-order symmetry energy coefficients
$C_4$, $C_6$, etc. which is not dealt here.  The expressions for $C_2$
or the higher-order symmetry energy coefficients so obtained are exact
within the precincts of chosen premises.  The second density
derivative of $C_2$ at $\rho_0$ can be calculated from Eq. (\ref{c2r1})
as,
\bea
\left(\frac{\partial^2 C_2}{\partial \rho^2}\right)_{\rho_0}=\tilde\alpha b \rho_0^{\tilde\alpha -2}
+y \rho_0^{-4/3}\left[\frac{25}{27}k_+\rho_0+\frac{25}{9}k_-\rho_0-\frac{10}{27}\right].
\eea
With the help of Eq. (\ref{c2r2}) expressions for $K_{sym}^0$ reads,
\bea
\label{aksym1}
K_{sym}^0&=&9\rho_0^2 \left(\frac{\partial^2C_2}{\partial \rho^2}\right)_{\rho_0}
=9 \tilde\alpha b \rho_0^{\alpha}+\frac{9}{5}E_F^0\left[\frac{25}{27}\cdot 2\left(\frac{m}{m^*_0}-1
\right)+\frac{25}{9}k_-\rho_0-\frac{10}{27}\right]\nonumber\\
&=&9\tilde\alpha \left[\frac{L_0}{3}-C_2^0+E_F^0\left\{-\frac{1}{3}k_-\rho_0
+\frac{1}{9}(3-2\frac{m}{m^*_0})\right\}\right]+E_F^0\left[\frac{10}{3}\left(\frac{m}{m^*_0}-1
\right)+5 k_-\rho_0-\frac{2}{3}\right]\nonumber\\
&=&3\tilde\alpha\left[L_0-3C_2\left(\rho_0\right)\right]+\frac{2}{3}E_F^0\frac{m}{m^*_0}
\left(5-3\tilde\alpha\right)+E_F^0(3\tilde\alpha-4)+E_F^0(k_-\rho_0)(5-3\tilde\alpha)\nonumber\\
&=&-3\tilde\alpha[3C_2^0-L_0]+E_F^0 \Big [(3\tilde\alpha -4)
+\left(\frac{2}{3}\frac{m}{m_0^*}+k_-\rho_0\right)(5-3\tilde\alpha)\Big ].
\eea
Using the penultimate step of the Eq. (\ref{aksym1}), expression for
$k_-\rho_0$ can be written as,
\bea
\label{kminus}
k_-\rho_0=\frac{K_{sym}^0-3\tilde\alpha\left[L_0-3C_2\left(\rho_0\right)\right]-\frac{2}{3}E_F^0\frac{m}{m^*_0}
\left(5-3\tilde\alpha\right)-E_F^0(3\tilde\alpha-4)}{E_F^0(5-3\tilde\alpha)}.
\eea
The third density derivative of $C_2$ at $\rho_0$ can be calculated from
Eq. (\ref{c2r1}) as,
\bea
\left(\frac{\partial^3 C_2}{\partial \rho^3}\right)_{\rho_0}=\tilde\alpha(\tilde\alpha-2)b\rho_0^{\tilde\alpha-3}
-y\rho_0^{-7/3}\left[\frac{25}{81}k_+\rho_0+\frac{25}{27}k_-\rho_0-\frac{40}{81}\right].
\eea
Eventually utilizing the value of $k_-\rho_0$ from Eq. (\ref{kminus}), 
the symmetry element $Q_{sym}^0$ is given by,
\bea
\label{aqsym1}
Q_{sym}^0&=&27\rho_0^3\left(\frac{\partial^3 C_2}{\partial \rho^3}\right)_{\rho_0}
=27\tilde\alpha(\tilde\alpha-2)b\rho_0^{\tilde\alpha}-27y\rho_0^{2/3}\left[\frac{25}{81}k_+\rho_0+
\frac{25}{27}k_-\rho_0-\frac{40}{81}\right]\nonumber\\
&=&9\tilde\alpha(\tilde\alpha-2)[L_0-3C_2(\rho_0)]-E_F^0\frac{m}{m^*_0}
\left\{6\tilde\alpha(\tilde\alpha-2)+\frac{10}{3}\right\}+
E_F^0[9\tilde\alpha(\tilde\alpha-2)+6]\nonumber\\
&&-E_F^0(k_-\rho_0)[9\tilde\alpha(\tilde\alpha-2)+5]\nonumber\\
&=&9\tilde\alpha(\tilde\alpha-2)[L_0-3C_2(\rho_0)]-E_F^0\frac{m}{m^*_0}
\left\{6\tilde\alpha(\tilde\alpha-2)+\frac{10}{3}\right\}+
E_F^0[9\tilde\alpha(\tilde\alpha-2)+6]\nonumber\\&&
-\frac{E_F^0[9\tilde\alpha(\tilde\alpha-2)+5]\left[K_{sym}^0-
3\tilde\alpha\left[L_0-3C_2\left(\rho_0\right)\right]
-\frac{2}{3}E_F^0\frac{m}{m^*_0}\left(5-3\tilde\alpha\right)-
E_F^0(3\tilde\alpha-4)\right]}{E_F^0(5-3\tilde\alpha)}\nonumber\\
&=&15 \tilde\alpha [3C_2^0-L_0]+K_{sym}^0(3\tilde\alpha -1)
+E_F^0(2-3\tilde\alpha ). 
\eea
While exploring the standard Skyrme EDFs, exactly the same correlated
structure was found between $K_{sym}^0$ or $Q_{sym}^0$ and $[3C_2^0-L_0]$
as in Eqs. (\ref{aksym1}) and (\ref{aqsym1}).

\section{Results and discussion}
\begin{figure}[]{}
\centering
\includegraphics[width=0.7\textwidth]{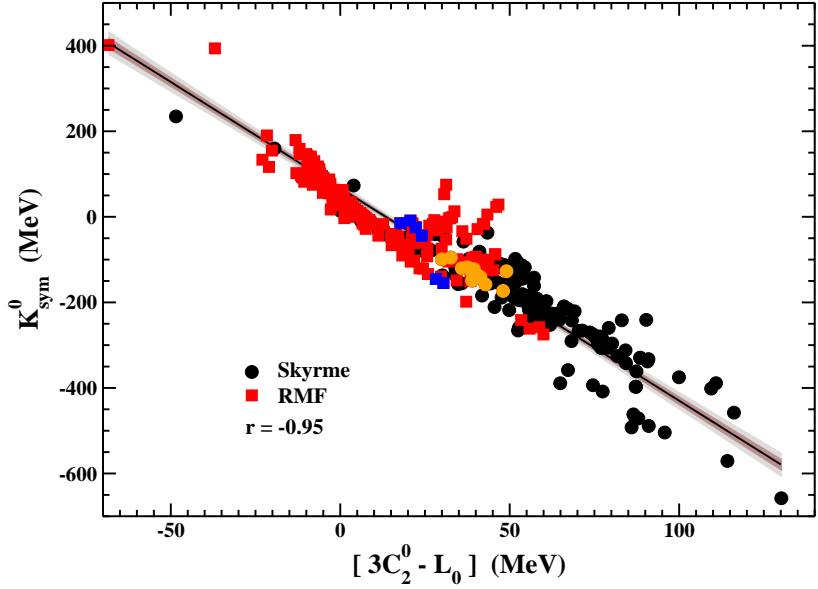}
\caption{\label{ch6_fig5}
The correlation between $K_{sym}^0$ and $[3C_2^0-L_0] $ as obtained from
500 EDFs \cite{Dutra12, Dutra14}. The black circles correspond to
the Skyrme-inspired EDFs, the red squares refer to those obtained from RMF
models. The models consistent with all the constraints demanded by Dutra
et al. are highlighted by orange circles for Skyrme EDFs \cite{Dutra12}
and blue squares for RMF EDFs \cite{Dutra14}.  The inner (outer) colored
regions around the best-fit straight line  through these points depict
the loci of 95$\% $ confidence (prediction) bands of the regression
analysis. }
\end{figure}

Eq. (\ref{aksym1}) throws a  hint that there is a strong likelihood that
$K_{sym}^0$ calculated with different EDFs may be linearly correlated
to $[3C_2^0-L_0]$. This is realized from
the correlated structure of $K_{sym}^0$ with $[3C_2^0-L_0]$ as
displayed in Fig.\ref{ch6_fig5} for five hundred energy density
functionals \cite{Dutra12, Dutra14} that have been in use to explain
nuclear properties.  
The results as presented in Fig. \ref{ch6_fig5}
span both the Skyrme-inspired nonrelativistic (black circles) EDFs which
tend to have negative values for $K_{sym}^0$ and also the relativistic
mean-field EDFs (red squares) that tend to have larger, sometimes positive
values for $K_{sym}^0$.  Skyrme (orange circles) and RMF (blue squares)
models chosen by Dutra et. al. \cite{Dutra12, Dutra14} which were found
to satisfy specific constraints on nuclear matter and neutron star
properties are highlighted. The linear correlation as observed seems
to be nearly universal and intrinsic to an EDF consistent with nuclear
properties. The correlation coefficient is seen to be $r=-0.95$.
\begin{figure}[]{}
\centering
\includegraphics[width=0.7\textwidth]{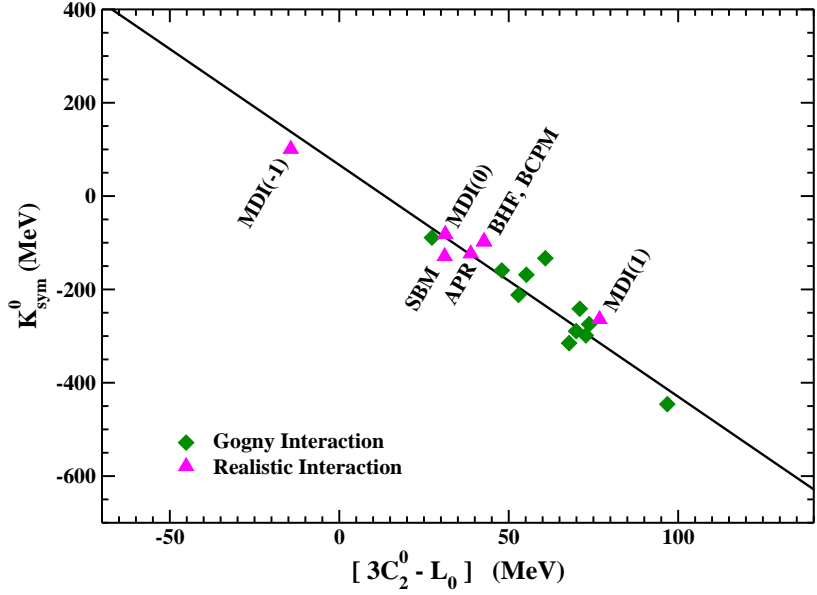}
\caption{\label{ch6_fig6}
The correlation line between $K_{sym}^0$ and $[3C_2^0-L_0]$ obtained
from the Skyrme-RMF models in Fig.  (\ref{ch6_fig5}) is depicted.
The magenta triangles are the results obtained from EDFs with realistic
interactions, MDI(0), MDI(1), MDI(-1) \cite{Chen09}, APR \cite{Akmal98},
BHF \cite{Taranto13}, BCPM \cite{Baldo13} and SBM \cite{Agrawal17},
respectively. The green diamonds represent results from a few Gogny
interactions \cite{Sellahewa14}.}
\end{figure}
The near-universality in the correlation is brought into sharper focus in
Fig. (\ref{ch6_fig6}), where results corresponding to EDFs obtained from
several realistic interactions (magenta triangles) and a few finite-range
Gogny interactions (green diamonds) are displayed. They lie nearly on
the correlation line. The linear regression analysis yields
\bea
\label{corr1}
K_{sym}^0=d_1[3C_2^0-L_0] +d_2 ,
\eea
with $d_1 =-4.97 \pm 0.07$ and $d_2 =66.80 \pm 2.14$ MeV. This is a
robust correlation among the symmetry energy elements. Incidentally,
from the density-dependent M3Y (DDM3Y) interaction, a similar
kind of relation between these symmetry elements can be observed
\cite{Dong12}.  The correlation between the $K_{sym}^0$ and
$L_0$ values from different effective forces and realistic
interactions has also been considered in previous literature
\cite{Yoshida06,Danielewicz09,Chen09,Vidana09,Ducoin11,Dong12,Santos14}.
The results have shown relatively varying degrees of correlation
(c.f. Figs (\ref{ch6_fig2}) and (\ref{ch6_fig3})). In particular, the
correlation between $K_{sym}^0$ and $L_0$ from all the 500 EDFs (see
Fig. (\ref{ch6_fig4})) is not as strong as the correlated structure of
$K_{sym}^0$ with $[3C_2^0-L_0]$.

\begin{figure}[]{}
\centering
\includegraphics[width=0.7\textwidth]{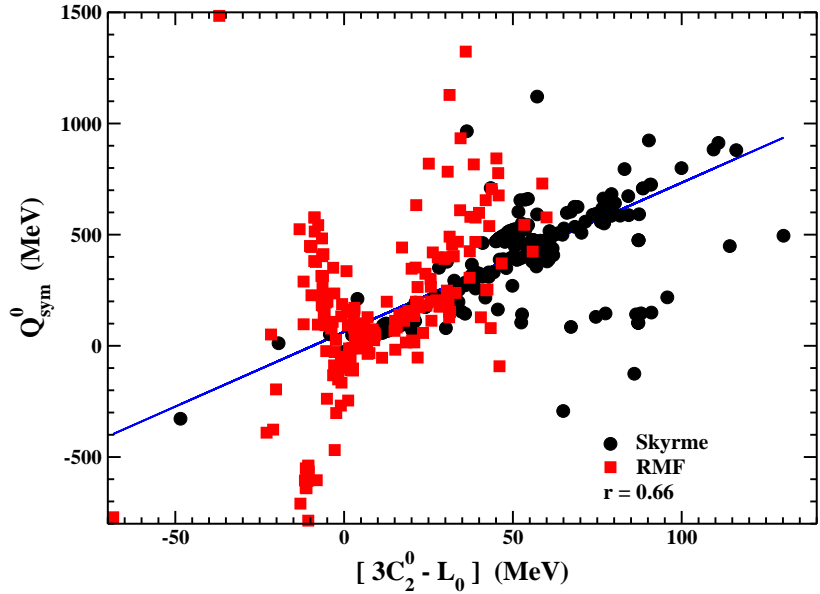}
\caption{\label{ch6_fig7}
The correlation between $Q_{sym}^0$ and $[3C_2^0-L_0] $ as obtained from
500 EDFs \cite{Dutra12, Dutra14}. The black circles correspond to
the Skyrme-inspired EDFs, the red squares refer to those obtained from RMF
models.}
\end{figure}
Incidentally, a very similar correlated structure as $K_{sym}^0$ is
also anticipated for $Q_{sym}^0$, suggested by Eq. (\ref{aqsym1}). In
Fig. (\ref{ch6_fig7}), values of $Q_{sym}^0$ are plotted
as a function of $[3C_2^0-L_0]$ for the same 500 models as in
Fig. (\ref{ch6_fig5}). However, the correlation between $Q_{sym}^0$ with
$[3C_2^0-L_0]$ is not as good as that for $K_{sym}^0$. The
correlation coefficient is merely 0.66. One of the possible reason
behind this is propagation of errors from $K_{sym}^0$ in the right
hand side of Eq. (\ref{aqsym1}). Moreover, all three terms in the
RHS of Eq. (\ref{aqsym1}) have similar contributions to the value of
$Q_{sym}^0$ in terms of magnitude. Isoscalar properties like $K_0$ or
$\frac{m^*_0}{m}$ may still possess some variation across the plethora
of mean-field models compiled in Refs. \cite{Dutra12,Dutra14}. This can
cause reasonable variation in the value of $\tilde\alpha$ or $E_F^0$
across different models, which might be screening the correlation between
$Q_{sym}^0$ and $[3C_2^0-L_0]$.
\begin{figure}[]{}
\centering
\includegraphics[width=0.7\textwidth]{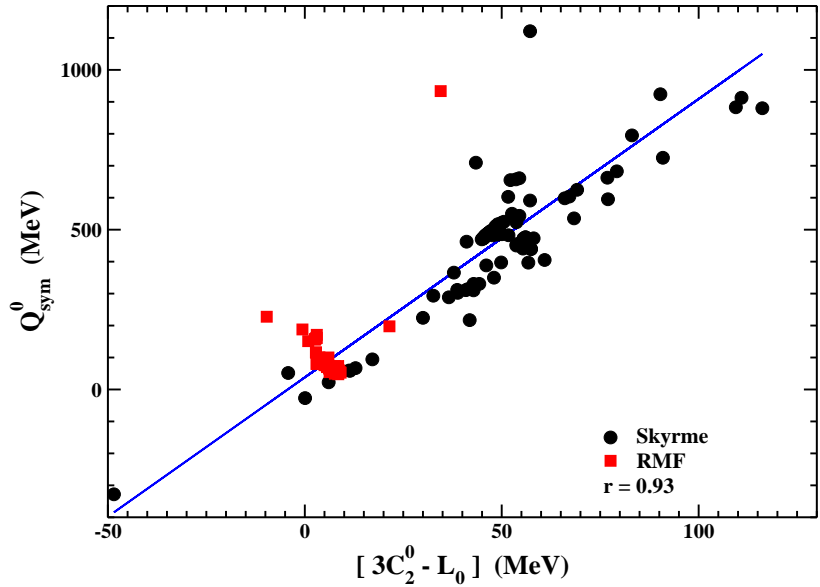}
\caption{\label{ch6_fig8}
The correlation between $Q_{sym}^0$ and $[3C_2^0-L_0] $ as obtained from
$\sim$ 200 EDFs chosen from \cite{Dutra12, Dutra14} with constraints
on $K_0=230\pm30$ and $\frac{m^*_0}{m}=0.75\pm0.1$. The black
circles correspond to the Skyrme-inspired EDFs, the red squares refer
to those obtained from RMF models.}
\end{figure}
To limit the variation in $\tilde\alpha$ and $E_F^0$, we restrict the
values of $K_0=230\pm30$ MeV and $\frac{m^*_0}{m}=0.75\pm0.1$ \cite
{De15,Mondal17b}. This set of constraints are followed by $\sim110$
Skyrme models and $\sim80$ RMF models given in Refs. \cite {Dutra12,
Dutra14}. In Fig. \ref{ch6_fig8}, $Q_{sym}^0$ is plotted as a function
of $[3C_2^0-L_0]$ for these $\sim200$ models. The correlation improved
drastically for these constrained set of models with correlation
coefficient $r=0.93$. This again points out to the universal nature of the
correlated structures, which were proposed by the analytical relations.

From accumulated experimental data over several decades and their
theoretical analyses, there seems to be a broad consensus about the values
of some of the nuclear constants. The saturation density $\rho_0$ of SNM,
its energy per nucleon $e_0$ and its incompressibility coefficient $K_0$
are taken as a subset of the constants characterizing symmetric nuclear
matter.  The nucleon effective mass $m_0^*$ for SNM at $\rho_0$ is also
taken as an input datum though its value is not as certain as $e_0$
or $\rho_0$. Two more nuclear constants related to asymmetric nuclear
matter (ANM) are further considered. They are the nuclear symmetry energy
coefficients $C_2(\rho)$ at $\rho_0$ and at a somewhat lesser density
$\rho_1$ (= 0.1 fm$^{-3}$), ``the crossing density''.  There is less room
for uncertainty in the symmetry energy coefficient $C_2^0$ which has been
determined from exploration of nuclear masses \cite{Moller12, Jiang12}.
With the realization that the nuclear observables related to average
properties of nuclei constrain the nuclear EDFs better at around the
average density of terrestrial atomic nuclei \cite{Khan12}, the so-called
``crossing density'' \cite{Brown13} assumes a special significance. The
symmetry energy $C_2^1$ ($= C_2(\rho_1)$) at that density, in Skyrme EDFs
is seen to be strongly correlated to the Giant Dipole Resonance (GDR) in
spherical nuclei and is now fairly well constrained \cite{Trippa08}. From
the apparently universal, EDF-independent correlation between the
isovector observables, the isovector elements $L_0$, $K_{sym}^0$, etc. can
now be threaded to the above-mentioned nuclear constants as shown below.

With $m_0^*$ as input, $k_+$ is known. From given values of $e_0,\ \rho_0$
and $K_0$  for SNM, $\tilde\alpha$ can be  calculated as Eq. (\ref{alpha})
\cite{De15}.  The symmetry energy $C_2(\rho_1 )$ can be expressed as
\bea
\label{c2r3}
C_2(\rho_1 )= C_2^0-L_0\epsilon+\frac{1}{2}K_{sym}^0 \epsilon^2
-\frac{1}{6}Q_{sym}^0 \epsilon^3+\cdots,
\eea
where $\epsilon = \frac{(\rho_0 -\rho_1 )}{3\rho_0}$.  From
Eqs.~(\ref{aqsym1}), (\ref{corr1}) and (\ref{c2r3}), ignoring terms beyond
$\epsilon^3$,  which are negligible, $L_0$, $K_{sym}^0$ and $Q_{sym}^0$
are calculated with known values of $C_2^0$ and $C_2^1$.  The constant
$k_-$ then follows from Eq. (\ref{aksym1}). From Eq. (\ref{mstar6}),
the nucleon effective mass splitting at saturation density to leading
order in $\delta$ is given as
\bea
\label{mstar8}
\Delta m^*_0=\left(\frac{m_n^*-m_p^*}{m}\right)_{\rho_0}
\simeq -k_-\rho_0\left(\frac{m_0^*}{m}\right)^2\delta,
\eea
where  the approximation $ (m_n^*\cdot m_p^*)\simeq (m_0^*)^2$ is made.

Comparing Eqs. (\ref{aksym1}) and (\ref{corr1}) one would expect $|d_1|$
to be close to $3\tilde{\alpha}$. With the input values of the isoscalar
nuclear constants $e_0$, $\rho_0$ and $K_0$, $3\tilde{\alpha}$ is seen to
be 3.54 as opposed to $\sim 5$ for $|d_1|$. The reason for this change
seems to be two-fold, (a) all 500 EDFs employed in Fig. \ref{ch6_fig5}
have different values for $\tilde{\alpha}$, and (b) the RMF models
are also included in the fit which have no explicit counterpart of
$\tilde{\alpha}$.

In summary, the values of $L_0$, $K_{sym}^0$, $Q_{sym}^0$ and
$\Delta m_0^*$ can be calculated in terms of empirically known
nuclear constants namely, $\rho_0$, $e_0$, $K_0$, $C_2^0$, $C_2^1$
and $\frac{m_0^*}{m}$ using Eqs.~(\ref{aqsym1})--(\ref{mstar8}).
From the diverse theoretical endeavours like the liquid drop type models
\cite{Moller12, Myers98,Myers96}, the microscopic ab-initio or variational
calculations \cite{Akmal97, Baldo13}, or different Skyrme or RMF models --
all initiated to explain varied experimental data, a representative set
of the input nuclear constants for SNM is chosen with $\rho_0=0.155 \pm
0.008$ fm$^{-3}$ and $e_0=-16.0\pm0.2$ MeV. From microscopic analysis
of isoscalar giant monopole resonances (ISGMR), the value of $K_0$
is constrained as $230\pm40$ MeV \cite{Khan12}. Analyzing the compact
correlation between the 'experimental' double-differences of symmetry
energies of finite nuclei and their mass number, Jiang et. al. \cite{
Jiang12} find $C_2^0=32.1\pm0.3$ MeV. This value is included in the chosen
set of nuclear constants. For $C_2^1$, the value $C_2^1=24.1\pm0.8$
MeV as quoted from microscopic analysis of GDR in $^{208}$Pb \cite{
Trippa08} is taken. There is an overall consistency of this $C_2^1$
value with those from the best-fit Skyrme EDFs \cite{Brown13} and with
that given in \cite{Dong12}.  For the nucleon effective mass, a value
of $\frac{m_0^*}{m} =0.70 \pm 0.05 $ is taken, this is consistent with
the empirical values obtained from many analyses \cite{Jaminon89,Li15b}.

The values of the symmetry energy elements calculated from
Eqs.~(\ref{aqsym1})--(\ref{mstar8}) using the values of input
nuclear constants as mentioned come out to be $L_0=60.3\pm14.5$
MeV, $K_{sym}^0=-111.8\pm71.3$ MeV, $Q_{sym}^0 =296.8\pm73.6$
MeV and $\Delta m^*_0=(0.17\pm0.24) \delta $. The value of $L_0$ is
remarkably close to its global average 58.9$\pm $ 16 MeV \cite{Li13},
obtained from analyses of terrestrial experiments and astrophysical
observations. The value of $L$ at $\rho_1$ is calculated to be
$49.3\pm4.2$ MeV. From dipole polarizability in $^{208}$Pb an empirical
value of $L=47.3\pm7.8$ MeV was obtained at $\rho\simeq0.11$ fm$^{-3}$
\cite{Zhang14a}.  There is no experimental value for $K_{sym}^0$
or $Q_{sym}^0$ to compare. However, the symmetry incompressibility
$K_\delta$ defined at the saturation density of nuclear matter at
asymmetry $\delta \Big($ $K_\delta =K_{sym}^0 -6L_0 -\frac {Q_0
L_0}{K_0}$, where $Q_0=27\rho_0^3\left(\frac{\partial^3e}{\partial
\rho^3}\right)_{\rho_0}$\Big) has been extracted from breathing
mode energies of Sn-isotopes \cite{Li07}. Corrected for the
nuclear surface term, $K_\delta$ is quoted to be $\simeq -350$ MeV
\cite{Pearson10}. This is in close agreement with the calculated
value $K_\delta =-378.6\pm17.0$ MeV; $Q_0$ has been calculated from
Eq.~(\ref{ener4}) to be $-364.7\pm27.7$ MeV corresponding to $\delta =0
$ \cite{De15} with the input nuclear constants mentioned.  
\begin{figure}[]{}
\centering
\includegraphics[width=0.7\textwidth]{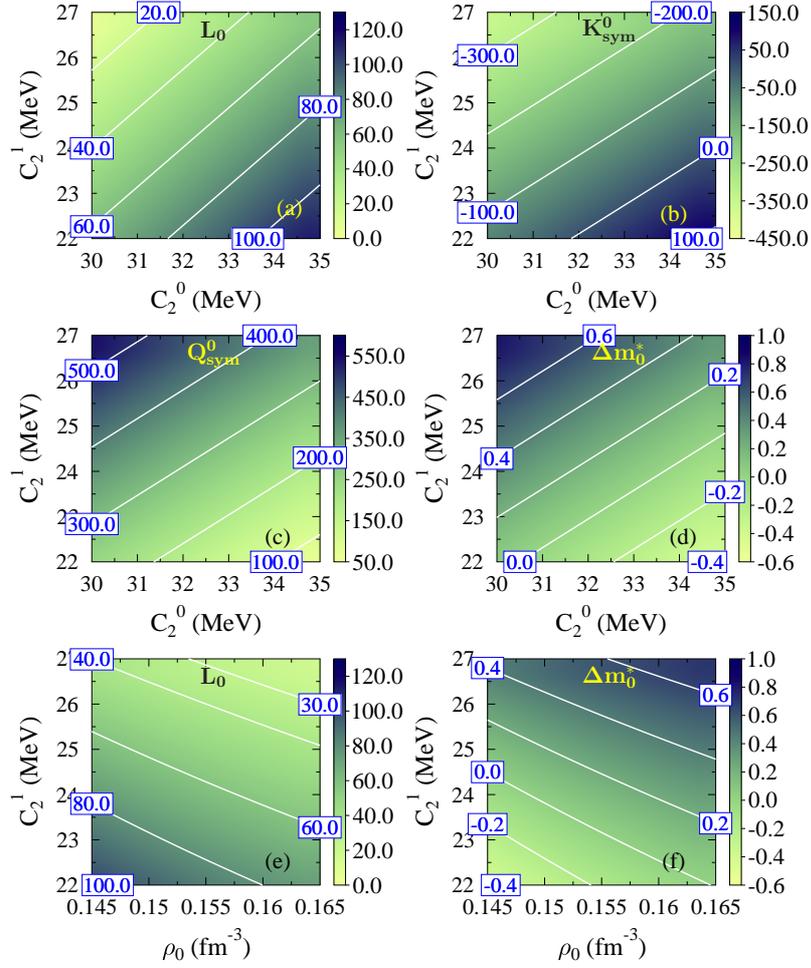}
\caption{\label{ch6_fig9}
Contours of constant $L_0$, $K_{sym}^0$, $Q_{sym}^0$
and $\Delta m_0^*$ in color shades (as indicated on the right side
of each panel) as functions of the input nuclear constants $C_2^0$,
$C_2^1$ and $\rho_0$ depicting the interdependence between various 
symmetry energy elements. The values of $L_0$, $K_{sym}^0$ and 
$Q_{sym}^0$ are in units of MeV and those for $\Delta m_0^*$ are 
in units of the free nucleon mass. For details, see text.} 
\end{figure}

The set of nuclear constants what is chosen in the present work is a
conservative set; depending on possible new experimental inputs, their
values may however change somewhat which would affect the calculated
values of the density derivatives of the symmetry energy coefficients. The
evaluated isovector elements are seen to be quite sensitive to the input
quantities $C_2^0$, $C_2^1$ and $\rho_0$. There is still some variance in
the choice of these input nuclear constants \cite{Zhang13,Gang08,Liu10}
besides the ones we have chosen. The aforesaid sensitivity can be gauged
from the displayed six panels in Fig. \ref{ch6_fig9}. In the upper four
panels (a)-(d), the contours of constant $L_0$, $K_{sym}^0$, $Q_{sym}^0$
and $\Delta m_0^*$ are shown in the $C_2^0-C_2^1$ plane in color shades,
the white lines within the panels are the loci of constant isovector
elements as marked when all other input elements are left unchanged.  With
increase in $C_2^1$, $L_0$ and $K_{sym}^0$ are seen to decrease whereas
$Q_{sym}^0$ and $\Delta m_0^*$ are found to increase. The opposite is
observed for an increase in $C_2^0$. This points out the interdependence
between different symmetry energy elements. The change in $\rho_0$
has also a sizeable effect on the isovector elements.  All other inputs
remaining intact, an increase in $\rho_0$ decreases $L_0$ and $K_{sym}^0$
and increases $Q_{sym}^0$ and $\Delta m_0^*$.  Only glimpses of these
changes are shown in panels (e) and (f), where contours of constant $L_0$
and $\Delta m_0^*$ are drawn in the $\rho_0-C_2^1$ plane. The isovector
elements as studied here are seen to be nearly insensitive to changes in
$e_0$ and $m_0^*$ (not shown here). Similarly, $K_0$ has little effect on
these isovector elements except on $\Delta m_0^*$.  An increase of $K_0$
by, e.g., $\sim $ 30 MeV is seen to push $\Delta m_0^*$ drastically in
the negative domain.  Uncertainties in the input nuclear constants bear
signature on the uncertainties in the calculated isovector elements.

\section{Summary}
To sum up, without  reference to any specific nuclear interaction,
with only a few reasonable approximations, analytic expressions for
the density derivatives of the symmetry energy coefficient $C_2(\rho )$
at the saturation density in terms of  empirical nuclear constants are
found out.  The symmetry observables are seen to be sensitive to the
values of the input nuclear constants, particularly to $C_2^0$, $C_2^1$
and $\rho_0$; precise values of these constants are thus required to
narrow down the uncertainties in the density dependence of the symmetry
energy. In doing the calculations, a correlated structure connecting
the different symmetry energy elements emerged. The consonance of
these structures with those inherent in the plethora of EDFs based on
relativistic and non-relativistic mean-field indicates a universality
in the correlated structure in the symmetry energy coefficients. This
helps further in a better realization of the information content of the
isovector observables.

%% file: chap7.tex
\chapter{Summary and Future outlook}\label{ch7}
In this thesis work, we start from a general introduction to the 
nuclear symmetry energy arising from the asymmetry in the neutron-proton 
content of a nuclear system. Symmetry energy plays crucial roles in 
binding and shaping the finite nuclei as well as neutron stars. 
As the densities associated with finite nuclei and neutron stars are 
widely apart from each other, a microscopic description of symmetry 
energy over a wide range of density is very important. In this respect, 
importance of precise characterization of the properties of infinite 
nuclear matter, specially those which determine the density dependence 
of symmetry energy, is pointed out in the present work. As nuclear matter 
is not accessible in the laboratory, connecting the properties of 
nuclear matter to the observables of finite nuclei and neutron stars 
is very important. To this purpose, mean-field models both relativistic 
as well as non-relativistic are used in the present thesis work. 
In Chapter \ref{ch2}, details on the calculation of ground state 
properties e.g binding energy and charge radii of spherical nuclei are given 
in the mean-field formalism. The particulars are discussed both 
for a relativistic mean-field model and a non-relativistic one based 
on Skyrme force. A formal introduction to the infinite nuclear matter 
properties is also given using both the relativistic and 
non-relativistic frameworks. 

Throughout this thesis work, one of the primary motivation was to 
explore correlations of symmetry energy parameters to the properties 
of finite nuclei and neutron stars. Correlation between two quantities 
can be investigated in two ways: firstly, by exploring the two 
concerned quantities from a set of models or secondly, exploring 
them by means of a single model through a covariance analysis. The 
ingredients of optimizing the parameters of a model and eventually 
performing the covariance analysis is given in Chapter \ref{ch3}. 
By covariance analysis one can also calculate uncertainties in 
various quantities of interest, which gives a clear idea of relevance 
of proposing a new theoretical model. While performing the covariance 
analysis one obtains derivatives of different experimental observables 
of interest with respect to the model parameters, which can be further 
used to study the sensitivity of particular observables to different 
model parameters. Details of this sensitivity analysis is also given 
in Chapter \ref{ch3}.

Binding energies of finite nuclei are the most accurately known 
experimental quantities in nuclear physics. Information on these 
precisely known quantities are exploited in the literature to constrain 
the symmetry energy coefficient quite tightly. However, mean-field models 
obtained by fitting binding energies and charge radii of closed shell 
spherical nuclei show a wide variation in the slope of symmetry 
energy. In Chapter \ref{ch4}, it was identified that slope of symmetry 
energy can be constrained in a narrow range if the binding energies of 
extremely asymmetric nuclei (neutron number twice to that of protons) 
are included in the fit data to optimize the model parameters of a 
relativistic mean-field model. A 
sensitivity analysis was performed further to show quantitatively 
how the experimental data on binding energies of highly asymmetric 
nuclei help to constrain the value of different symmetry energy 
parameters. 

To minimize the energy of an asymmetric nucleus, where the number of 
neutrons is higher than the protons, neutrons are pushed towards the 
surface giving rise to a neutron skin thickness. Droplet Model can account 
for this conclusion, which further suggests that slope of symmetry 
energy should be correlated to neutron skin thickness of a heavy 
nucleus. Exploration of microscopic mean-field models testify for the 
existence of this correlation. In Chapter \ref{ch5}, we point out 
that there might be a hint of model dependence in this correlation. 
Droplet Model provides a prescription for decomposing the neutron 
skin thickness into a bulk and surface part. The degree of model 
dependence in the correlation between slope of symmetry energy and 
neutron skin thickness of a heavy nucleus can be reduced if one 
looks for the correlation involving bulk part of the neutron skin 
thickness instead of total neutron skin thickness. An effective 
value of slope parameter is also suggested for a heavy nucleus in 
Chapter \ref{ch5}, which might be identified better with the 
experimental information on neutron skin thickness of a heavy nucleus. 

In chapter \ref{ch6}, we start with some basic equations of statistical 
mechanics and arrive at the energy density functional of infinite 
nuclear matter with some reasonable assumptions. Analytical relations 
for different symmetry energy parameters are derived further, which 
show a dependence of higher order of symmetry energy parameters 
on the lower order ones. These inter-relationships are verified using 
500 different mean-field models existing in the literature. Specially 
the correlation between curvature parameter of symmetry energy with 
linear combination of symmetry energy coefficient and its slope 
parameter is found to be a universal one. 

In the present thesis work, parameters controlling the density 
dependence of symmetry energy are constrained by looking into 
different perspectives. A special attention is given to binding 
energies of highly asymmetric nuclei to constrain the slope parameter 
of the symmetry energy. This is realized by applying covariance 
analysis on a relativistic mean-field model. Similar analyses should 
be performed with other type of mean-field models e.g. non-relativistic 
mean-field models based on Skyrme or Gogny force. It might clarify the 
robustness of the conclusions made in the present work with a particular 
variant of mean-field model. 

The analytical relations we have derived using a simplistic model in 
Chapter \ref{ch6} is followed by well tested Skyrme formalism. 
The quantities like isovector splitting of nucleon effective mass, 
which acquires a large range of values across different theoretical 
models, can be measured experimentally in near-future. The model we 
propose can be tested to explain the isovector splitting of nucleon 
effective mass in a simple way. One may also think of extending this 
formalism to explain the properties of finite nuclei, which might 
provide an alternative view of the finite nuclei in comparison to the 
modern mean-field models.